\documentclass[%
superscriptaddress,
preprint,
 amsmath,amssymb,
 aps,
]{revtex4-1}

\usepackage{graphicx}
\usepackage{dcolumn}
\usepackage{bm}

\usepackage[colorlinks=true,linkcolor=blue,citecolor=blue]{hyperref}

\usepackage{orcidlink}
\newcommand{\orcidJuan}{0000-0003-3756-5016}
\newcommand{\orcidJorge}{0009-0009-7860-0608}
\newcommand{\orcidAlberto}{0000-0002-9681-103X}
\newcommand{\orcidLuis}{0000-0001-7176-5674}
\newcommand{\orcidSalvador}{0000-0001-5265-7360}

\begin{document}


\title{Highly-enhanced propagation of long-range kinks in heterogeneous media}

\author{\textcopyright Jorge A. Gonz\'{a}lez\orcidlink{\orcidJorge}
}
\affiliation{Department of Physics, Florida International University, Miami, Florida 33199, U.S.A.}

\author{Alberto Bellor\'{i}n\orcidlink{\orcidAlberto}
}
\affiliation{Escuela de F\'{\i}sica, Facultad de Ciencias, Universidad Central de Venezuela, AP 47586, Caracas 1041-A, Venezuela}

\author{Luis E. Guerrero\orcidlink{\orcidLuis}
}
\affiliation{Departamento de F\'{\i}sica, Universidad Sim\'{o}n Bol\'{\i}var, Apartado 89000, Caracas 1080-A, Venezuela}

\author{Salvador Jim\'{e}nez\orcidlink{\orcidSalvador}
}
\affiliation{Departamento de Matem\'{a}tica Aplicada a las TIC, ETSI Telecomunicaci\'{o}n, Universidad Polit\'{e}cnica de Madrid, 28040-Madrid, Spain}

\author{Juan F. Mar\'in\orcidlink{\orcidJuan}}
 \email{j.marinm@utem.cl}
\affiliation{Departamento de Física, Facultad de Ciencias Naturales, Matemática y del Medio Ambiente, Universidad Tecnológica Metropolitana, Las Palmeras 3360, Ñuñoa 780-0003, Santiago, Chile.}%




\date{\today}

\begin{abstract}
We investigate a field-theoretical model that describes the interaction between kinks and antikinks and between kinks and other heterogeneous fields and impurities. We show that the long-range kink can tunnel through a barrier created by heterogeneous fields and impurities even when the energy of the center of mass of the kink is less than the height of the energy barrier. We also study the conditions under which the kink can pass freely through a disordered medium. We introduce the concept of ``effective translational symmetry''. We compare our results with those from recent papers published in this journal, where the Bogomol’nyi-Prasad-Sommerfield property is discussed.

\vspace{1cm}
Preprint accepted in \textbf{Journal of High Energy Physics}, September 2024.

\end{abstract}

\maketitle


\section{Introduction}
\label{Sec:Intro}

Self-dual impurity models have gained significant interest in the recent literature \cite{Adam2019, Adam2019-2, Adam2018, Adam2019-3}. In Ref.~\cite{Adam2019}, the $\phi^4$ model is coupled to an impurity in a way that preserves one-half of the so-called Bogomol’nyi-Prasad-Sommerfield (BPS) property. Although the impurity breaks the translational invariance, Adam \textit{et al.}  argue that this symmetry is somehow restored in the BPS sector, where the energy of the soliton-impurity solution does not depend on their mutual distance. Furthermore, they discuss the existence of certain ``generalized translational symmetry'' that provides a motion on moduli space that transforms one BPS solution into another. This avenue of research aims to formulate a model that allows for domain walls that do not get stuck in impurities.

The existence of the generalized translational symmetry would imply that the binding energy between the kink and the impurity is zero, and the kink may be translated freely through the wire, although it changes its shape during this translation. The impurity is coupled to different terms in the Lagrangian in a non-conventional way. Therefore, soliton-impurity solutions with zero binding energy will exist, which results in a zero static force between these solitons and the impurity. A natural question is whether it is possible to realize such a symmetry generalization experimentally.

In the system discussed in Refs.~\cite{Adam2019, Adam2019-2}, infinitely many energetically equivalent solutions describe a kink at an arbitrary distance from the impurity. However, and this is a crucial point, when the kink moves, it is no longer a BPS state. Adam \textit{et al.} argue that at low speed, the domain wall dynamics may be approximated by a sequence of BPS states (i.e., a geodesic motion on the moduli space) describing an adiabatic motion. Even if this model cannot be realized experimentally, the mere formulation of the problem is an important step.

Another recent development is the explosion of papers dedicated to kinks with power-law tails (see, for example, Refs.~\cite{Christov2019, Campos2021, Kumar2021, Khare2021,  Khare2022-2}). Gonz\'{a}lez and Estrada-Sarlabous \cite{Gonzalez1989} investigated the behavior of kink-antikink long-range interaction forces for the first time. After that, several papers presented applications of the original results \cite{Guerrero1997, Gonzalez1994, Mello1998, Gonzalez2002}. A whole ``long-range'' movement is gaining strength throughout the physics community \cite{ Gonzalez1994, Gonzalez1996-2, Gonzalez1996-3, Guerrero1997, Guerrero1998, Gonzalez1998, Mello1998, Gonzalez1999-2, Gonzalez2002, Bazeia2018, Manton2019, Christov2019, dOrnellas2020, Andrade2020, Campos2021, Kumar2021, Khare2021, Khare2022, Khare2022-2, Manton2023, Bazeia2023, Bazeia2023-2, Andrade2023}. Many researchers are discovering new properties and applications of long-range solitons \cite{Campos2021, Kumar2021, Khare2021,  Khare2022-2}.

This article will show that long-range kinks can pass freely through a barrier that will be effectively ``transparent'' if some conditions are provided. Also, we will find the conditions under which the long-range kinks can move through an utterly disordered medium as if there are no obstacles.  Finally, we will introduce the concept of ``effective translational symmetry'' based on our findings.  Thus, we will be combining ideas from these two current developments: (i) how to design a wire where impurities are effectively transparent for topological defects and (ii) kinks with long-range interactions. The article is organized as follows. In Section~\ref{Sec:The models}, we briefly review the models of long-range kinks that will be the focus of this study. In Section~\ref{Sec:Implications} we overview the implications of the long-range property in the physics of quantum kinks, Kac-Baker interactions, non-local Josephson electrodynamics, and kink tunneling. In Section~\ref{Sec:NatureTunneling}, we explain all the theoretical aspects of the phenomenon of long-range kink tunneling. The behavior of the kink is sensitive to the presence of a heterogeneous field $F(x)$ and its topological properties: the intervals of $x$ where $F(x)<0$, $F(x)>0$, the points where $F(x)=0$, etc. The interaction of the long-range kinks with the negative structures of $F(x)$ is crucial for understanding long-range tunneling. Among all the theoretical results presented in this Section, two models are investigated in deeper detail: one with long tails and another without long tails. In Section~\ref{Sec:Tunneling}, we show in numerical simulations that the tunneling of kinks can be highly enhanced by increasing their long-range character. We show in Section~\ref{Sec:Disorder} that long-range kinks can travel even through a disordered distribution of impurities. In Section~\ref{Sec:Applications}, we discuss experimental and technological applications. In Section~\ref{Sec:Discussion}, we give a final discussion of our results and conclude in Section~\ref{Sec:Conclusions}.

\section{The models}
\label{Sec:The models}

\subsection{Modeling long-range kinks in interaction with other fields}
\label{Sec:Models}

\begin{figure}
    \centering
    \scalebox{0.26}{\includegraphics{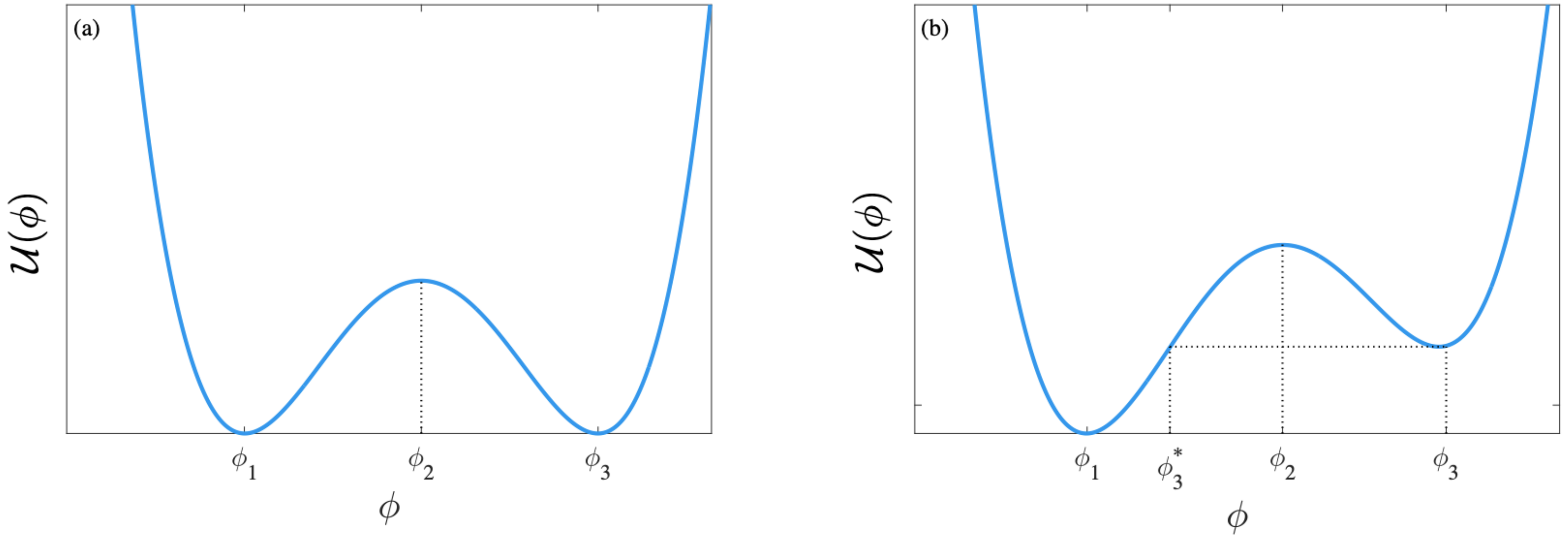}}
    \caption{Bistable potentials supporting kink-antikink solutions connecting \textbf{(a)} two degenerate stable states $\phi_1$ and $\phi_3$, and \textbf{(b)} one global minimum $\phi_1$ (true-vacuum state) and one metastable state $\phi_3$ (false-vacuum state). In both cases, stable states are separated by a barrier at $\phi=\phi_2$.}
    \label{Fig:01}
\end{figure}

In this section, we introduce the field equations that will be studied in the present article and briefly review their properties. Consider a real scalar field $\phi(x,\,t)$, whose dynamic is described by the Lagrangian density
\begin{equation}
    \label{Eq:01}
    \mathcal{L}=\frac{1}{2}\left(\frac{\partial\phi}{\partial t}\right)^2-\frac{1}{2}\left(\frac{\partial\phi}{\partial x}\right)^2-\mathcal{U}(\phi).
\end{equation}
The field equation will be
\begin{equation}
\label{Eq:02}
\frac{\partial^2\phi}{\partial t^2}-\frac{\partial^2\phi}{\partial x^2}=-\frac{\hbox{d} \mathcal{U}(\phi)}{\hbox{d}\phi}.
\end{equation}
The potential $\mathcal{U}(\phi)$ is an analytical function of $\phi$ having three local extrema: two minima at $\phi_1$ and $\phi_3$, and a maximum at $\phi_2$ with $\phi_1<\phi_2<\phi_3$, as shown in Fig.~\ref{Fig:01}. If the potential $\mathcal{U}(\phi)$ fulfills the condition $\mathcal{U}(\phi_1)=\mathcal{U}(\phi_3)$, Eq.~\eqref{Eq:02} has solutions of kink and antikink types which can move with any constant velocity $0\leq|v|\leq 1$.

The existence of a difference between $\mathcal{U}(\phi_1)$ and $\mathcal{U}(\phi_3)$ (i.e., $\Delta:=\mathcal{U}(\phi_1)-\mathcal{U}(\phi_3)\neq0$) is equivalent to the action of an effective external force on the solitons \cite{Gonzalez1987, Marin2021}. If $\Delta>0$ ($\Delta<0$), there is a force acting on the kink in the negative (positive) direction of the $x$-axis. When the initial state is that of an antikink, the existence of $\Delta>0$ ($\Delta<0$) implies the action of a force on the antikink in the positive (negative) direction of the $x$-axis.

\begin{figure}
    \centering
    \scalebox{0.26}{\includegraphics{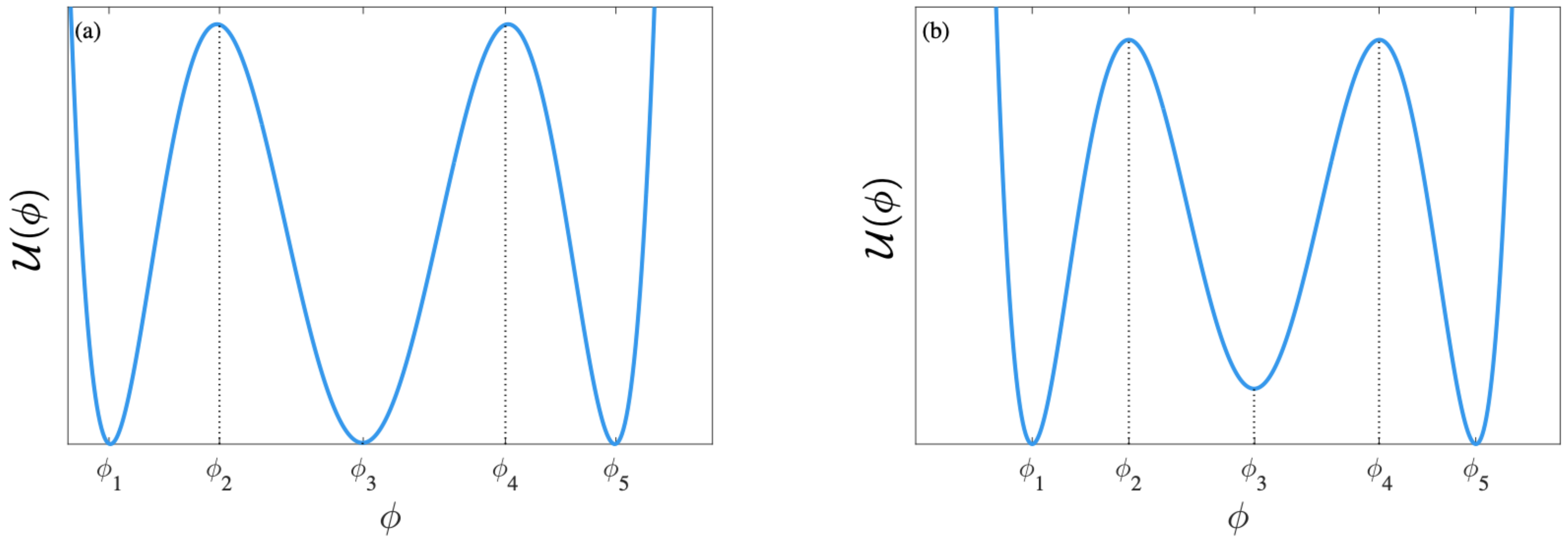}}
    \caption{Potentials with three stable states, $\phi_1$, $\phi_3$, and $\phi_5$, separated by two barriers at $\phi_2$ and $\phi_4$. In \textbf{(a)}, the minima are degenerated, whereas in \textbf{(b)}, the minimum at $\phi=\phi_3$ has more potential energy than the other two.}
    \label{Fig:02}
\end{figure}

\begin{figure}
    \centering
    \scalebox{0.26}{\includegraphics{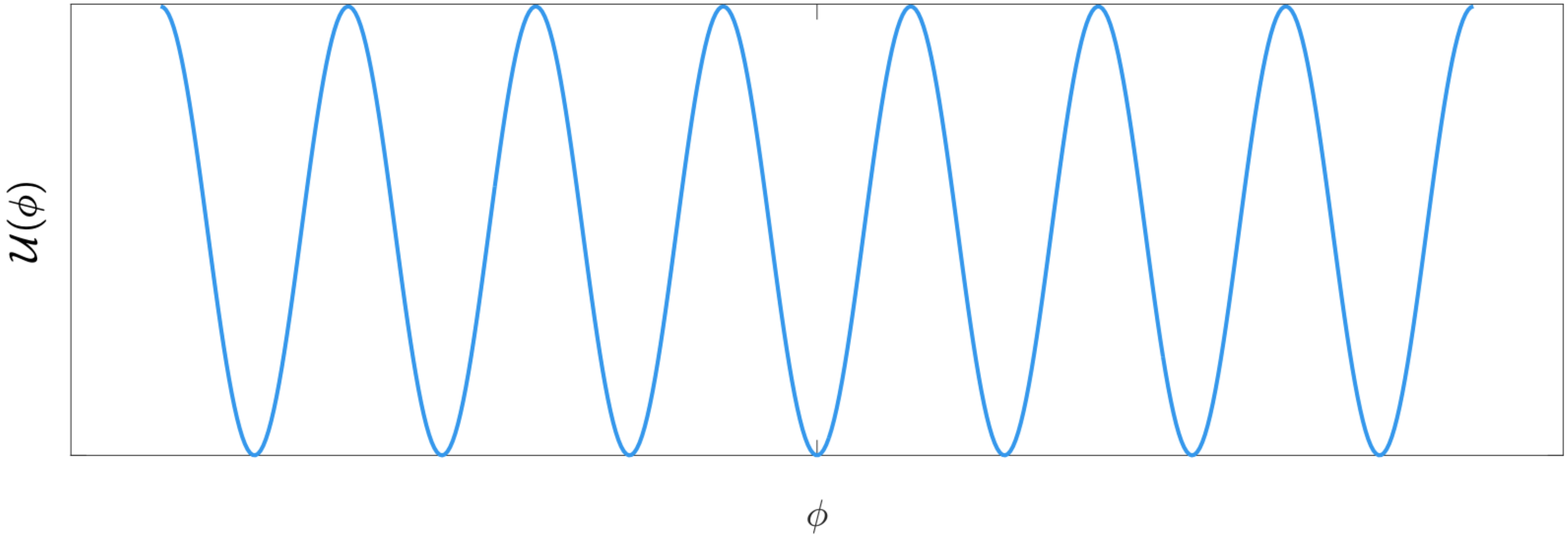}}
    \caption{Periodic sine-Gordon-like potential with multiple degenerate equilibrium states separated by multiple barriers.}
    \label{Fig:03}
\end{figure}

For some of the phenomena investigated in the present work, we will consider that $\mathcal{U}(\phi)$ has three minima and two maxima (see Fig.~\ref{Fig:02}). Some of our results are obtained for a sine-Gordon-like periodic potential $\mathcal{U}(\phi)$ (see Fig.~\ref{Fig:03}). All these potentials are known to support kink and antikink solutions connecting the equilibrium states of the underlying potential. We will also investigate the motion of kinks in heterogeneous media, which can be pursued considering the following modification of Eq.~\eqref{Eq:02},
\begin{equation}
    \label{Eq:03}
    \frac{\partial^2\phi}{\partial t^2}-\frac{\partial^2\phi}{\partial x^2}+\frac{\hbox{d} \mathcal{U}(\phi)}{\hbox{d}\phi}=F(x).
\end{equation}
where $F(x)$ accounts for the presence of heterogeneous fields. Similar results can be obtained for parametrically perturbed equations like the following,
\begin{equation}
\label{Eq:ParametricalModel}
\frac{\partial^2\phi}{\partial t^2}-\frac{\partial^2\phi}{\partial x^2}+\left[1+g(x)\right]\frac{\mbox{d}\mathcal{U}(\phi)}{\mbox{d}\phi}=0,
\end{equation}
where the role of the zeros of $F(x)$ in Eq.~\eqref{Eq:03} is played by the local maxima and minima of $g(x)$ \cite{Gonzalez2007}. Most of our results are general and can be applied to any equation of type~\eqref{Eq:03} if the potential $\mathcal{U}(\phi)$ possesses the properties discussed in this work. Our sound theoretical findings in general models will be illustrated mostly in numerical experiments of the following equations,
\begin{equation}
    \label{Eq:29}
    \frac{\partial^2\phi}{\partial t^2} +b\frac{\partial\phi}{\partial t} -\frac{\partial^2\phi}{\partial x^2}+\frac{1}{2}\phi(\phi^2-1)^{2n-1}=F(x),
\end{equation}
and
\begin{equation}
    \label{Eq:30}
    \frac{\partial^2\phi}{\partial t^2} +b\frac{\partial\phi}{\partial t}-\frac{\partial^2\phi}{\partial x^2}+2\sin^{2n-1}\left(\frac{\phi}{2}\right)\cos\left(\frac{\phi}{2}\right)=F(x),
\end{equation}
where $b\partial\phi/\partial t$ is a dissipative term and $b$ is the damping coefficient. The potential for Eq.~\eqref{Eq:29} is $\mathcal{U}(\phi)=(1/8n)(\phi^2-1)^{2n}$ and for Eq.~\eqref{Eq:30} is $\mathcal{U}(\phi)=(2/n)\sin^{2n}(\phi/2)$. When $n=1$, Eq.~\eqref{Eq:29} will recover the $\phi^4$ equation, and Eq.~\eqref{Eq:30} will be equivalent to the sine-Gordon model.

The mass of a kink can be calculated using the formula
\begin{equation}
\label{Eq:MassKink}
m=\int_{\phi_1}^{\phi_3}\mbox{d}\phi\sqrt{2\mathcal{U}(\phi)}.
\end{equation}
For the potential $\mathcal{U}(\phi)=(1/8n)(\phi^2-1)^{2n}$, the kink's mass can be calculated using the explicit formula
\begin{equation}
    \label{Eq:40}
    M_k(n)=\frac{1}{2\sqrt{n}}\left[\frac{2^{2n+1}(n!)^2}{(2n+1)!}\right].
\end{equation}
As $n$ is increased, the kink's mass is decreased. The numerical values of the kink's mass both for potential $\mathcal{U}(\phi)=(1/8n)(\phi^2-1)^{2n}$ and $\mathcal{U}(\phi)=(2/n)\sin^{2n}(\phi/2)$ are shown in Table~\ref{Tab:01} and Table~\ref{Tab:02}.

\begin{table}[]
    \centering
    \begin{tabular}{||r|r||r|r||r|r||}
    \hline\hline
      $n$   & $m$ & $n$   & $m$ & $n$   & $m$ \\
      \hline\hline
        1   & 0.666667 & 15 & 0.0576543 & 29 & 0.0301714 \\
        \hline
        2   & 0.377124  & 16 & 0.0541319 & 30 &0.0291779 \\
        \hline
        3   & 0.263932 & 17 & 0.0510152 & 31 & 0.0282479 \\
        \hline
        4   & 0.203175 & 18 & 0.0482379 & 32 & 0.0273752 \\
        \hline
        5   & 0.165204 & 19 & 0.0457475 & 33 & 0.0265549 \\
        \hline
        6   & 0.139210 & 20 & 0.0435016 & 34 & 0.0257823 \\
        \hline
        7   & 0.120291 & 21 & 0.0414659 & 35 & 0.0250535 \\
        \hline
        8   & 0.105903 & 22 & 0.0396123 & 36 & 0.0243646 \\
        \hline
        9   & 0.0945911 & 23 & 0.0379173 & 37 & 0.0237127 \\
        \hline
        10   & 0.0854638 & 24 & 0.0363614 & 38 & 0.0230947 \\
        \hline
        11   & 0.0779436 & 25 & 0.0349282 & 39 & 0.0225081 \\
        \hline
        12   & 0.0716403 & 26 & 0.0336037 & 40 &0.0219506 \\
        \hline
        13   & 0.0662805 & 27 & 0.032376 & 41 & 0.0214201 \\
        \hline
        14   & 0.0616671 & 28 & 0.0312348 &  & \\
        \hline
    \end{tabular}
    \caption{Kink's mass for $\mathcal{U}(\phi)=(1/8n)(\phi^2-1)^{2n}$, following the formula of Eq.~\eqref{Eq:MassKink}.}
    \label{Tab:01}
\end{table}

\begin{table}[]
    \centering
    \begin{tabular}{||r|r||r|r||r|r||}
    \hline\hline
      $n$   & $m$ & $n$   & $m$ & $n$   & $m$ \\
      \hline\hline
        1   & 8 & 15 & 0.657394 & 29 & 0.342775 \\
        \hline
        2   & 4.44288  & 16 & 0.616948 & 30 & 0.331444\\
        \hline
        3   & 3.0792 & 17 & 0.58119 & 31 & 0.320838 \\
        \hline
        4   & 2.35619 & 18 & 0.549349 & 32 & 0.310891 \\
        \hline
        5   & 1.90811 & 19 & 0.520816 & 33 & 0.301541 \\
        \hline
        6   & 1.60319 & 20 & 0.495101 & 34 & 0.292737 \\
        \hline
        7   & 1.38227 & 21 & 0.471805 & 35 & 0.284433 \\
        \hline
        8   & 1.21485 & 22 & 0.450603 & 36 & 0.276587 \\
        \hline
        9   & 1.0836 & 23 & 0.431224 & 37 & 0.269162 \\
        \hline
        10   & 0.977936 & 24 & 0.413443 & 38 & 0.262126 \\
        \hline
        11   & 0.891047 & 25 & 0.397071 & 39 & 0.255448 \\
        \hline
        12   & 0.818335 & 26 & 0.381946 & 40 & 0.249101 \\
        \hline
        13   & 0.756594 & 27 & 0.367931 & 41 & 0.243063 \\
        \hline
        14   & 0.703515 & 28 & 0.354907 &  & \\
        \hline
    \end{tabular}
    \caption{Kink's mass for $\mathcal{U}(\phi)=(2/n)\sin^{2n}(\phi/2)$, following the formula of Eq.~\eqref{Eq:MassKink}.}
    \label{Tab:02}
\end{table}

The behavior of the kink in the tails (i.e., the asymptotic behavior of kinks at infinity) governs how the kink interacts with other structures and fields placed at some distance from the kink. The most studied systems in the literature, such as the $\phi^4$ and the sine-Gordon models, have the property that the tails of kink have an exponential behavior at infinity. Such exponential behavior is deeply rooted in the fact that the kink is connecting \emph{Morse critical points}: in the Taylor expansion of the potential $\mathcal{U}(\phi)$ in a neighborhood of the critical point, the first term different from zero is of the second degree. Nevertheless, when the first term different from zero in the Taylor expansion is of higher order, kink solutions tend in a slower way to the critical values, giving rise to the phenomenon of long-range interactions between kinks \cite{Gonzalez1989}.

\subsubsection{Long-range kink-antikink interactions}

\begin{figure}
    \centering
    \scalebox{0.35}{\includegraphics{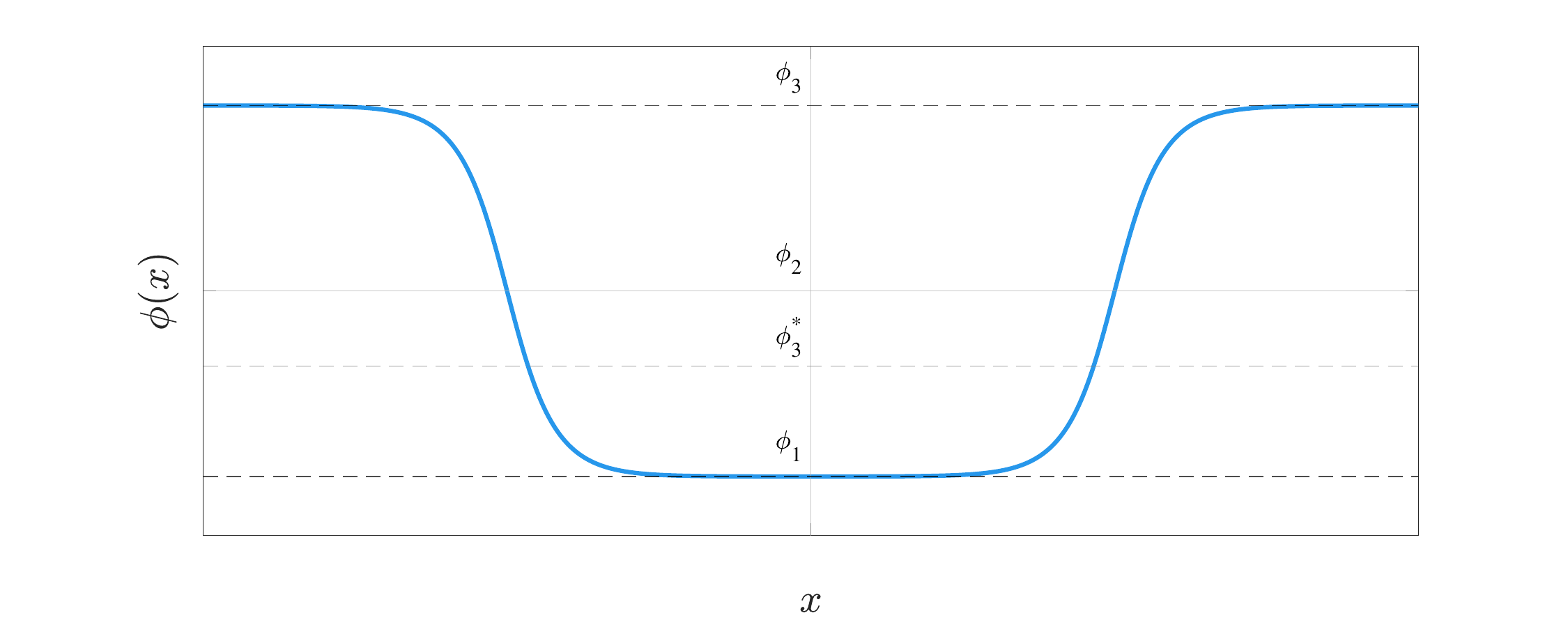}}
    \caption{Antikink-kink equilibrium state forming a bell-like solution of Eq.~\eqref{Eq:01}. The function $\Delta(d)$ defined in the text, where $d$ is the distance separating the inflection points, gives a measure of the kink-antikink interaction force.}
    \label{Fig:04}
\end{figure}

Let the degree of the first term different from zero in the Taylor expansion of $\mathcal{U}(\phi)$ in the neighborhood of the minima $\phi_1$ and $\phi_3$ be $2n$ (with $n>1$). Then, for large values of $x$, it is known that the solution fulfills the relation \cite{Gonzalez1989}
\begin{equation}
\label{Eq:10}
\phi-\phi_j\sim x^k,
\end{equation}
where $k=1/(1-n)$, with $j=1,3$. For $\Delta\neq0$, Eq.~\eqref{Eq:02} has stationary states in the form of bell-shaped solutions, as shown in Fig.~\ref{Fig:04}, which are kink-antikink equilibrium states \cite{Gonzalez1987}.
These bell-like solutions are known as sphalerons and play an important role in many branches of physics (see, e.g., Ref.~\cite{Klinkhamer1984}). These states exist due to a balance between the mutual attracting force between the kink and the antikink, which, at a certain critical distance, is balanced by the effective external force (generated by the fact that $\Delta\neq0$) acting on the kink and antikink in opposite directions \cite{Gonzalez1987, Gonzalez1989}. Similar arguments explained the formation and stability of similar bell-like solutions in one \cite{Gonzalez2006} and two spatial dimensions \cite{GarciaNustes2017, Marin2018, Castro-Montes2020, Marin2021}. The bell solutions are unstable equilibrium states. However, the fact that this equilibrium configuration exists (at least mathematically) allows calculating the distance $d$ between the inflection points of the function $\phi(x)$ describing the bell, which can be obtained from the formula \cite{Gonzalez1987, Gonzalez1989}
\begin{equation}
   \label{Eq:09}
d=\sqrt{2}\int_{\phi_2}^{\phi_3^*}\frac{\mbox{d}\phi}{\sqrt{\mathcal{U}(\phi)-\mathcal{U}(\phi_3)}}.
\end{equation}
Equation~\eqref{Eq:09} implicitly contains the function $\Delta(d)$, which measures the attracting interaction force between the kink and the antikink.

We are interested in bell-like solutions like that shown in Fig.~\ref{Fig:04}, which represent a stationary state of a kink and an antikink, where these excitations have the same asymptotic behavior as the free kinks and antikinks existing when $\Delta=0$. We assume that the perturbation of the potential $\mathcal{U}(\phi)$, which produced the difference $\Delta\neq0$, holds without alterations of the order $2n$ of the minima. Without loss of generality, we may suppose that $\phi_1=0$ and $\phi_3=1$ (this can be obtained by making an affine transformation $\phi\to(\phi-\phi_1)/(\phi_3-\phi_1)$). In these new coordinates, the difference $\mathcal{U}(\phi_1)-\mathcal{U}(\phi_3)$ is $\Delta_1\sim\Delta$. We will obtain an evaluation of the integral in Eq.~\eqref{Eq:09} for small $\Delta_1$ by considering the integral
\begin{equation}
\label{Eq:11}
d^*=\sqrt{2}\int_{\phi_2}^{\phi_3^*}\frac{\mbox{d}\phi}{\sqrt{-p_{4n}(\phi)}},
\end{equation}
where $p_{4n}(\phi)$ is the polynomial of degree $4n$ in the variable $\phi$ interpolating $\mathcal{U}(\phi)$ $2n$ times in $\phi_1$ and $\phi_3$, respectively, and once in $\phi_3^*$, that is
\begin{align*}
    p_{4n}^{(j)}(0)=0\quad&\text{for }j=0,\,1,\,\ldots,\,2n-1,\\
    p_{4n}^{(j)}(1)=0\quad&\text{for }j=1,\,2,\,\ldots,\,2n-1,\\
    p_{4n}(\phi_3^*)=0\quad&\text{for }p_{4n}(1)=\Delta_1.
\end{align*}
These conditions determine uniquely the polynomial $p_{4n}(\phi)$, which also fulfills
\begin{equation}
    \label{Eq:P6**2}
    p_{4n}(\phi)=\phi^{2n}(\phi_3^*-\phi)p_{2n}(\phi),
\end{equation}
with
\begin{equation}
    \label{Eq:P6**3}
    \phi_3-\phi_3^*=\mathcal{O}\left[\left(-\Delta_1\right)^{\frac{1}{2n}}\right],
\end{equation}
\begin{equation}
    \label{Eq:P6**4}
    \phi_2-\frac{1}{2}=\mathcal{O}\left[\left(-\Delta_1\right)^{\frac{1}{2n}}\right],
\end{equation}
and
\begin{equation}
    \label{Eq:P6**5}
    p_{2n}(\phi)-\left(1-\phi\right)^{2n-1}=\mathcal{O}\left[\left(-\Delta_1\right)^{\frac{1}{2n}}\right]q_{2n}(\phi),
\end{equation}
where $q_{2n}(\phi)$ is a polynomial in $\phi$ of degree $2n$. Thus, for $0<-\Delta_1\ll 1$, the integral in Eq.~\eqref{Eq:11} is
\begin{equation}
    \label{Eq:P6**6}
d^* \simeq K\int_{1/2}^{\phi_3^*}\frac{\mbox{d}\phi}{\sqrt{\left(\phi_3^*-\phi\right)(1-\phi)^{2n-1}}}=d^{**},
\end{equation}
where $K$ is a constant. Consider the hyperelliptic curve $\psi^2=(\phi_3^*-\phi)(1-\phi)^{2n-1}$. After introducing the new coordinates $u:=1/(1-\phi)$ and $\nu:=\psi/(1-\phi)^n$, we obtain the curve $\nu^2=(\phi_3^*-1)u+1$. We can write the integral in Eq.~\eqref{Eq:P6**6} in terms of the new coordinates as
\begin{equation}
    \label{Eq:P6**7}
d^{**}=K\int_{1/2}^{\phi^*_3}\frac{\mbox{d}\phi}{\psi(\phi)}=K\int_{2}^{1/(1-\phi^*_3)}\frac{u^{n-2}}{\nu}\mbox{d}u.
\end{equation}
The above integral can be calculated analytically,
\begin{equation*}
    \int_{2}^{1/(1-\phi^*_3)}\frac{u^{n-2}}{\nu}\mbox{d}u=\left\{\begin{array}{ll}
    \displaystyle \ln\left|\frac{(1-\phi^*_3)u}{(1+\nu)^2}\right|,\quad&\text{for } n=1,\\
    \displaystyle\frac{2\nu}{(1-\phi^*_3)^{n-1}} \sum_{k=0}^{n-2}(-1)^k {{n-2}\choose{k}}\left[\frac{\nu^{2n-4-2k}}{2n-3-2k}\right],\quad&\text{for }n>1.
    \end{array}
    \right.
\end{equation*}
Using Eq.~\eqref{Eq:P6**3}, we get

\begin{equation}
\label{Eq:12}
    d^{**}=\left\{\begin{array}{ll}\displaystyle\mathcal{O}\left[-\ln(-\Delta_1)\right],\quad&\text{for }n=1,\\
    \displaystyle\mathcal{O}\left[ (-\Delta _{1})^{\frac{1-n}{2n}}\right],\quad&\text{for }n>1.
    \end{array}\right.
\end{equation}

For potentials $\mathcal{U}(\phi)$ in which the minima are parabolic, the interaction force decreases exponentially with the increase of the distance. On the other hand, when the minima of $\mathcal{U}(\phi)$ behave as $\mathcal{U}(\phi)\sim(\phi-\phi_i)^{2n}$, $n>1$, the interaction force decreases with distance as a power law
\begin{equation}
    \label{Eq:13}
    \Delta\sim d^{\frac{2n}{1-n}}.
\end{equation}
Kink-antikink interactions have a long-range character in potentials where $n>1$. The interaction force behaves as $\Delta\sim d^{-2}$ when $n\to\infty$.

\subsubsection{Long-range kink-kink interaction}

\begin{figure}
    \centering
    \scalebox{0.35}{\includegraphics{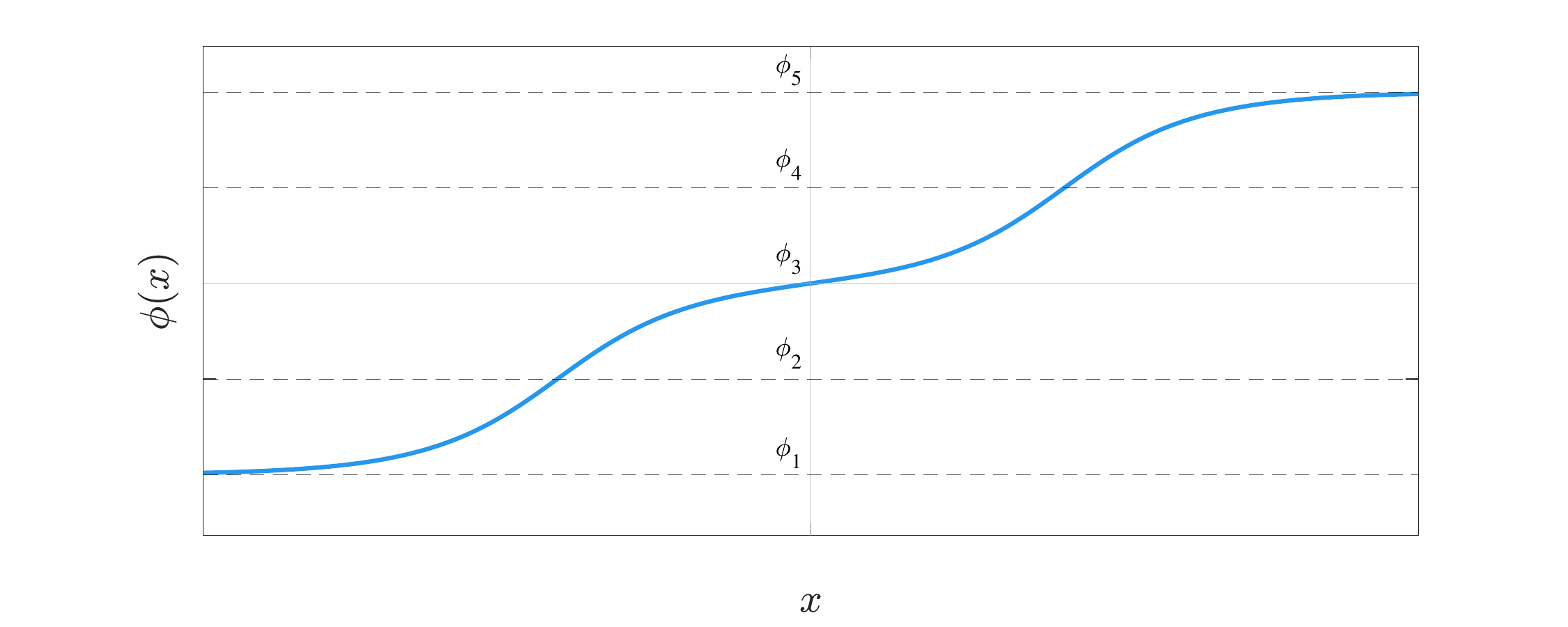}}
    \caption{Kink-kink equilibrium state of Eq.~\eqref{Eq:01} when the underlying potential have three minima or more. The function $\Delta(d)$ gives a measure of the kink-antikink interaction force.}
    \label{Fig:05}
\end{figure}

When we have a potential with three minima or more, as shown in figures~\ref{Fig:02} and~\ref{Fig:03}, kink-kink structures are possible (see Fig.~\ref{Fig:05}). In the case of Fig.~\ref{Fig:02}a, the two kinks are repealing each other, and they will just move away. However, the situation depicted in Fig.~\ref{Fig:02}b leads to the existence of a kink-kink stationary structure, which is stable. Define
\begin{subequations}
    \label{Eq:14}
    \begin{equation}
        \label{Eq:14a}
        \Delta_{13}:=\mathcal{U}(\phi_1)-\mathcal{U}(\phi_3),
    \end{equation}
        \begin{equation}
        \label{Eq:14b}
        \Delta_{35}:=\mathcal{U}(\phi_3)-\mathcal{U}(\phi_5).
    \end{equation}
\end{subequations}
We assume that $\vert\Delta\vert=\vert\Delta_{13}\vert=\vert\Delta_{35}\vert$. Let us call kink$_{13}$ the soliton that connects the vacua $\phi_1$ and $\phi_3$ ($\phi(x)\to\phi_3$ as $x\to\infty$ and $\phi(x)\to\phi_1$ as $x\to-\infty$), and let us call kink$_{35}$ the soliton that connects the vacua $\phi_3$ and $\phi_5$ ($\phi(x)\to\phi_5$ as $x\to\infty$ and $\phi(x)\to\phi_3$ as $x\to-\infty$). The fact that $"\Delta_{13}  < 0$ and $\Delta_{35} > 0$ means that there exists the equivalent of an external force acting on the kink$_{13}$ to the ``right'' and there exists an external force acting on the kink$_{35}$ to the ``left''. Recall that kink$_{13}$ and kink$_{35}$ repel each other with a force that depends on the distance between their centers of mass. There is always a distance $d$ at which the two forces acting on kink$_{13}$ and the two acting on kink$_{35}$ cancel out.

Function $\Delta(d)$ will provide the law that governs the kink-kink interactions. To calculate $\Delta(d)$, we need a modified potential $\mathcal{U}(\phi)$ like the one shown in Fig.~\ref{Fig:02}b. The stable stationary kink-kink structure shown in Fig.~\ref{Fig:05} due to the potential from Fig.~\ref{Fig:02}b will provide the function $\Delta(d)$. The stationary kink-kink structure is a solution to the equation
\begin{equation}
    \label{Eq:15}
    \frac{\mbox{d}^2\phi}{\mbox{d}x^2}-\frac{\mbox{d}\mathcal{U}}{\mbox{d}\phi}=0.
\end{equation}
This expression can be integrated,
\begin{equation}
    \label{Eq:16}
    \frac{1}{2}\left(\frac{\mbox{d}\phi}{\mbox{d}x}\right)^2-\mathcal{U}\left(\phi(x)\right)=\Delta.
\end{equation}
Thus, the equation that connects $\Delta$, $d$, and the shape of the kink-kink stationary structure is the following,
\begin{equation}
    \label{Eq:17}
d=\frac{1}{\sqrt{2}}\int_{\phi_3}^{\phi_4}\frac{\mbox{d}\phi}{\sqrt{\mathcal{U}(\phi)+\left|\Delta\right|}}.
\end{equation}
We recall that the original potential behaves asymptotically as $\mathcal{U}(\phi)\sim\phi^{2n}$ as $\phi\to\phi_3=0$. Therefore,
\begin{equation}
\label{Eq:P8**1}
    \left|\Delta\right|^{(1-n)/2n}\sim d,\quad\text{for }n>1,
\end{equation}
and
\begin{equation}
    \label{Eq:P8**2}
    d=\mathcal{O}\left[-\ln\left(|\Delta|\right)\right],\quad\text{for }n=1.
\end{equation}
Thus, for $n=1$, the kink-kink repulsive force decays exponentially with distance. For $n>1$,
\begin{equation}
\label{Eq:P8**3}
    \Delta\sim d^{2n/1-n}.
\end{equation}
In potentials where $n>1$, the kink-kink repulsive force decreases with distance as the power law given in Eq.~\eqref{Eq:13}. For the kink-kink repulsion, the force behaves as $\Delta\sim d^{-2}$ when $n\to\infty$ as before. The coefficient is larger for the kink-kink interaction than for the kink-antikink interaction. It depends on the specific potential $\mathcal{U}(\phi)$. As d'Ornellas has shown \cite{dOrnellas2020}, the modified $\mathcal{U}(\phi)$ method can yield the exact coefficient. So when we use the concrete potential in the integral \eqref{Eq:17}, we can obtain the correct coefficients. We have checked all these calculations using the collective-coordinate calculation of the acceleration and the modified potential techniques. Since our formulas are general, our calculations do not have contradictions after considering other particular cases \cite{Manton2019, Christov2019, dOrnellas2020}.

The tunneling problem can be solved using energy and net-force considerations: the net force acting on the long-range kink decides if the kink will move through the barrier. Indeed, tunneling is a non-equilibrium process that depends on the net force and how the long-range kink interacts with the force.

\begin{figure}
    \centering
    \scalebox{0.4}{\includegraphics{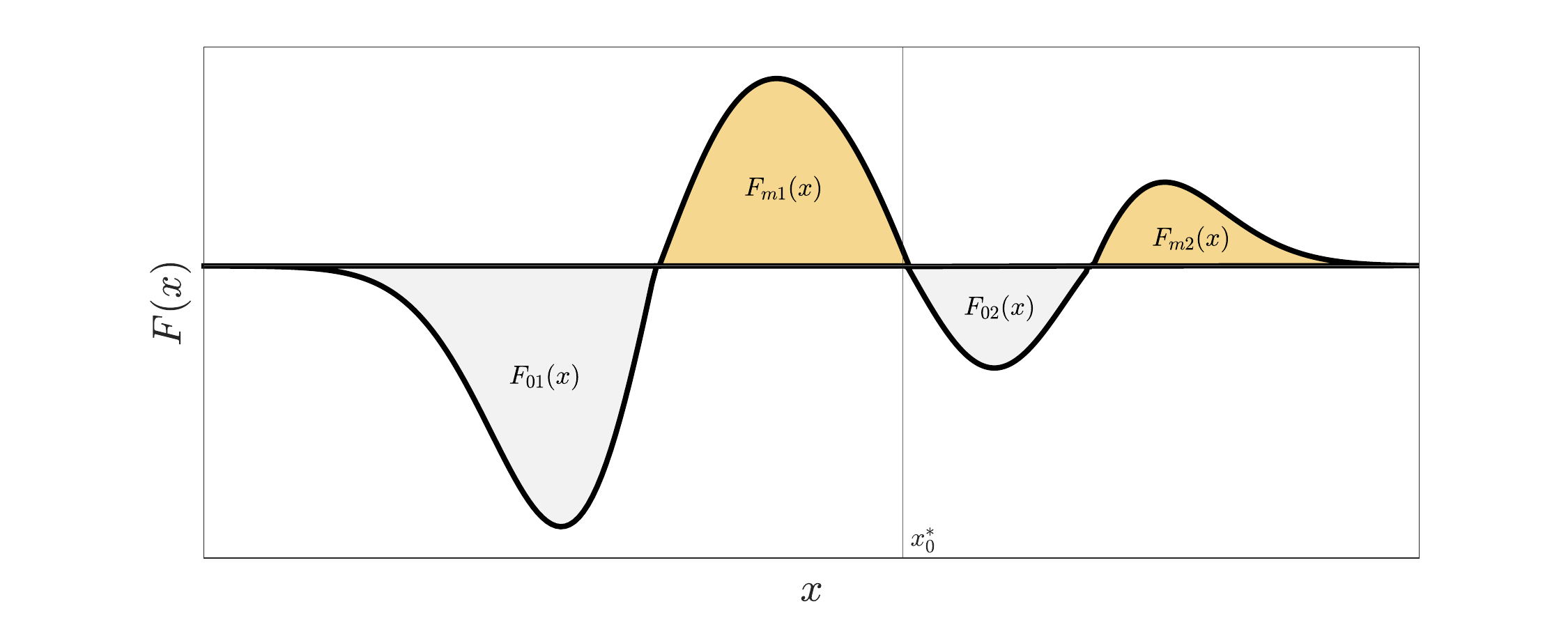}}
    \caption{General topological form of the force $F(x)$ in the kink-stability analysis.}
    \label{Fig:ForceShape}
\end{figure}

The general topological form of the force $F(x)$ for which the kink-stability problem must be solved is schematized in Figure~\ref{Fig:ForceShape}. It can be proved that the stability condition of the translational mode for the equilibrium position $x_0^*$ depends on the properties of the zones $F_{01}(x)$, $F_{m1}(x)$, $F_{02}(x)$, and $F_{2m}(x)$.  The general conditions for long-range kink tunneling can be expressed as a competition between the negative zones, $F_{01}(x)$ and $F_{02}(x)$, and the positive zones, $F_{m1}(x)$ and $F_{m1}(x)$.

\begin{figure}
    \centering
    \scalebox{0.4}{\includegraphics{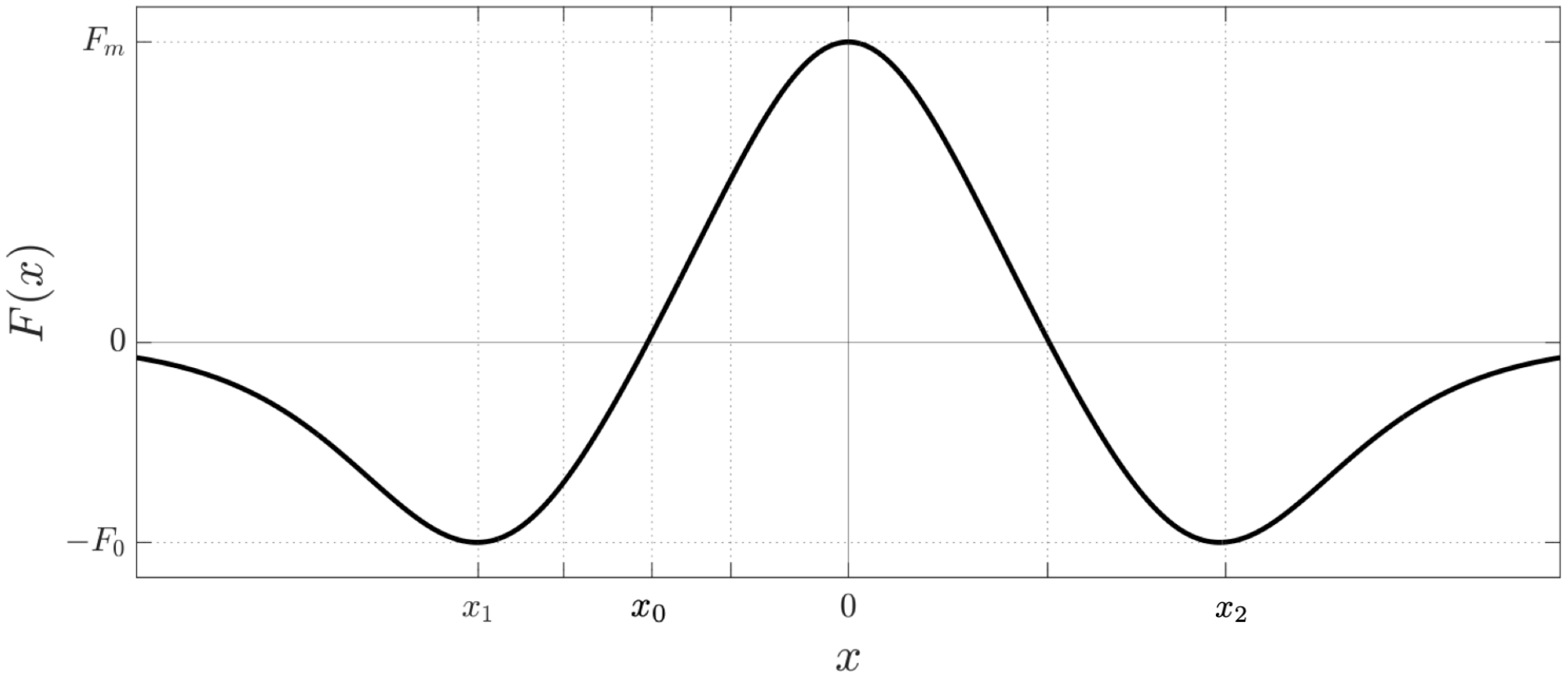}}
    \caption{Typical structure of the heterogeneous force for observing long-range kink tunneling.}
    \label{Fig:Illustrating}
\end{figure}

The conditions for kink tunneling can be obtained using several methods, such as stability theory, energy considerations, positive work conditions, and the formalism of equilibrium-nonequilibrium transitions. All these methods yield similar results. The most powerful long-range kink tunneling condition is the following,
\begin{equation}
\label{Eq:36}
E_P(x_0)>E_P(x_z),
\end{equation}
where $x_z$ is any point in the interval $x_0<x_z<x_2$ and $x_0$ is the initial position of the kink (see Fig.~\ref{Fig:Illustrating}). The function $E_P(X)$ is defined as
\begin{equation}
\label{Eq:37}
    E_P(X)=-\int_{-\infty}^{+\infty}\mbox{d}x\,F(x)\phi_s(x-X).
\end{equation}

The expression $E_P(X)$ is the potential energy of the kink due to the interaction with the field $F(x)$. This formula is obtained using the term in the Hamiltonian that describes the interaction between the fields $\phi(x,\,t)$ and $F(x)$.

\begin{figure}
    \centering
    \scalebox{0.4}{\includegraphics{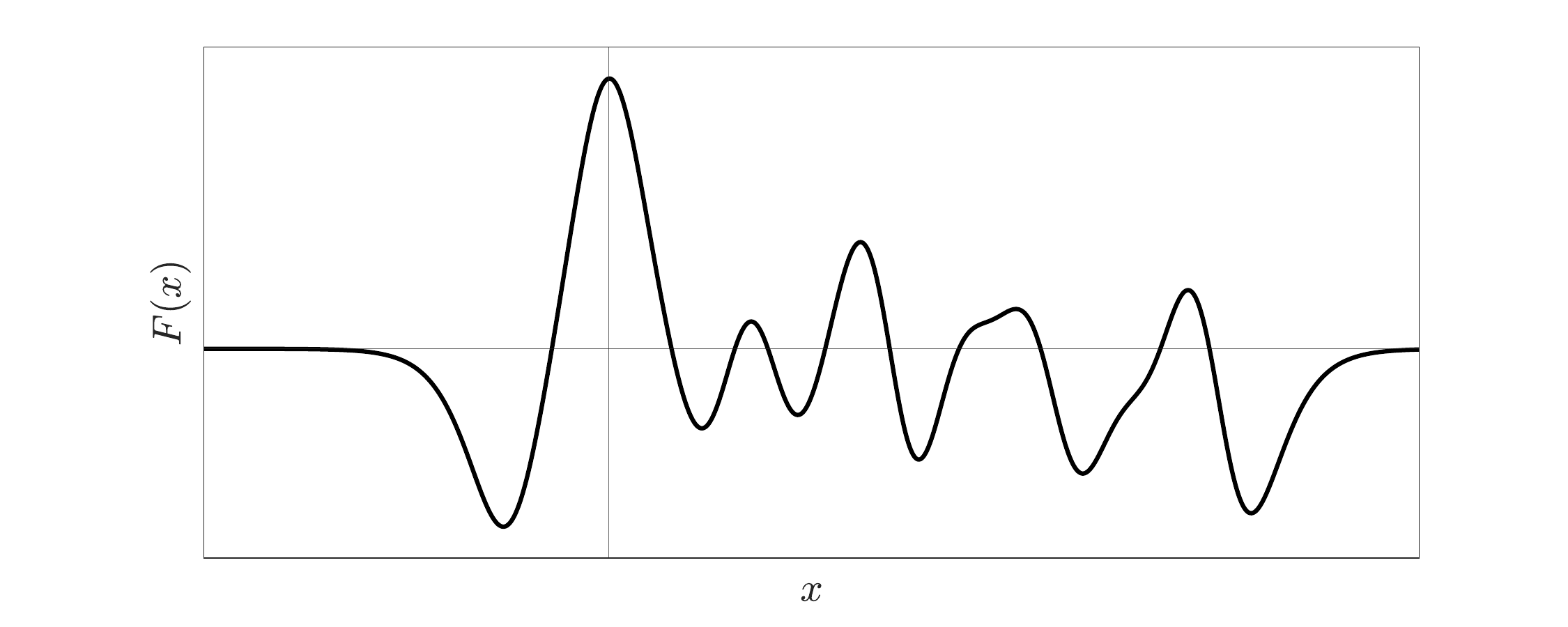}}
    \caption{Illustration of a randomly disordered force with a distribution of maxima and minima. Tunneling of long-range kinks is possible if the condition of Eq.~\eqref{Eq:36} is satisfied for every structure with a negative minimum, a positive maximum, and a negative minimum as in Fig.~\ref{Fig:Illustrating}.}
    \label{Fig:IllustratingTwo}
\end{figure}

Our findings show that a long-range kink of arbitrarily low energy can penetrate any finite barrier if the kink extent is sufficiently long. In Fig.~\ref{Fig:Illustrating}, we can see the typical $F(x)$ structure for observing long-range kink tunneling. We remark that even when the total energy of the center of mass of the kink is less than the height of the energy barrier, kink tunneling can occur. Moreover, even if the total initial energy of the center of mass of the kink is precisely zero, long-range kink tunneling can happen. Consider a randomly disordered $F(x)$ as that shown in Fig.~\ref{Fig:IllustratingTwo}. If for every basic block structure containing a negative minimum, a positive maximum, and another negative minimum (as in Fig.~\ref{Fig:Illustrating}), the tunneling condition of Eq.~\eqref{Eq:36} is satisfied, then the long-range kink can pass freely through the disordered zone.

\section{Implications in other systems}
\label{Sec:Implications}

\subsection{The quantum long-range kink}

The results discussed above suggest that quantum tunneling of a kink can be enhanced if the system parameters are changed so that it will support long-range kinks. First, the long-range kink tunneling phenomenon would shorten the wall and reduce the height of the barrier. In some situations, long-range kink tunneling can suppress the barrier completely. Second, the kink mass will be smaller. Following Eq.~\eqref{Eq:40}, the kink mass decreases as $n$ increases, so the longer the extent of the kink is, the more quantum the kink is.

\subsection{Kac-Baker interactions}

Another example of physical systems relevant to ours is the kink dynamics governed by the sine-Gordon equation with the Kac-Baker long-range interaction potential \cite{Baker1961, Kac1963, Sarker1981, Ishimori1982, Pokrovsky1983,  Remoissenet1985, Ferrer1989}. For instance, the Hamiltonian of a system of particles of mass $m$ placed on a one-dimensional lattice can be the following
\begin{equation}
\label{Eq:05}
H=\frac{1}{2}m\sum_{i}\dot{\phi}_i^2+\sum_{i}\mathcal{U}(\phi_i)+\frac{1}{2}\sum_{i\neq j}V_{ij}\left(\phi_i-\phi_j\right)^2,
\end{equation}
where each particle is lying on an on-site potential $\mathcal{U}(\phi_i)$, where $i$ and $j$ are the lattice points, and $V_{ij}$ is the Kac-Baker potential,
\begin{equation}
\label{Eq:06}
V_{ij}=\frac{c(1-r)}{2r}r^{|i-j|},
\end{equation}
where $c$ is a constant and $r$ is the number of neighboring interactions \cite{Kittel2018}. Using the results from Refs.~\cite{Woafo1993}, we conclude that the kink's extent goes to infinity as $r$ (that defines the range of the interaction) is increased. Indeed, as $r\to1$, $L_K^2\to1/(1-r)^2$. These kinks will behave as long-range kinks. Note that we do not have to change the sine-Gordon substrate potential $\mathcal{U}(\phi)$ to observe long-range features.

\subsection{Non-local Josephson electrodynamics}

Long-range kinks exist in the context of non-local Josephson electrodynamics \cite{Abdumalikov2006}. Indeed, the long-range kinks studied in this article are also solutions to non-local integrodifferential equations governing the dynamics of long ultranarrow Josephson junctions like the following
\begin{equation}
    \label{Eq:04}
    \frac{\partial^2\phi}{\partial t^2}+ b\,\frac{\partial\phi}{\partial t}-\frac{\partial}{\partial x}\int\mbox{d}x'Q(x,\,x')\frac{\partial\phi}{\partial x'}+\sin\phi=F(x),
\end{equation}
written here in dimensionless form. The function $Q(x,\,x')$ is the non-locality kernel,
\begin{equation}
\label{Eq:19}
    Q(x)=\left(\frac{1}{\pi\lambda_L}\right)\mbox{K}_0\left(\frac{|x|}{\lambda_L}\right),
\end{equation}
where $\lambda_L$ is the London penetration depth, and $\mbox{K}_0$ is the modified Bessel function. The kink solution is
\begin{equation}
    \label{Eq:20}
\phi(x,\tilde\lambda_J)=2\arctan\left(\frac{x}{\tilde\lambda_J}\right)+\pi,
\end{equation}
which is a long-range kink. There are experimental situations where the dynamics of fluxons in ultranarrow long Josephson junctions are described by equation~\eqref{Eq:04} \cite{Ivanchenko1990, Gronbech-Jensen1990, Gronbech-Jensen1995, Gronbech-Jensen2002, Abdumalikov2006}. Sometimes, the interaction between coupled long Josephson junctions is so strong that the nonlocality leads to phenomena where the fluxons behave like Coulomb particles in the presence of an external field with a potential like $V(r)\sim1/r$ \cite{Ivanchenko1990}. We believe the fluxons can have this behavior in stacks of ultrathin-long Josephson junctions.

We hypothesize that the strong interaction between the junctions in stacks of ultranarrow Josephson junctions will lead to Ivanchenko’s non-local electrodynamics. In this case, the interaction energy between the Josephson vortices behaves as follows \cite{Ivanchenko1990},
\begin{equation}
    \label{Eq:21}
    \mathcal{E}_{\mbox{\footnotesize int}}\simeq\frac{8\pi\sigma_1\sigma_2}{\gamma(\Delta x)},
\end{equation}
where $\Delta x$ is the distance between the vortices. The above result means these topological objects will behave as our long-range kinks when $n\gg1$. The experimental and technological significance of this observation will be discussed in Section~\ref{Sec:Applications}.

\subsection{Kink tunneling}
\label{Sec:KinkTunneling}

\begin{figure}
    \centering
    \scalebox{0.35}{\includegraphics{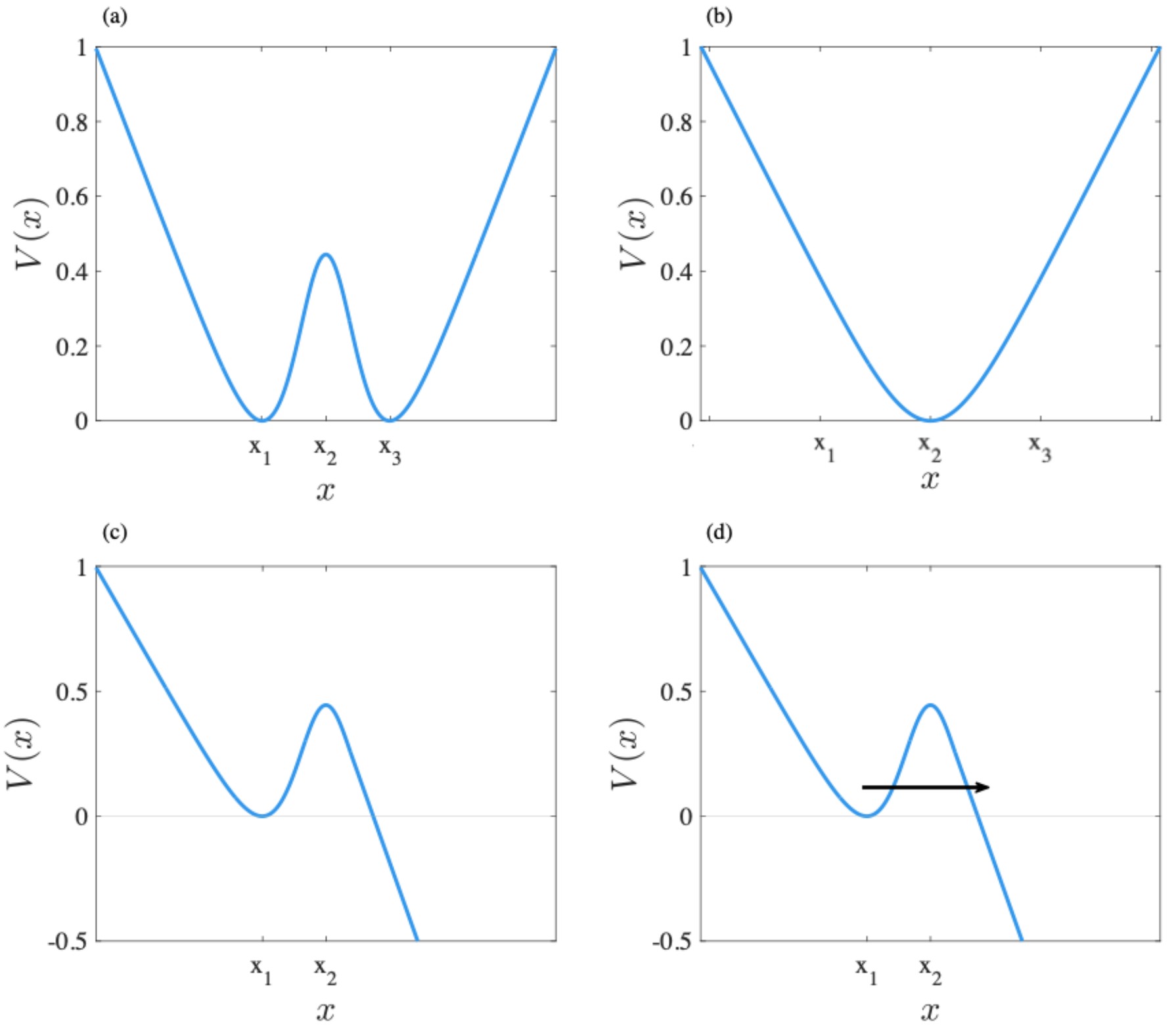}}
    \caption{Bistable and monostable potential wells. These potentials are created by the perturbation $F(x)$. They are not related to the potential $\mathcal{U}(\phi)$.}
    \label{Fig:06}
\end{figure}

Kink tunneling was discovered in the perturbed $\phi^4$ theory \cite{Gonzalez1999}, but it can occur in other models. The main idea in kink tunneling is that a kink is not a point particle \cite{Gonzalez1987, Gonzalez1992}. Consider the general heterogeneous Klein-Gordon equation~\eqref{Eq:03} and imagine a kink moving in a potential landscape created by $F(x)$. The field $F(x)$ can create potential wells and barriers for the motion of the kink \cite{Gonzalez2007}. For instance, when $F(x)$ has three zeros, there is the possibility that the kink will be moving in a bistable potential, as shown in Fig.~\ref{Fig:06}(a). 

The collective coordinate methods (CCM) \cite{McLaughlin1978, Scott2003} have been a very useful technique utilized in the solution of many problems \cite{Bishop1980, Kivshar1989, Sanchez1998, Goodman2002, Goodman2007}. In the context of our Eq.~\eqref{Eq:03}, CCM would yield a correct result if the amplitude of $F(x)$ is very small and $F(x)$ has a single zero. Otherwise, many wrong conclusions can appear. For example, the stability of the equilibria for the kink \cite{Gonzalez1992, Gonzalez1996, Gonzalez1998, Gonzalez1999}, soliton breakup due to instability of the shape mode \cite{Gonzalez1996-2}, formation of kink-kink structures \cite{GarciaNustes2012}, and soliton explosions \cite{Gonzalez1996, Gonzalez1996-3}. Many of these phenomena are unthinkable in the world of CCM. The appearance or disappearance of soliton shape modes can depend on the behavior of $F(x)$ \cite{Gonzalez2002-2}. For a discussion of spectral walls, see Ref.~\cite{Adam2019-3}. During soliton collisions, internal modes can change and (in particular) can disappear into the continuum as the solitons approach each other \cite{Adam2019-3}. This means that the collective coordinate dynamics do not correspond to the real dynamics of the process. Recent developments in the area of collective coordinates techniques can be found in Refs.~\cite{Manton2021, Pereira2021, Adam2021-2}. In some cases of $F(x)$, the collective coordinate techniques can predict that the kink is trapped inside the potential well on the left of Fig.~\ref{Fig:06}(a). However, the actual dynamic of the kink can be entirely different \cite{Gonzalez1992, Gonzalez1998, Gonzalez1996, Gonzalez1987}. 

Imagine there is a stationary solution $\phi_k(x)$ of Eq.~\eqref{Eq:03} in such a way that the kink center of mass is at point $x=0$, which is usually a zero of $F(x)$ (or a zero of $F(x)$ is close to $x=0$). To know if the point $x=0$ is a stable equilibrium for the kink (or not), we must solve the complete stability problem using the exact solution $\phi_k(x)$ to the exact nonlinear partial differential equation~\eqref{Eq:03}. For this purpose, we express the spatiotemporal dynamics as
\begin{equation}
    \label{Eq:26}
    \phi(x,\,t)=\phi_k(x)+f(x)e^{\lambda t}.
\end{equation}
The complete spectral problem is
\begin{equation}
    \label{Eq:27}
    \hat{L}f=\Gamma f,
\end{equation}
where
\begin{equation}
    \label{Eq:28}
    \hat{L}=-\frac{\mbox{d}^2}{\mbox{d} x^2}+\left.\frac{\mbox{d}^2\mathcal{U}(\phi)}{\mbox{d}\phi^2}\right|_{\phi=\phi_k(x)},\quad\Gamma=-\lambda^2.
\end{equation}
The translational invariance is lost in Eq.~\eqref{Eq:03}, but the nodeless mode $f_o(x)\sim\mbox{d}\phi_k(x)/\mbox{d}x$ is still the translational mode! The eigenvalue $\Gamma_0=-\lambda^2\neq0$ corresponding to the translational mode, i.e., the Goldstone mode, is not zero anymore, but it will decide the stability of the equilibrium position \cite{Gonzalez1992, Gonzalez1996, Gonzalez1998, Castro-Montes2020, Marin2021}.

There are situations where there are three equilibrium positions for the kink, as in Fig.~\ref{Fig:06}(a). The kink will be trapped near the point $x=x_1$, not moving to the right. Under certain conditions, although there is an $F(x)$ in Eq.~\eqref{Eq:03} creating obstacles for the motion of the kink, the kink does not feel the barrier as in Fig.~\ref{Fig:06}(a). The potential $V(x)$ for the kink will be like that shown in Fig.~\ref{Fig:06}(b). Thus, the equilibrium position $x=0$ is now stable.

Imagine now that we construct a potential $V(x)$ similar to that shown in Fig.~\ref{Fig:06}(a) with the difference that at point $x=x_3$ there is no local minimum, and the function $V(x)$ will continue decreasing monotonically, as shown in Fig.~\ref{Fig:06}(c). If the condition for the stability of point $x=x_2$ is satisfied, then the following phenomenon will occur. A kink at the bottom of the left potential well (at point $x=x_1$) will move to the right. It will cross the barrier, and it will escape the potential well despite the fact that its center of mass has less energy than the barrier (see Fig.~\ref{Fig:06}(d)). Of course, as for any tunneling, the height and the width of the barrier play an important role in this phenomenon. 

\section{The nature of long-range kink tunneling}
\label{Sec:NatureTunneling}

Let us consider the dynamics of a kink in the following equation
\begin{equation}
\frac{\partial^{2}\phi}{\partial t^{2}}-\frac{\partial^{2}\phi}{\partial
x^{2}}+\frac{dU(\phi)}{d\phi}=F, \label{eqNS1}%
\end{equation}
where $F$ is a constant field. We will assume that $\left\vert F\right\vert <F_{crit}$, in such a way that
two minima of $U(\phi)$ are not destroyed. That is, the kink solution exists. $F_{crit}$ can be found from the condition that the algebraic equation $dU(\phi)/d\phi=F$ must still have three real solutions \cite{Gonzalez1987,Gonzalez1992}. If $F$ is negative ($F<0$), then there is a force pushing the kink to the \textquotedblleft right\textquotedblright. The kink will undergo a positive acceleration. If $F>0$, then the force is acting on the kink in the negative direction.

\begin{figure}
    \centering
    \includegraphics[width=0.49\linewidth]{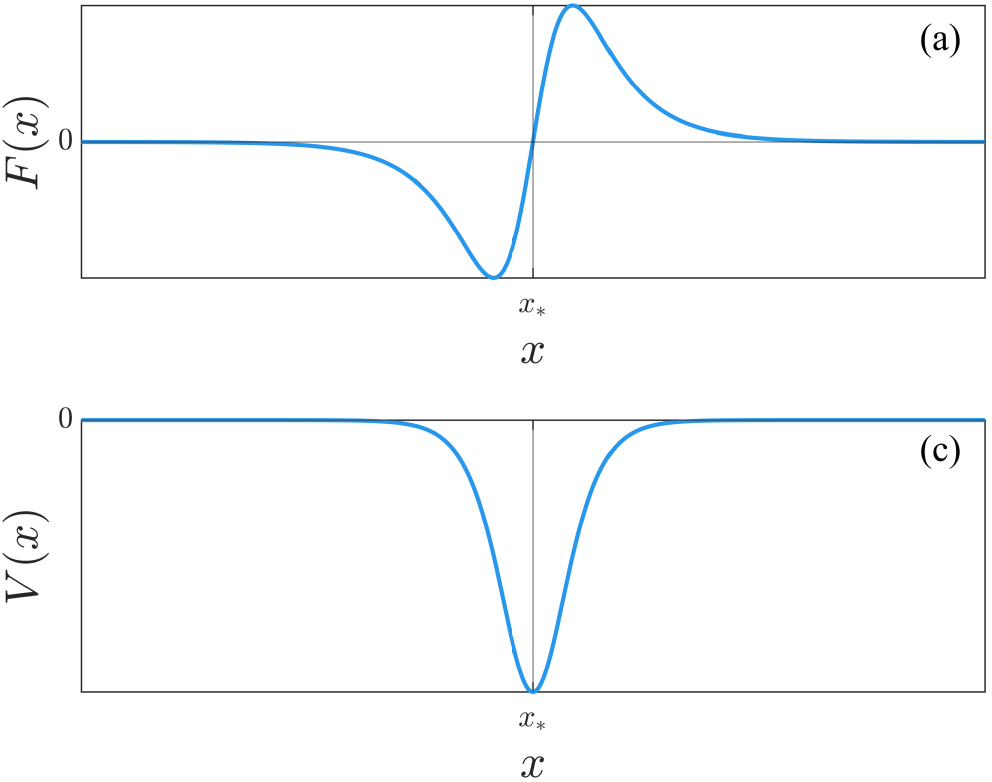}\ \includegraphics[width=0.49\linewidth]{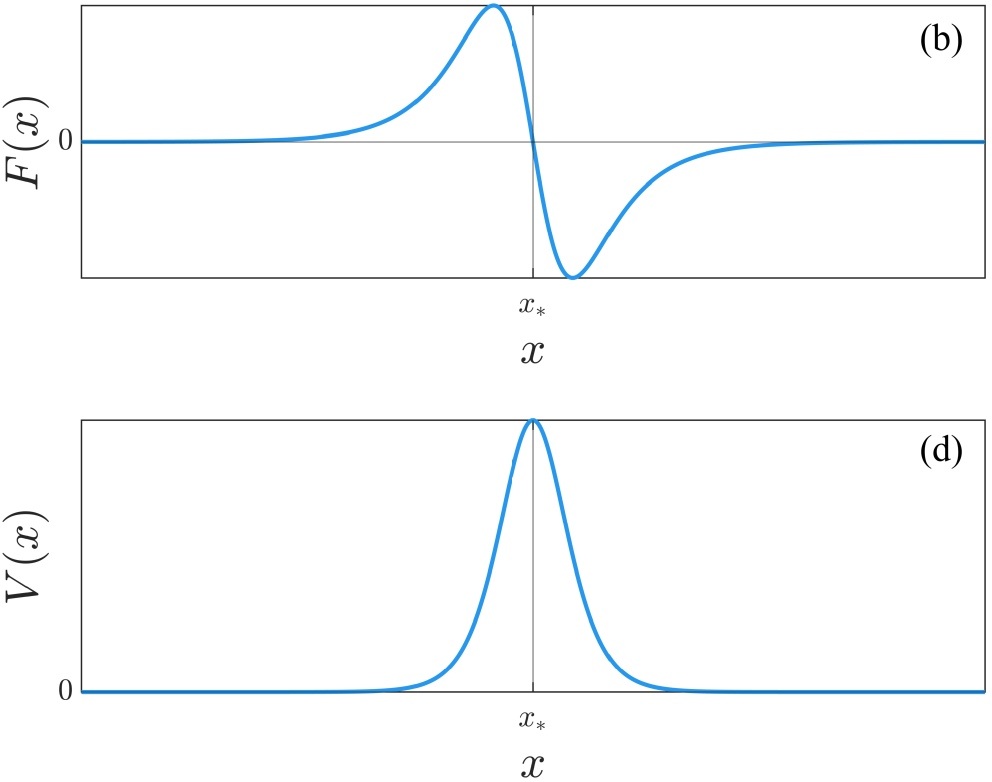}
    \caption{\textbf{(a)} Heterogeneous field $F(x)$ that creates a stable equilibrium for the kink. \textbf{(b)} Inhomogeneous field that creates an unstable equilibrium for the kink.  \textbf{(c)} Potential $V(x)$ created by field $F(x)$ in (a). \textbf{(d)} Potential $V(x)$ created by field $F(x)$ in (b).}
    \label{NS1}
\end{figure}

Let us imagine now that $F(x)$ is space-dependent, as shown in Fig. \ref{NS1} \cite{Gonzalez1992,Gonzalez1998}. It is reasonable to expect that a zero of the function $F(x)$ (define $x_{\ast}$, such that $F(x_{\ast})=0$) would imply the existence of an equilibrium position for the kink. If $F(x)$ has only one zero $x_{\ast}$, then we can be sure that in the neighborhood of $x_{\ast}$ there is an equilibrium for the kink. Is this equilibrium stable or unstable?

Observe Fig. \ref{NS1}(a). If the position of the kink is slightly shifted to the \textquotedblleft left\textquotedblright\ ($x_{cm}<x_{\ast}$), then the force
($F(x)<0$) will try to push the kink back to the equilibrium position. When the center of mass of the kink is moved slightly to the \textquotedblleft right\textquotedblright, then a positive $F(x)$ will pull the kink back to the equilibrium position. Thus, the stability condition is
\begin{equation}
\left(  \frac{dF(x)}{dx}\right)  _{x=x_{\ast}}>0. \label{eqNS2}%
\end{equation}
This condition is confirmed by rigorous mathematics and physics \cite{Gonzalez1992, Gonzalez1996, Gonzalez1998}. If
\begin{equation}
\left(  \frac{dF(x)}{dx}\right)  _{x=x_{\ast}}<0, \label{eqNS3}%
\end{equation}
then $x=x_{\ast}$ is an unstable equilibrium for the kink (see Fig. \ref{NS1}(b)). If the initial position of the center of mass of the kink is slightly outside the equilibrium position, then the kink will move away from $x=x_{\ast}$, and it will not return. CCM confirms this result. Notice that the existence of a stable (unstable) equilibrium would imply the presence of a potential well (barrier),  as shown in figures \ref{NS1}(c) and \ref{NS1}(d).

When $F(x)$ has several zeros (say, $x_{\ast1}$, $x_{\ast2}$, $x_{\ast3}$...) and the distance between them is sufficiently long for the $\phi^{4}$-kink and the sine-Gordon kink, the stability of the created equilibrium can still be determined using the condition given by inequality \eqref{eqNS2}. The CCM can lead to correct conclusions in these cases also. However, for a generic set of zeros of a function $F(x)$, the CCM will fail in predicting the right dynamics of the kink \cite{Gonzalez1996, Gonzalez1998, Gonzalez1999, Gonzalez2008}.

In general, the rigorous method for the stability investigation must be based on the spectral problem of Eq.~\eqref{Eq:27} with $\widehat{L}=\left(  -\partial^{2}/\partial x^{2}+\partial^{2}U(\phi)/\partial\phi^{2}\right)  _{\phi=\phi_{s}(x)}$ and $\phi_{s}(x)$ is the exact solution to equation \eqref{eqNS1} representing a stationary kink, whose center of mass is at equilibrium point $x_{cm}=x_{\ast}$, in the presence of $F(x)$. The eigenvalue $\Gamma_{0}$ corresponding to the translational mode $f_{0}(x)\sim d\phi_{s}(x)/dx$ will decide the stability of the equilibrium point $x_{\ast}$ \cite{Gonzalez1996, Gonzalez1998}.

Let us discuss the equation for the $\phi^{4}$-theory:
\begin{equation}
\frac{\partial^2\phi}{\partial t^2}-\frac{\partial^2\phi}{\partial x^2}-\frac{1}{2}\phi+\frac{1}{2}\phi^{3}=F(x). \label{eqNS5}
\end{equation}
As a model function for $F(x)$, we will take
\begin{equation}
F_{1}(x)=\frac{1}{2}A\left(  A^{2}-1\right)  \tanh\left(  Bx\right)  +\frac
{1}{2}A\left(  4B^{2}-A^{2}\right)  \frac{\sinh\left(  Bx\right)  }{\cosh
^{3}\left(  Bx\right)  }. \label{eqNS6}%
\end{equation}
A system with an external force like \eqref{eqNS6} has been shown to exhibit a pitchfork bifurcation \cite{Gonzalez1996, Gonzalez1998}. The results can be generalized to other topologically equivalent systems (this
method has been developed in \cite{Gonzalez1998, Gonzalez1999-2}). Here, the most important thing is that changing parameters $A$ and $B$, $F_{1}(x)$ can have one zero or three zeros. In other words, the potential for
the motion of the kink can have one minimum (a stable equilibrium) or two minima (a bistable potential). Another important property of $F_{1}(x)$ is that the exact kink solution can be found, and the stability problem can be solved exactly.

It is possible to find situations where $F_{1}(x)$ has three zeros (say $x_{\ast1}$, $x_{\ast2}$, $x_{\ast3}$; $x_{\ast1}<x_{\ast2}<x_{\ast3}$) in such a way that $\left(  dF_{1}(x)/dx\right)_{x=x_{\ast1}}>0$, $\left(  dF_{1}(x)/dx\right)_{x=x_{\ast2}}<0$, $\left(  dF_{1}(x)dx\right)_{x=x_{\ast3}}$ $>0$. So, there must be a barrier at point $x=x_{\ast2}$, which is an unstable equilibrium. However, under certain conditions, due to the interaction of the kink with the other zeros of $F_{1}(x)$, the kink \textquotedblleft feels\textquotedblright\ the point $x=x_{\ast2}$ as a stable equilibrium. Hence, the kink does not feel the barrier. This can happen because the kink is an extended object \cite{Gonzalez1992, Gonzalez1998}. The $\phi^{4}$-kink can interact with the other zeros of $F(x)$ when the distance between the zeros is comparable with the size of the kink.

The $\phi^{4}$-kink interacts with an antikink with short-range forces. In the same way, a $\phi^{4}$-kink can interact with localized concentrations of the field $F(x)$ only with short-range forces. When the distance between the structures of $F(x)$ is very long, then the kink will feel them as what they are: the potential wells will be wells, and the barriers will be barriers. In that case, the kink cannot move through a barrier if it has less kinetic energy than the height of the energy barrier.

Let us present a system where the tunneling of the $\phi^{4}$-kink can be observed. Define $F(x)$ the following way
\begin{subequations}
    \label{Eq:4.6}
\begin{equation}
F(x)=F_{1}(x),\text{ for }x<x_{\ast\ast},\label{eqNS7}%
\end{equation}
\begin{equation}
F(x)=c,\text{ for }x>x_{\ast\ast},\label{eqNS8}%
\end{equation}    
\end{subequations}
where $x_{\ast\ast}$ is the point where $F_{1}(x)$ has a local minimum $\left[  \left(  dF(x)/dx\right)  _{x=x_{\ast\ast}}=0\right]$, and $c=F_{1}(x_{\ast\ast})$. The stability problem (where $\phi(x,t)=\phi_{s}(x)+f(x)\exp({\lambda t})$) of the equilibrium point $x_{\ast}=0$ is reduced to an eigenvalue problem of the same form as Eq.~\eqref{Eq:27}, where
\begin{equation}
    \label{Eq:4.7}
\widehat{L}=-\partial_{x}^{2}+\left(  \frac{3}{2}A^{2}-\frac{1}%
{2}-\frac{3A^{2}}{2\cosh^{2}\left(  Bx\right)  }\right),
\end{equation}
and $\Gamma=-\lambda^{2}$. The eigenvalues of the discrete spectrum are given by $\Gamma_{n}=-1/2+B^{2}\left(  \Lambda+2\Lambda n-n^{2}\right)  $, where $\Lambda\left(
\Lambda+1\right)=3A^{2}/2B^{2}$. If $A^{2}>1$ and $4B^{2}<1$, the field $F(x)$ possesses the desired properties, i.e., there is a zero of $F(x)$: $x_{\ast1}<0$ ($F(x_{\ast1})=0$), that would correspond to a stable equilibrium position. There is another zero of $F(x)$ at $x_{\ast2}=0$ that would correspond to an unstable
equilibrium position and would serve as a potential barrier. For $x>0$, the potential $V(x)$ is a monotonically decreasing function.

If $\Gamma_{0}<0$, the equilibrium point $x_{\ast2}=0$ is unstable, then the kink \textquotedblleft feels\textquotedblright\ the barrier. If the initial position of the kink is in the vicinity of point $x=0$ with the position of the center of mass situated on the left of the barrier ($X_{cm}\left(  t=0\right)<0$), and with zero initial velocity, the kink will not move to the right of point $x=0$. The kink will be trapped inside the potential well. On the other hand, if $\Gamma_{0}>0$, the kink will move to the right, through the barrier, even if its center of mass is placed in the minimum of the potential and its initial velocity is zero. Analytical calculations reveal that when $A^{2}>1$, $4B^{2}<1$, $2B^{2}\left(3A^{2}-1\right)  >1$, the potential $V(x)$ possesses a minimum at a point $x_{\ast1}<0$, a maximum at $x_{\ast2}=0$, and the kink can tunnel through the barrier.

\begin{figure}
    \centering
    \includegraphics[width=0.5\linewidth]{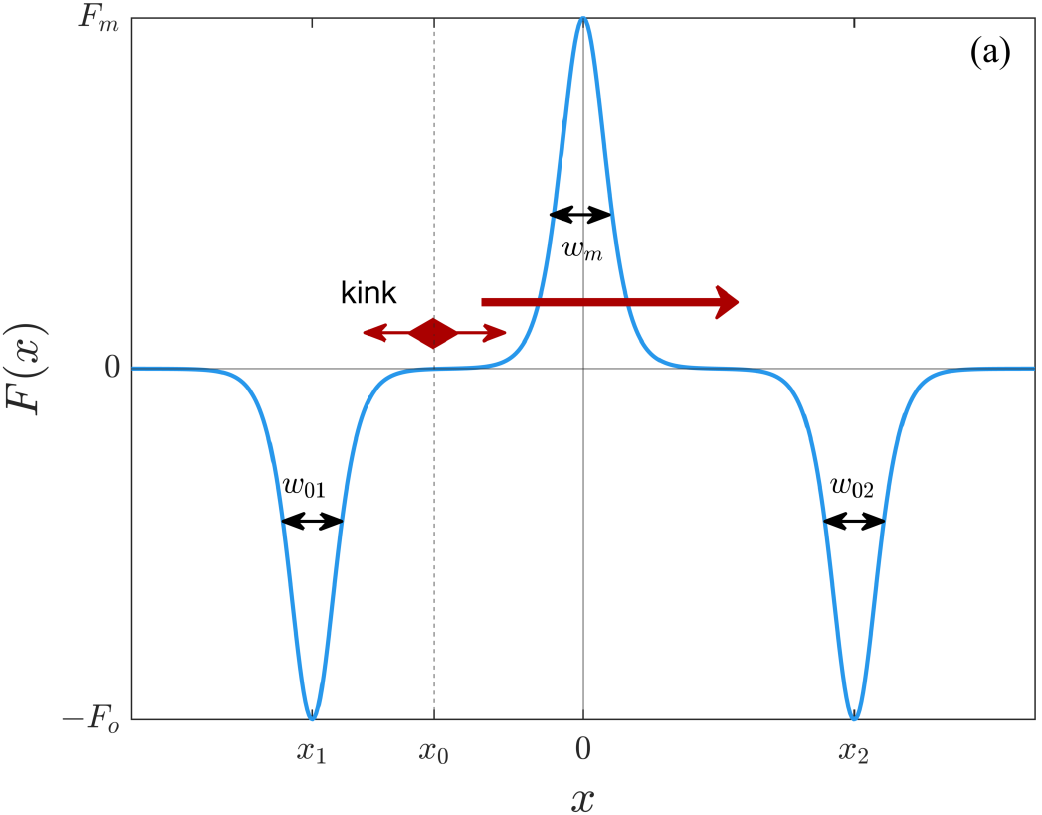}\includegraphics[width=0.5\linewidth]{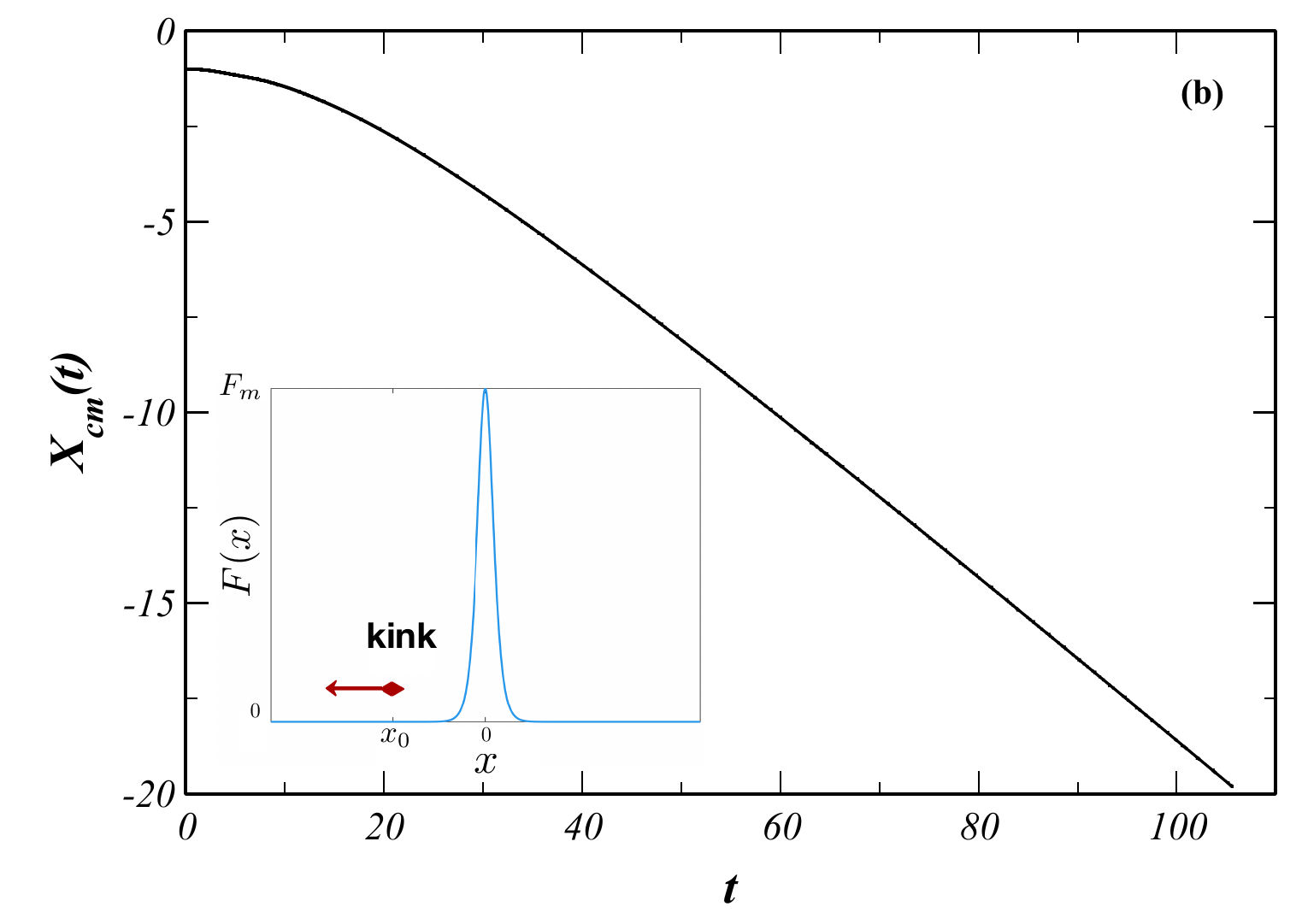}
    \caption{\textbf{(a)} Typical $F(x)$ for the observation of kink tunneling. \textbf{(b)} When $F(x)$ has only a positive maximum, the kink will be pushed to the \textquotedblleft left\textquotedblright .}
    \label{5NS}
\end{figure}

For a general $F(x)$ as that shown in Fig. \ref{5NS}, \ the condition $\Gamma_{0}>0$ yields
\begin{equation}
F_{0}>F_{m}\ \mbox{Sg}\left(  x_{1},x_{2},w_{m},w_{01},w_{02}\right)  ,\label{eqNS10}%
\end{equation}
where $\mbox{Sg}$ is a function that increases with $x_{1}$, $x_{2}$, $w_{m}$ and decreases with $w_{01}$ and $w_{02}$. Parameters $F_{0}$, $w_{01}$ and $w_{02}$ favor the tunneling. Parameters $F_{m}$, $x_{1}$, $x_{2}$, and $w_{m}$ work against the tunneling. 

\begin{figure}
    \centering
    \scalebox{0.5}{\includegraphics{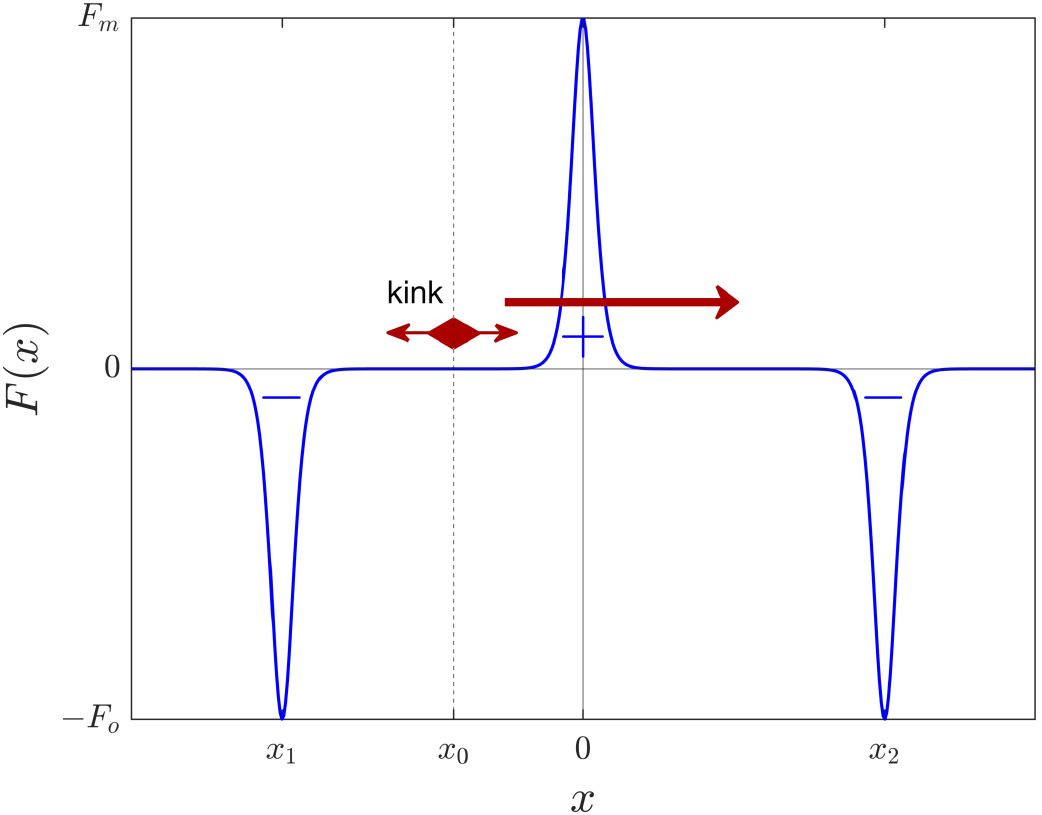}}
    \caption{The interaction of the kink with the localized bell-shape
structures of $F(x)$ is crucial for the kink tunneling phenomenon. The
positive bell-shape structure at point $x=0$ will try to push the kink to
the \textquotedblleft left\textquotedblright . The negative bell-shape
structures at points $x=x_{1}$ and $x=x_{2}$ will try to push the kink to
the \textquotedblleft right\textquotedblright . Imagine there is an 
\textquotedblleft A-zone\textquotedblright\ between points $x=x_{1}$ and $x=0
$, where $F(x)$ is exactly zero. If the initial position of a normal kink is
inside the \textquotedblleft A-zone\textquotedblright , and the initial
velocity is zero, then the kink will remain trapped inside the potential
well. A normal kink \textquotedblleft feels\textquotedblright\ only the
local properties of $F(x)$. On the other hand, a long-range kink can feel
the interaction with the negative structures at points $x=x_{1}$ and $x=x_{2}
$ even if the values of $\left\vert x_{1}\right\vert $ and $\left\vert
x_{2}\right\vert $ are large.}
    \label{6NS}
\end{figure}

\begin{figure}
    \centering
    \scalebox{0.5}{\includegraphics{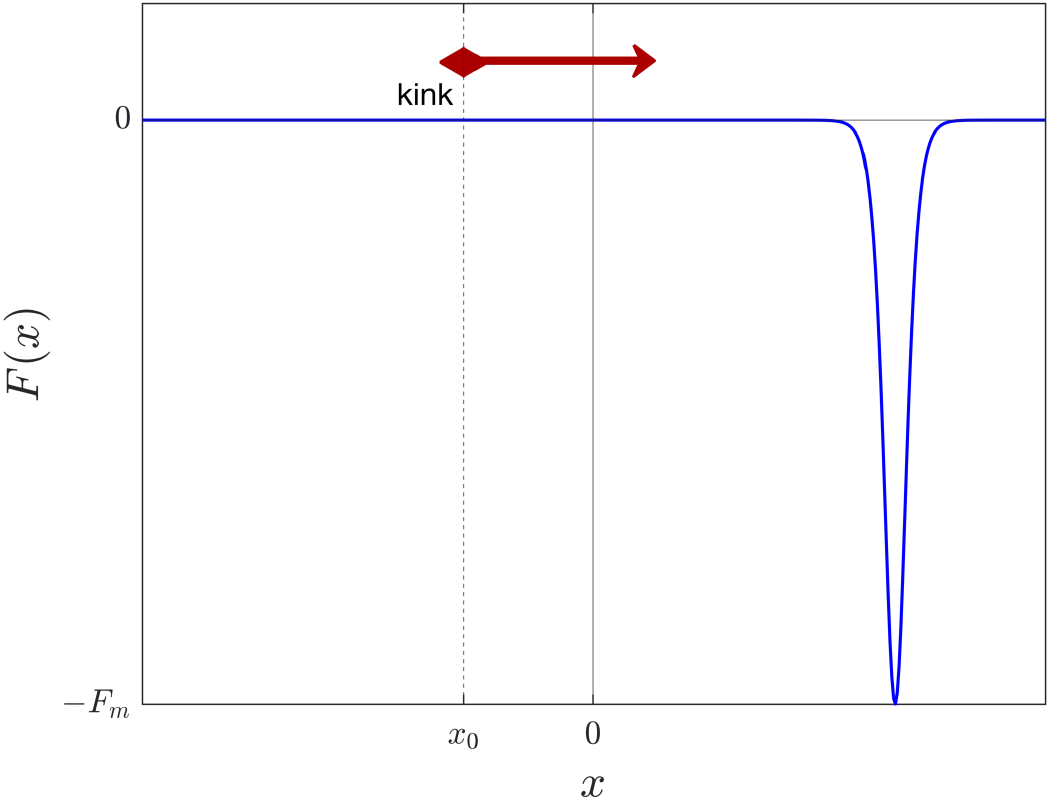}}
    \caption{A negative bell-like structure will push the kink to the
\textquotedblleft right\textquotedblright\ (see Fig. \ref{10NS}). However,
the distance between the kink and the localized structure is important (see
Fig. \ref{10NS}).}
    \label{9NS}
\end{figure}

\begin{figure}
    \centering
    \includegraphics[width=0.5\linewidth]{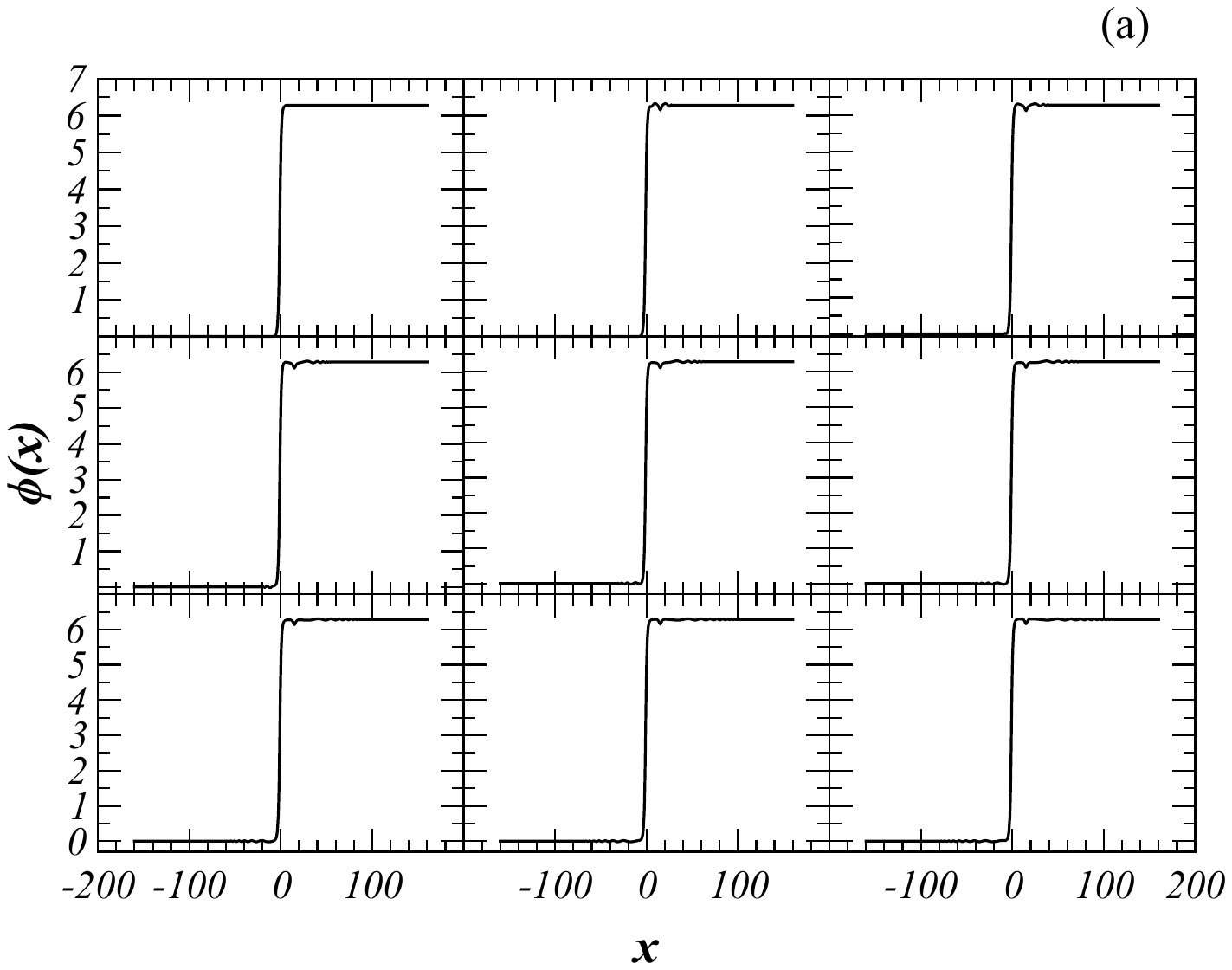}\includegraphics[width=0.5\linewidth]{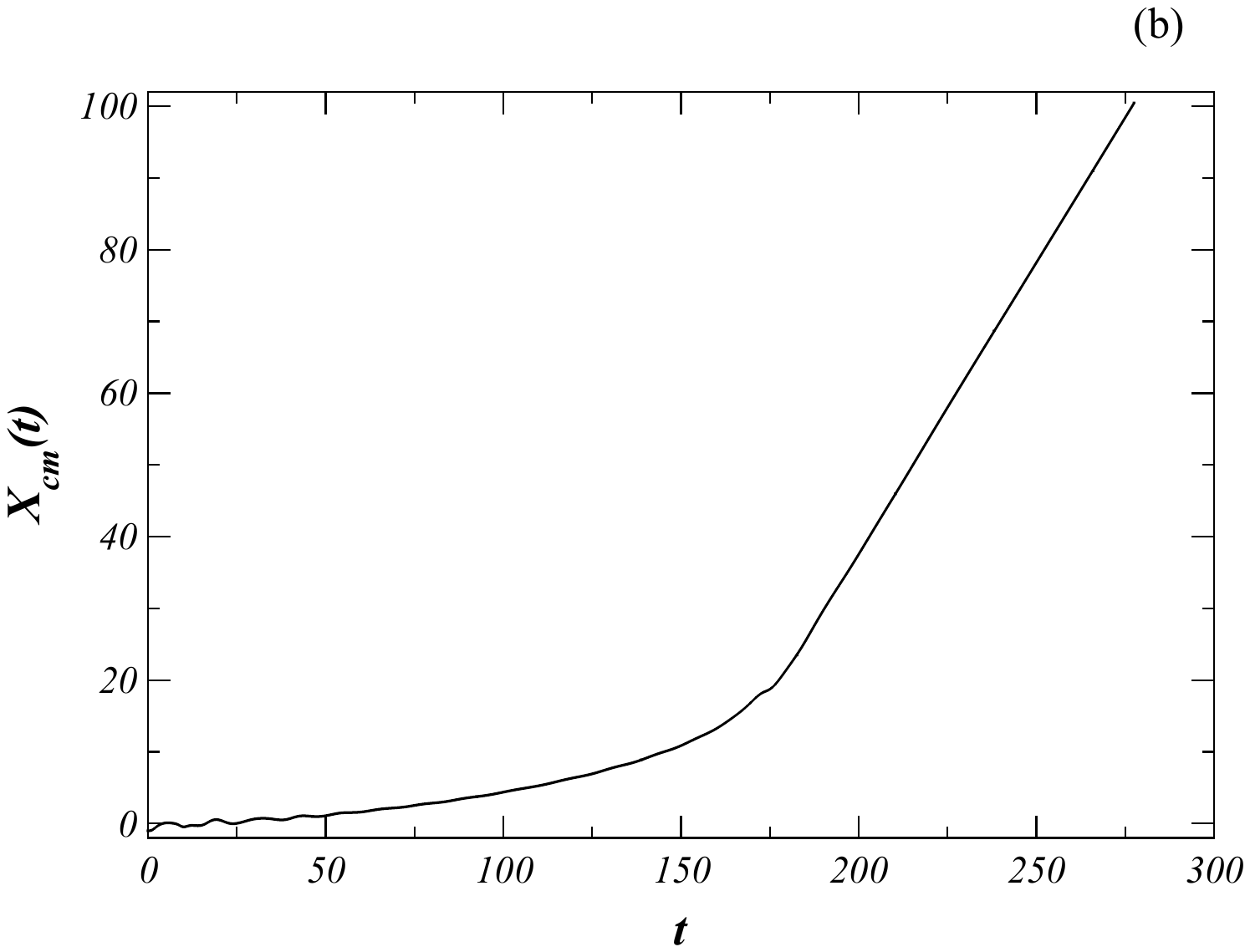}
    \caption{Dynamics of kinks due to the interaction with the $F(x)$ shown in Fig. \ref{9NS}. \textbf{(a)}  A normal kink is not moving because if the distance
between the kink and the localized structure is very long, the kink does not
feel the force. \textbf{(b)} The long-range kink does feel the force due to
the interaction with the negative bell-like localized structure at point $%
x=x_{2}$.
}
    \label{10NS}
\end{figure}

\begin{figure}
    \centering
    \scalebox{0.5}{\includegraphics{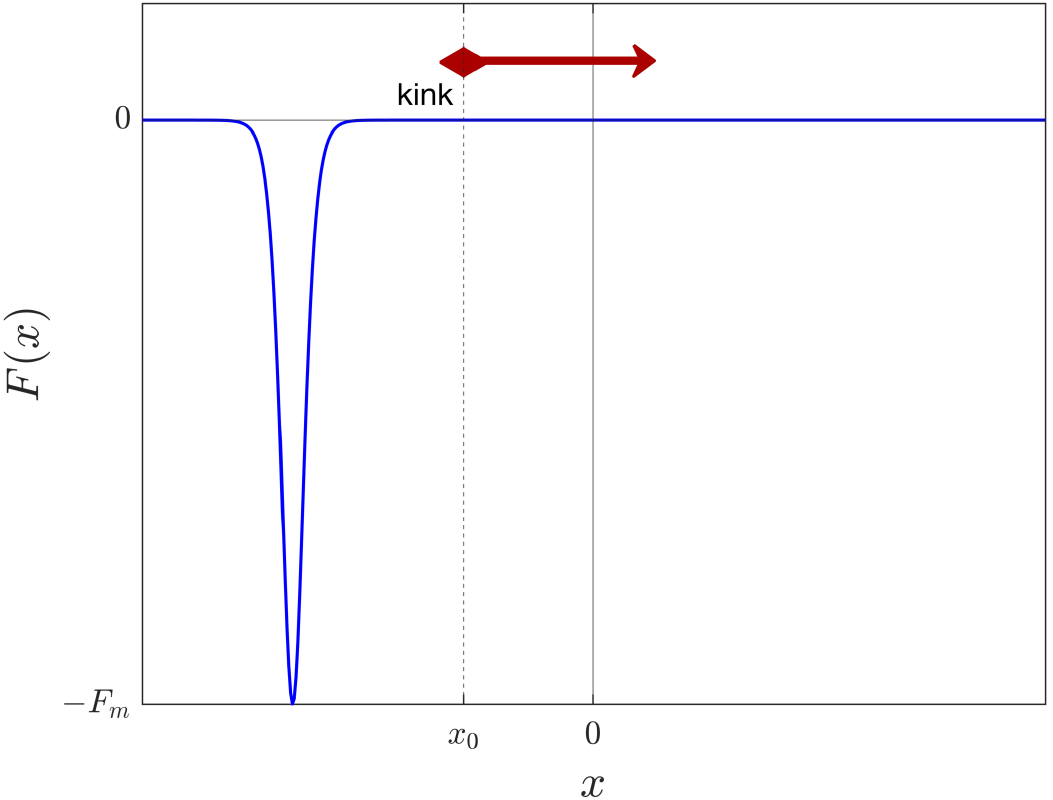}}
    \caption{The long-range kink can be moved to the \textquotedblleft
right\textquotedblright\ by the action of a negative bell-like localized
structure even if $\left\vert x_{1}\right\vert $ is large (see the dynamics
in Fig.~\ref{12NS}).
}
    \label{11NS}
\end{figure}

\begin{figure}
    \centering
    \includegraphics[width=0.5\linewidth]{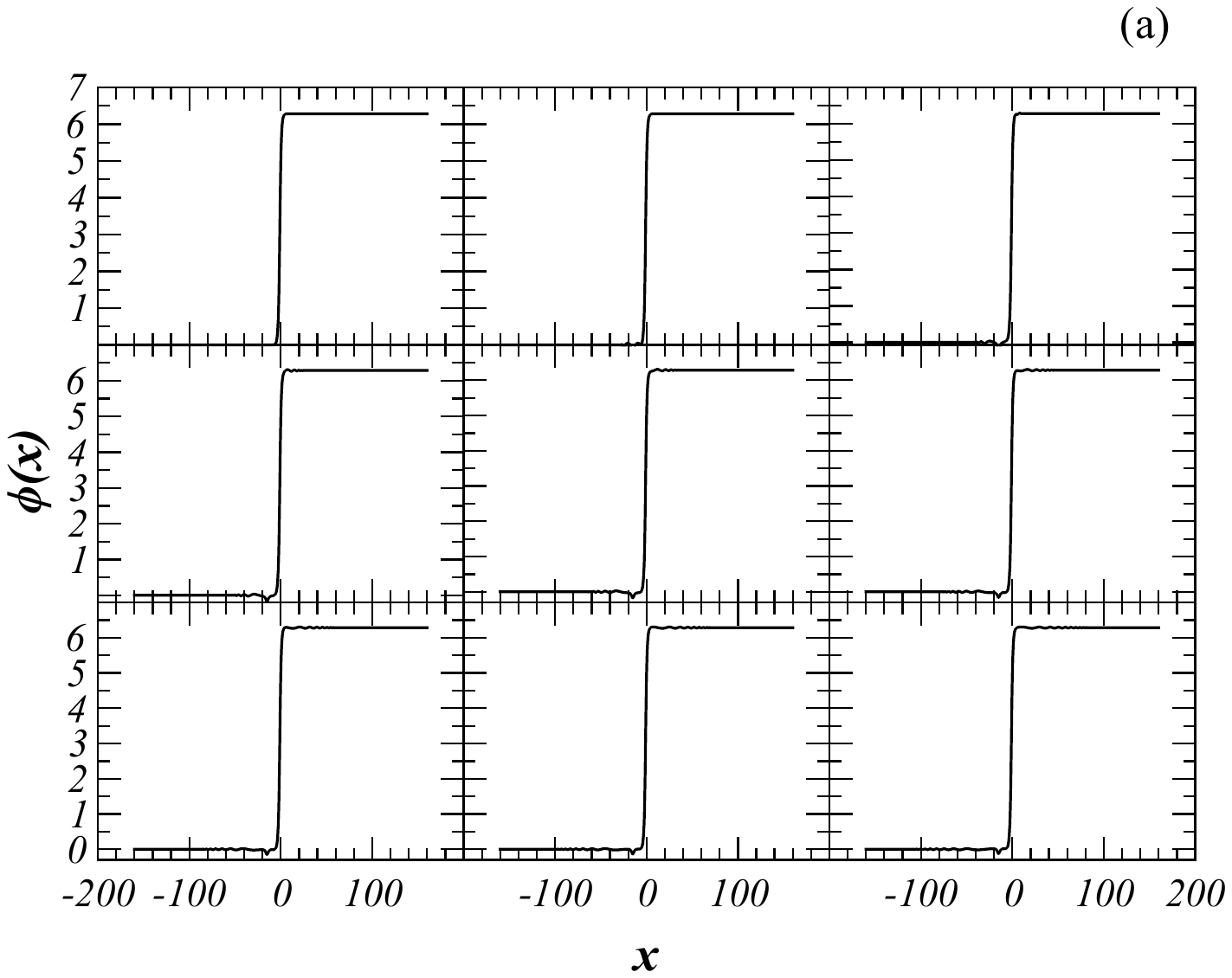}\includegraphics[width=0.5\linewidth]{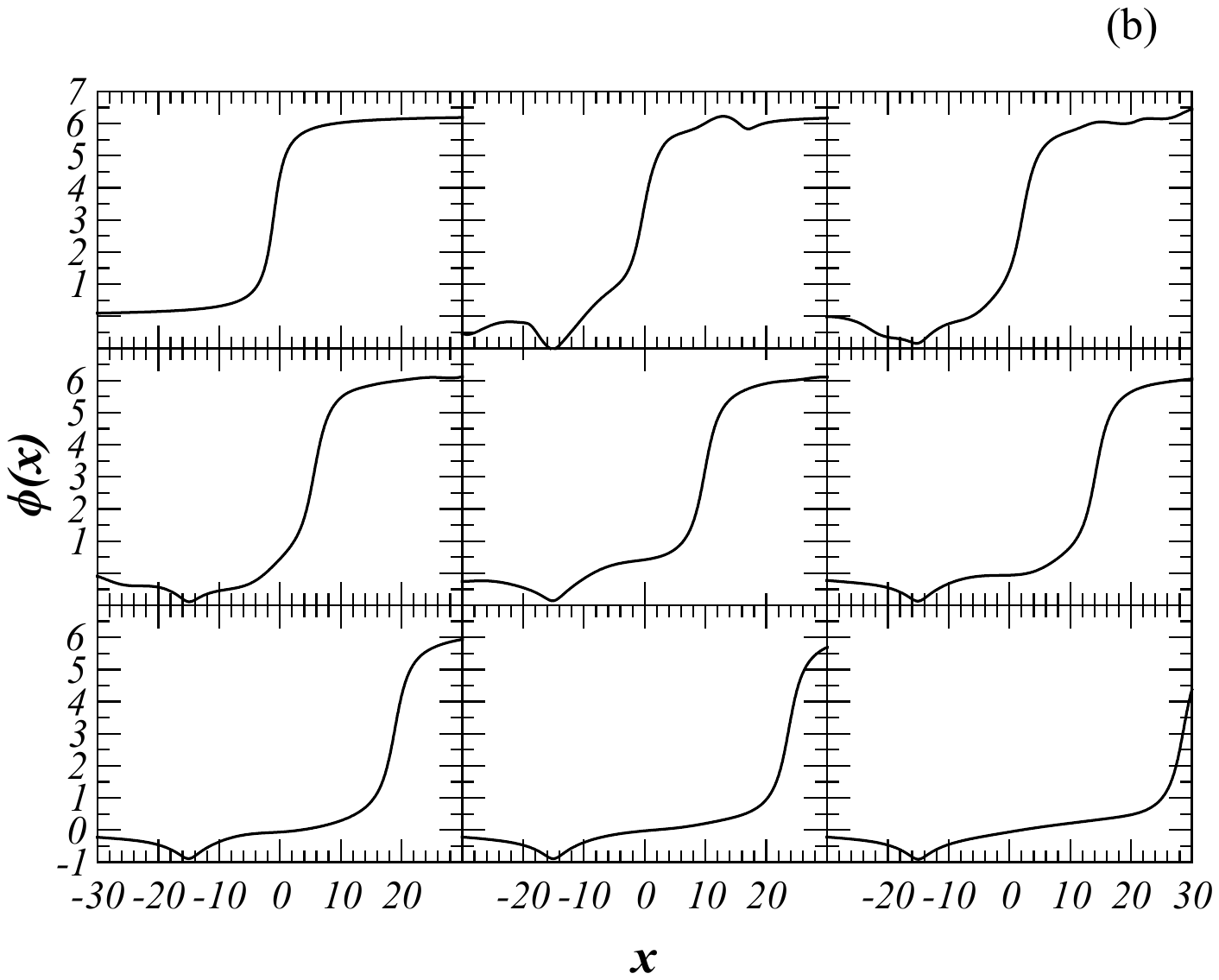}
    \caption{\textbf{(a)} The conventional kink is at rest. The negative
localized structure is too far away. \textbf{(b)} The long-range kink is
moving forward due to the interaction with the negative bell-like
configuration of $F(x)$ at point $x=x_{1}$.}
    \label{12NS}
\end{figure}

For extremely localized $F(x)$ fields, $\mbox{Sg}$ does not depend on $w_{m}$, $w_{01}$, $w_{02}$. We need to stress that the interaction of the kink with the positive and negative zones of $F(x)$ is key to kink tunneling (see Fig. \ref{6NS}). The competition between the negative and positive structures of $F(x)$ is crucial: Who is pushing harder? What happens when only one of the structures is acting alone? 
It is not a surprise that the barrier will try to repel the kink. See the motion of the kink due to the action of $F_{m} (x)=0.02\ \mbox{sech}^{2}(x)$ in Fig. \ref{5NS}(b). Conversely, a negative bell-like structure will push the kink to the ``right''
(see figures \ref{9NS} and \ref{10NS}). At first sight, one would not expect that the kink can interact with a structure that is immediately behind the barrier (the farthest border of the barrier). However, when we learn that if the width of the barrier is very large, the interaction with this negative structure is very small or non-existent, then it makes sense because tunneling should not occur for very wide barriers. Nevertheless, the most important fact is that conventional kinks do not feel the interaction with the negative bell-like structure in Fig. \ref{9NS}, whereas the long-range kink can be pulled by this structure. See the effects of the interaction on a $\phi^{4}$-kink, and on a long-range kink in Fig. \ref{10NS}. Most unexpected is the fact that the left wall of the potential well can play a role in kink tunneling. A negative structure on the left of the kink (see Fig.~\ref{11NS}) can push the long-range kink forward. However, if $\left\vert x_{1}\right\vert $ is large, then a normal kink would never feel the force exerted by the negative
structure placed at point $x_{1}$. Observe the effect of this negative structure on a normal and long-range kink in Fig.~\ref{12NS}. 

\begin{figure}
    \centering
    \includegraphics[width=0.5\linewidth]{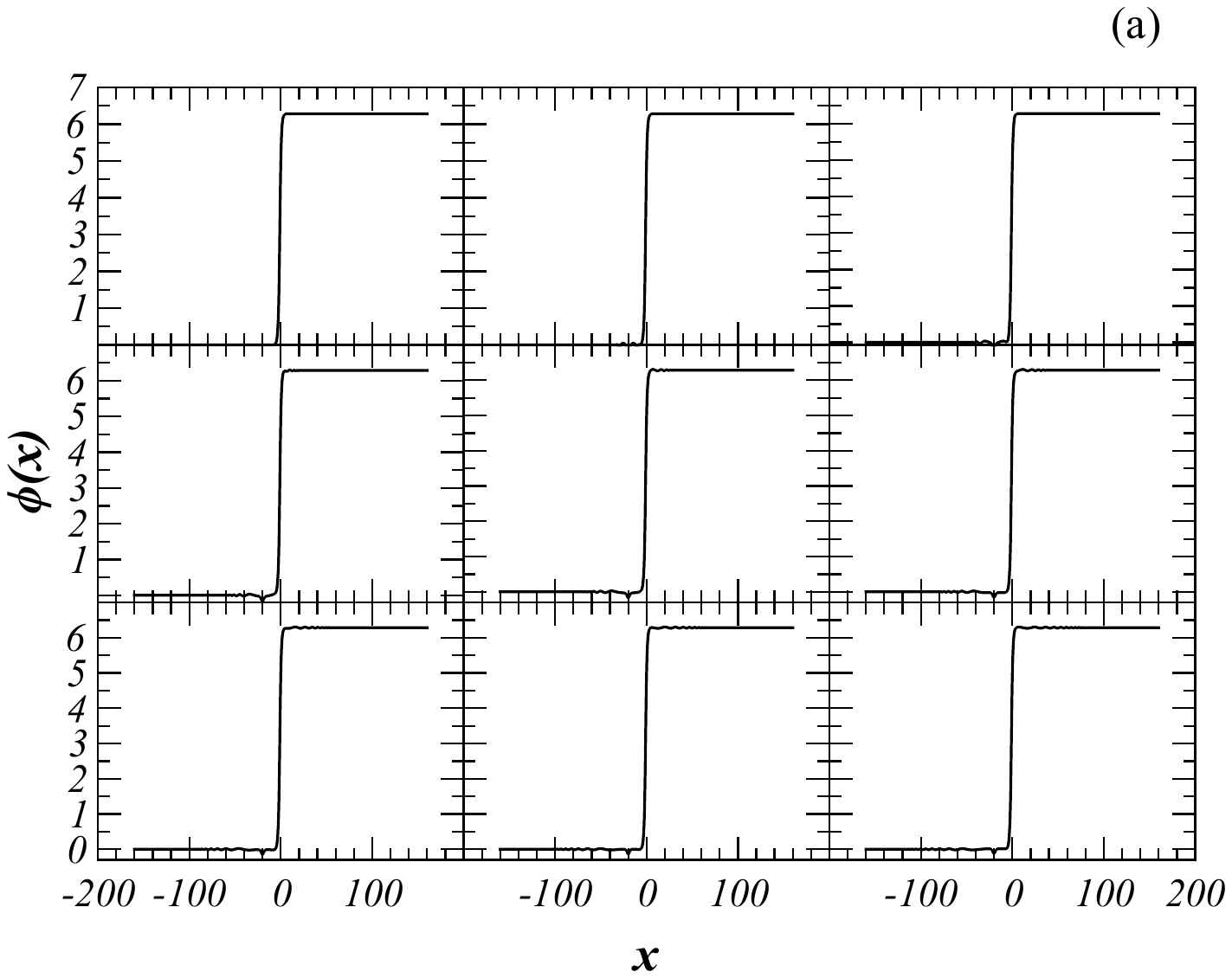}\includegraphics[width=0.5\linewidth]{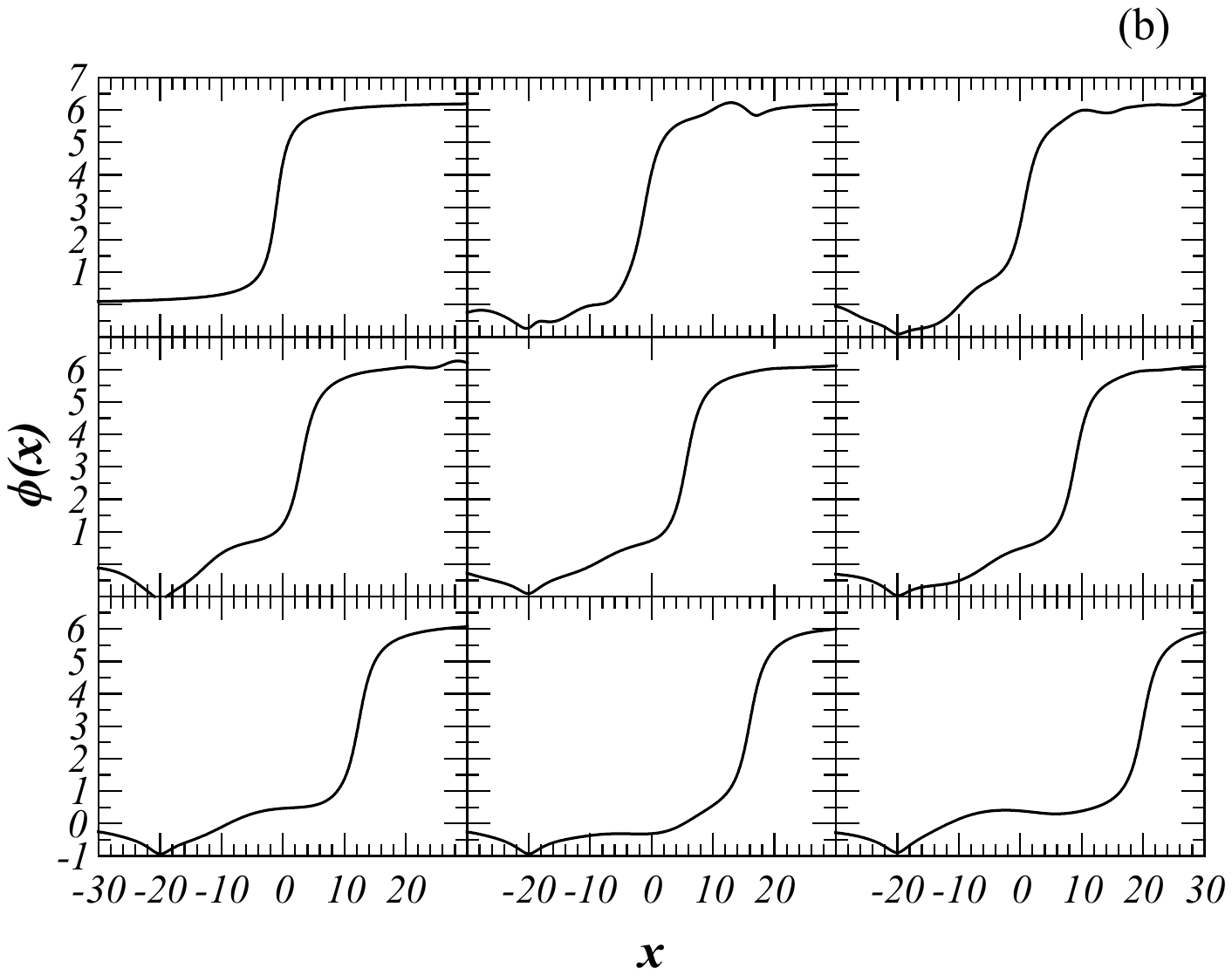}
    \caption{Action of a remote localized
field on a conventional kink \textbf{(a)} and on a long-range kink \textbf{%
(b)}. For the normal kink nothing is happening \textbf{(a)}. The long-range
kink is reacting to the action of a remote localized field whose position is
at point $x=-20$ (b).}
    \label{13NS}
\end{figure}

Can a conventional kink feel a remote localized field configuration? Let set $x_{1}=-20$, $x_{0}=-1$. Figure~\ref{13NS} shows that a long-range kink can interact with a remote localized field configuration. For the normal kink, nothing is happening.

Now, we will obtain the explicit tunneling condition for some typical long-range kinks. First, we will investigate the tunneling condition in equation \eqref{eqNS1} for two examples, one with $n=1$ and the second with $n=2$. Equation \eqref{eqNS1} is considered with the potential $U(\phi)=\left(
2/n\right)  \sin^{2n}(\phi/2)$. The field $F(x)$ is constructed with three extremely exponentially localized
bell-shaped functions $F_{1}(x)$, $F_{2}(x)$, and $F_{3}(x)$ that have the shape shown in Fig. \ref{6NS} ($F(x)=F_{1}(x)+F_{2}(x)+F_{3}(x)$). For $n=1$, the sine-Gordon kink will tunnel through the barrier if
\begin{equation}
F_{0}\left[  \frac{1}{\cosh(x_{1})}+\frac{1}{\cosh(x_{2})}\right]  >F_{m}.
\label{eqNS11}
\end{equation}
From inequality \eqref{eqNS11} it is clear that as $\left\vert x_{1}%
\right\vert $ and $\left\vert x_{2}\right\vert $ increase, for fixed $F_{0}$ and $F_{m}$, the left part of the equation decreases exponentially. This means that the tunneling condition \eqref{eqNS11} will not be satisfied already for some not very large values of $\left\vert x_{1}\right\vert $ and $\left\vert x_{2}\right\vert $.

Let us get the tunneling condition for $n=2$:
\begin{equation}
F_{0}\left[  \frac{1}{1+x_{1}^{2}/2}+\frac{1}{1+x_{2}^{2}/2}\right]  >F_{m}.
\label{eqNS12}
\end{equation}
All the tunneling conditions are obtained using the general tunneling condition \eqref{Eq:36}.

Now we can compare the tunneling conditions for a conventional kink and for a long-range kink, \eqref{eqNS11} and \eqref{eqNS12}. The function on the left of inequality \eqref{eqNS12} decreases as a power law as $\left\vert x_{1}\right\vert $ and $\left\vert x_{2}\right\vert $ increase. This means, for fixed $F_{0}$ and $F_{m}$ (using the same $F_{0}$ and $F_{m}$ in \eqref{eqNS11} and \eqref{eqNS12}), we conclude that tunneling is possible for much wider barriers when the kink is ($n=2$) long-range.

To give a concrete numerical example, let us consider the same $F(x)$ for the sine-Gordon kink and the long-range ($n=2$) kink. For $F_{0}=0.25$, $F_{m}=0.02$, $x_{1}=-6$, $x_{2}=6$, the calculations yield $0.25\left[  \cosh^{-1}(-6)+\cosh^{-1}(6)\right]<0.02$ for $n=1$ and $0.25\left[  [1+(0.5)(-6)^{2}]^{-1}+\right.$ $\left.[1+(0.5)(6)^{2}]^{-1}\right]  >0.02$, for $n=2$. This means the sine-Gordon kink
will not escape a 6-wide potential well tunneling through a 6-wide barrier. In contrast, the long-range ($n=2$) kink will tunnel through the 6-wide
barrier, thus escaping the 6-wide potential well. In the following we show numerical experiments for these two cases.

We are simulating the sine-Gordon equation for a conventional kink
\begin{equation}
\frac{\partial^{2}\phi}{\partial t^{2}}-\frac{\partial^{2}\phi}{\partial
x^{2}}+\sin\phi=F(x), \label{eqNS13}%
\end{equation}
with the initial conditions $\phi\left(  x,0\right)=4\arctan\left[
\exp\left(  x+1\right)  \right]  $, $\partial\phi/\partial t|_{\left(  x,t=0\right)}  =0$. Also, we are simulating the equation
\begin{equation}
\frac{\partial^{2}\phi}{\partial t^{2}}-\frac{\partial^{2}\phi}{\partial
x^{2}}+2\sin^{3}\left(  \frac{\phi}{2}\right)  \cos\left(  \frac{\phi}%
{2}\right)  =F(x),\label{eqNS14}%
\end{equation}
with the initial conditions $\phi\left(  x,0\right)=2\arctan\left[  \left(\sqrt{2}/2\right)  \left(  x+1\right)  \right]  +\pi$, $\partial\phi/\partial t|_{\left(  x,t=0\right)}  =0$. For both systems $F(x)=-F_{0}\ \mbox{sech}^{2}(x-x_{1})+F_{m}%
\ \mbox{sech}^{2}(x)-F_{0}\ \mbox{sech}^{2}(x-x_{2})$, with $F_{0}=0.25$, $F_{m}=0.02$, $x_{1}=-6$,and $x_{2}=6$.

\begin{figure}
    \centering
    \includegraphics[width=0.5\linewidth]{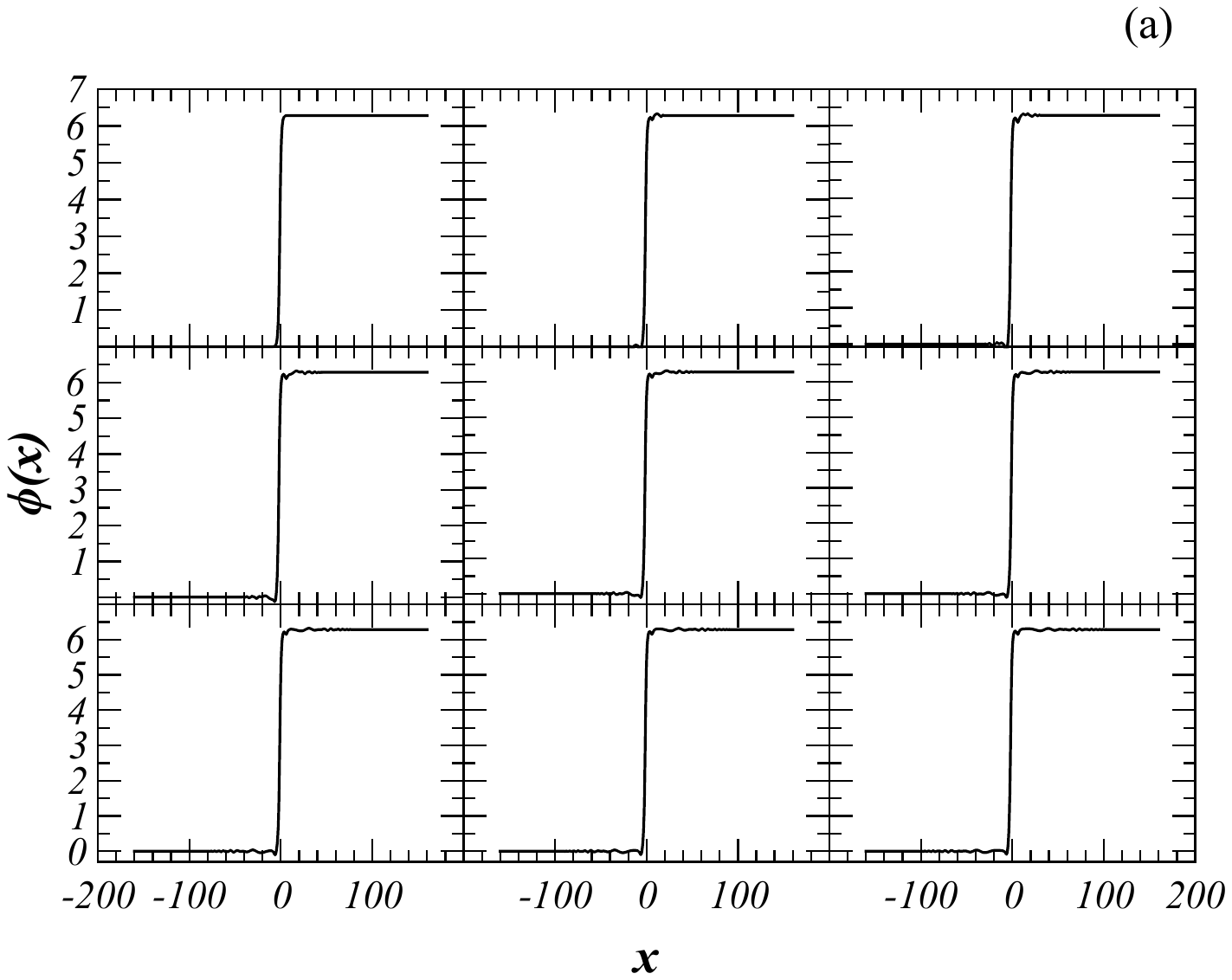}\includegraphics[width=0.5\linewidth]{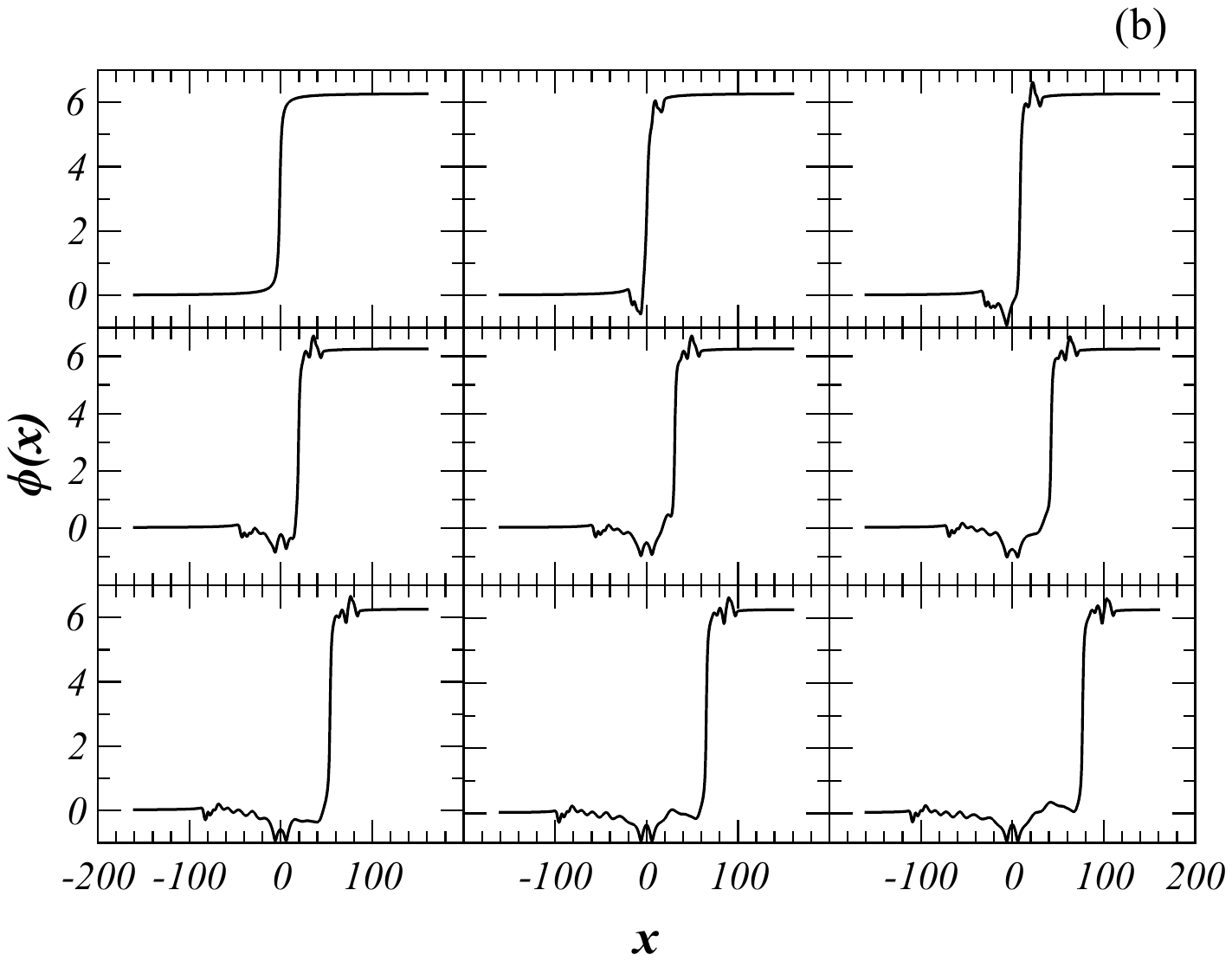}
    \caption{Spatiotemporal dynamics of kinks in a system with an $F(x)$ as
that shown in Fig. \ref{6NS}. \textbf{(a)} The sine-Gordon kink profile
dynamics (Eq.~\eqref{eqNS13}). The kink remains trapped inside the potential well on the left. 
\textbf{(b)} The long-range ($n=2$) kink is tunneling through a barrier that
is between the points $x=0$ and $x=6$ (Eq.~\eqref{eqNS14}). In both cases, the initial conditions
are constructed with the exact solutions to the respective nonlinear partial
differential equations. It is very clear that the initial long-range kink is
not excited.
}
    \label{14NS}
\end{figure}

\begin{figure}
    \centering
    \scalebox{0.5}{\includegraphics{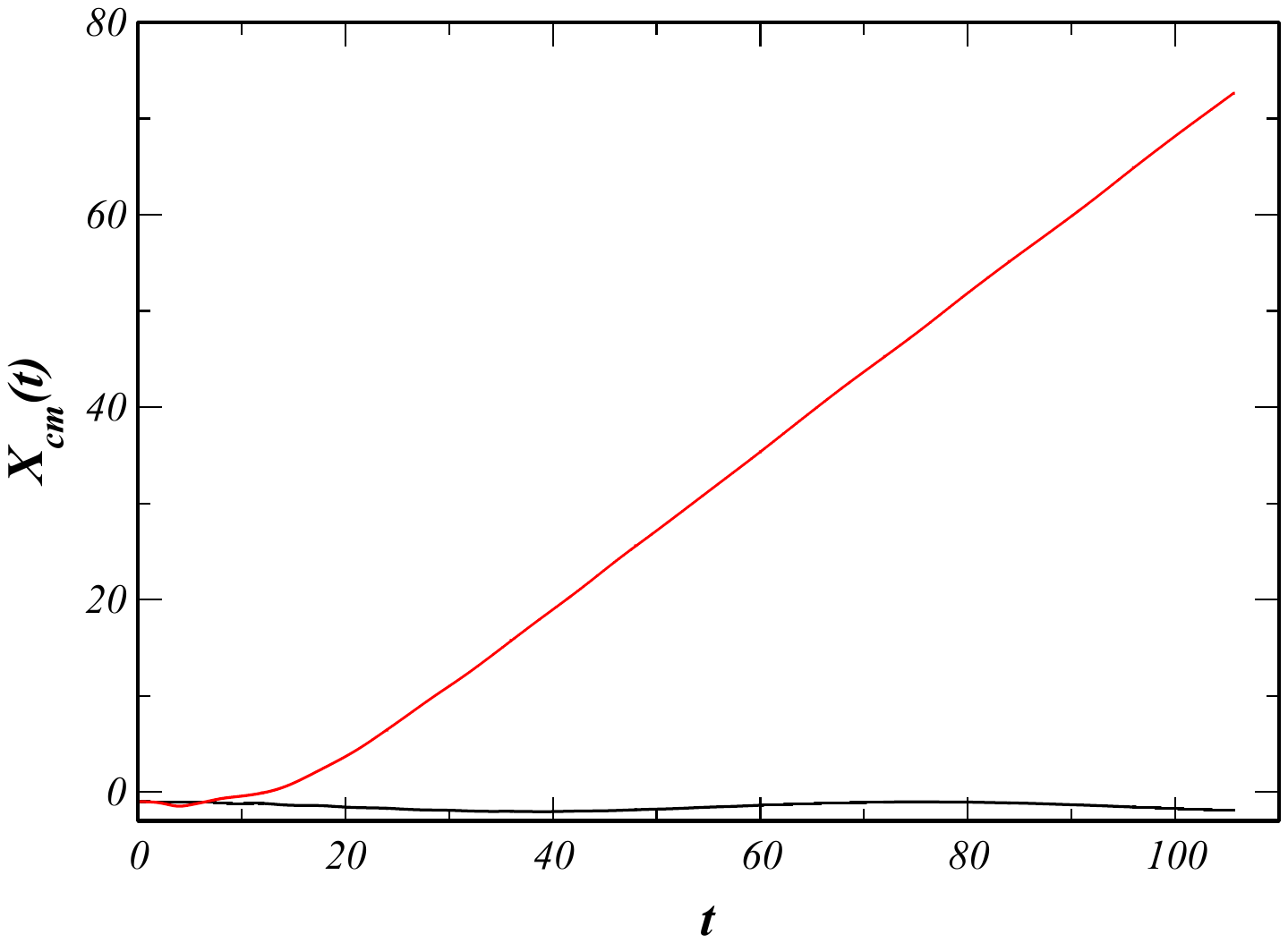}}
    \caption{Dynamics of the kink's center of mass. The black line is for the
sine-Gordon kink (Eq.~\eqref{eqNS13}). The red line is for the long-range ($n=2$) kink (Eq.~\eqref{eqNS14}). The
sine-Gordon kink never moves beyond the point $x=0$. The long-range kink is
moving decisively forward.
}
    \label{15NS}
\end{figure}

In both cases the initial conditions have been constructed with the exact stationary kink solutions to the complete nonlinear field equations, so the initial kinks are not excited. Of course, the heterogeneous external fields can create some deformations in the dynamical solutions, especially if the amplitude of the heterogeneous external fields is big. The results of the numerical experiments are shown in figures \ref{14NS} and \ref{15NS}.

For the equation
\begin{equation}
\frac{\partial^{2}\phi}{\partial t^{2}}-\frac{\partial^{2}\phi}{\partial
x^{2}}+\frac{1}{2}\phi\left(  \phi^{2}-1\right)  ^{2n-1}=F(x),\label{eqNS15}%
\end{equation}
the tunneling condition is the following:%
\begin{equation}
F_{0}%
\left\{  \frac{1}{\left[  1+\left(  \sqrt{2}-1\right)  \left\vert
x_{1}\right\vert \right]  ^{n/\left(  n-1\right)  }}+\frac{1}{\left[
1+\left(  \sqrt{2}-1\right)  \left\vert x_{2}\right\vert \right]  ^{n/\left(
n-1\right)  }}\right\}  >F_{m},\label{eqNS16}%
\end{equation}
where $n>1$. For $n=1$, the condition is
\begin{equation}
F_{0}\left[  \frac{1}{\cosh^{2}(0.5x_{1})}+\frac{1}{\cosh^{2}(0.5x_{2})}\right]  >F_{m}.\label{eqNS16b}%
\end{equation}
For the following $F(x)=-0.1\ \mbox{sech}^{2}(x+10)+0.02\ \mbox{sech}^{2}%
(x)-0.1\ \mbox{sech}^{2}(x-10)$, we obtain that the $\phi^{4}$-kink remains
trapped inside the potential well on the left. On the other hand, the
long-range ($n=9$) kink can tunnel through this very wide ($L_{B}=10$) barrier escaping the 10-wide potential well.

All the simulations that we have performed with Eq.~\eqref{eqNS15} are in agreement with the tunneling condition \eqref{eqNS16}. When $n$ is very large ($n\gg1$), the function on the left of inequality \eqref{eqNS16} decays with $\left\vert x_{1}\right\vert $ and $\left\vert
x_{2}\right\vert $ as
\begin{equation}
\frac{F_{0}}{\left(  \sqrt{2}-1\right)  }%
\left\{  \frac{1}{\left\vert x_{1}\right\vert }+\frac{1}{\left\vert
x_{2}\right\vert }\right\}  . \label{eqNS17}
\end{equation}
Hence, as the size of the potential well and/or the width of the barrier are increased, the function \eqref{eqNS17} decreases very slowly. The significance of this result is that even very wide barriers can be penetrated by long-range kinks.

\begin{figure}
    \centering
    \includegraphics[width=0.5\linewidth]{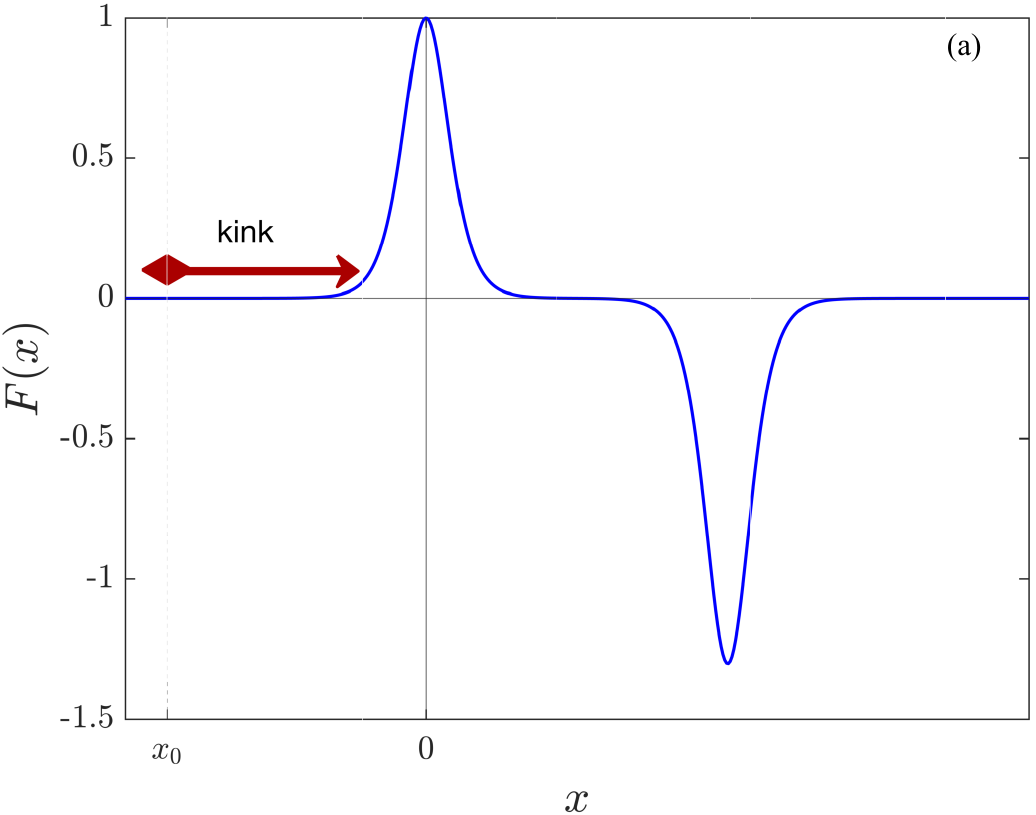}\includegraphics[width=0.5\linewidth]{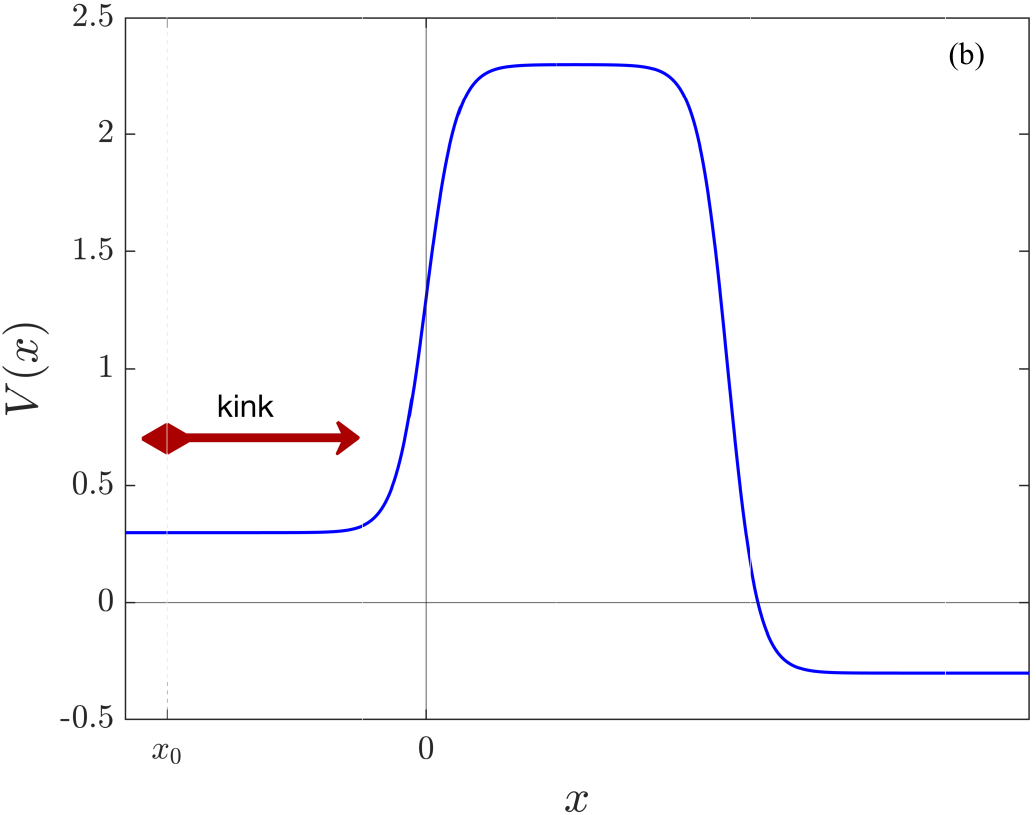}
    \caption{\textbf{(a)} $F(x)$ with a localized negative structure and one
localized positive structure.  \textbf{(b)} The $F(x)$ in \textbf{(a)}
generates a potential with a single barrier (without potential well).
}
    \label{16NS}
\end{figure}

The general tunneling condition \eqref{Eq:36} allows for the investigation of very different functions $F(x)$. For instance, it is possible to have an $F(x)$ that generates a system with a simple barrier (without a potential well). This can be appreciated in Fig. \ref{16NS}. When the tunneling condition \eqref{Eq:36} is satisfied, then a kink with an initial position $X_{cm}=x_{0}<0$ on the \textquotedblleft left\textquotedblright\ of the barrier can tunnel through the barrier. It is remarkable that the negative localized structure of $F(x)$ located at the point $x=x_{2}$ behind the barrier (actually, the farthest border of the barrier) can make the kink tunnel through the barrier.

If the barrier is sufficiently wide, shorter-range kinks cannot penetrate the barrier. They do not feel the interaction with the negative localized structure of $F(x)$ at the point $x=x_{2}$. On the other hand, long-range kinks do feel the pull of the negative structure at point $x=x_{2}$, even for large $x_{2}$. This phenomenon is another manifestation of the fact that long-range interactions play a fundamental role in long-range tunneling.

\begin{figure}
    \centering
    \includegraphics[width=0.5\linewidth]{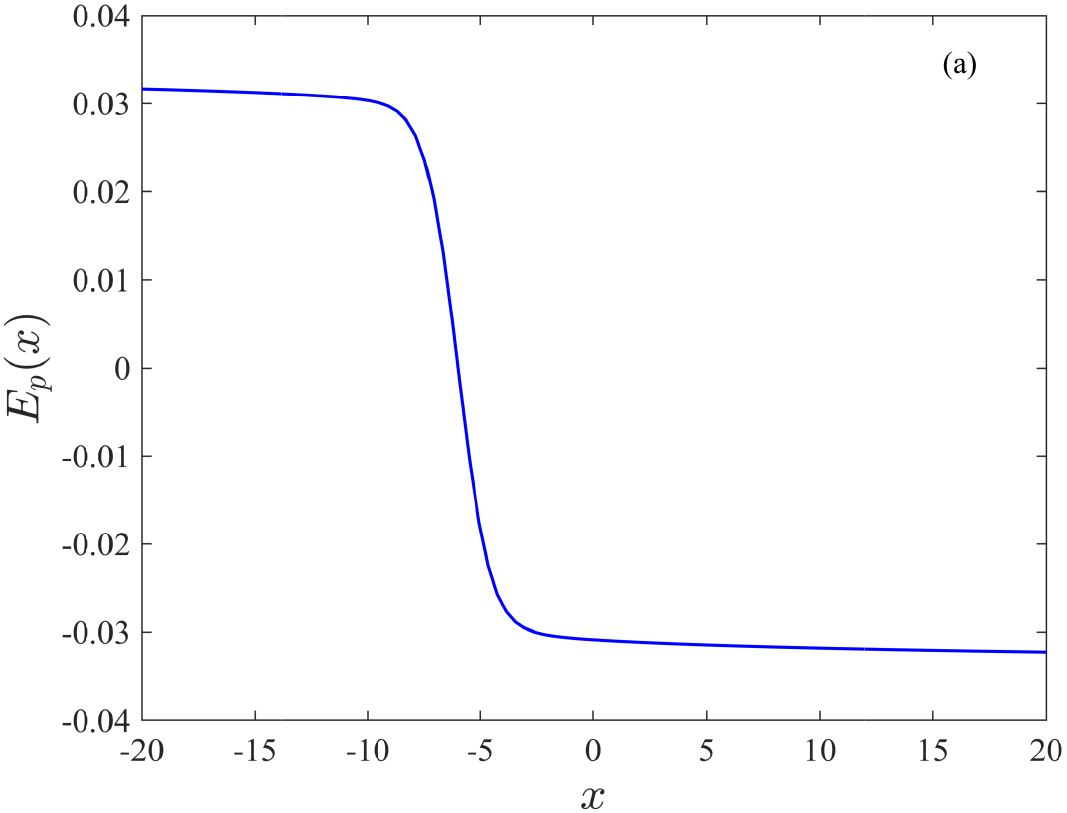}\includegraphics[width=0.5\linewidth]{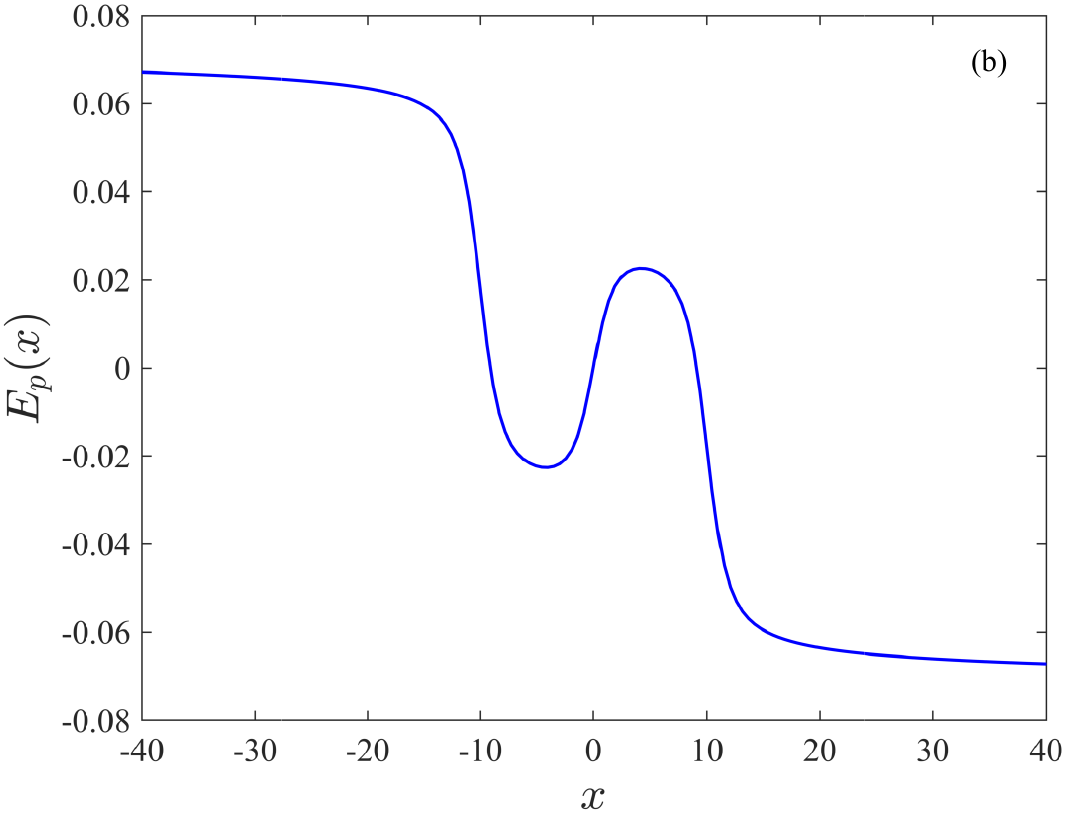}
    \caption{$E_{P}(x)$ can be seen as a superposition of three
components: an antikink-like structure at point $x=x_{1}$, another antikink-like structure at point $x=x_{2}$, and a kink-like structure at point $x=0$. \textbf{(a)} An example of the antikink-like structure at point $x=x_{1}$, generated by a single bell-like negative structure of $F(x)$ at point $x=x_{1}$. Even when $F(x)$ is exponentially localized for a system
with long-range kinks, \bigskip $E_{P}(x)$ will decay as a power-law for large $x$. This means that the kink will be moving forward under the action of a localized field for long distances and times. If $F(x)$ decays as a power-law, $E_{P}(x)$ will decay even slower. \textbf{(b)} Example when the components of $E_{P}(x)$ are antikink-like structures at points $x=x_{1}$
and $x=x_{2}$, and a kink-like structure at point $x=0$. However, the tunneling condition is not satisfied in this case.}
    \label{17NS}
\end{figure}

\begin{figure}
    \centering
    \includegraphics[width=0.5\linewidth]{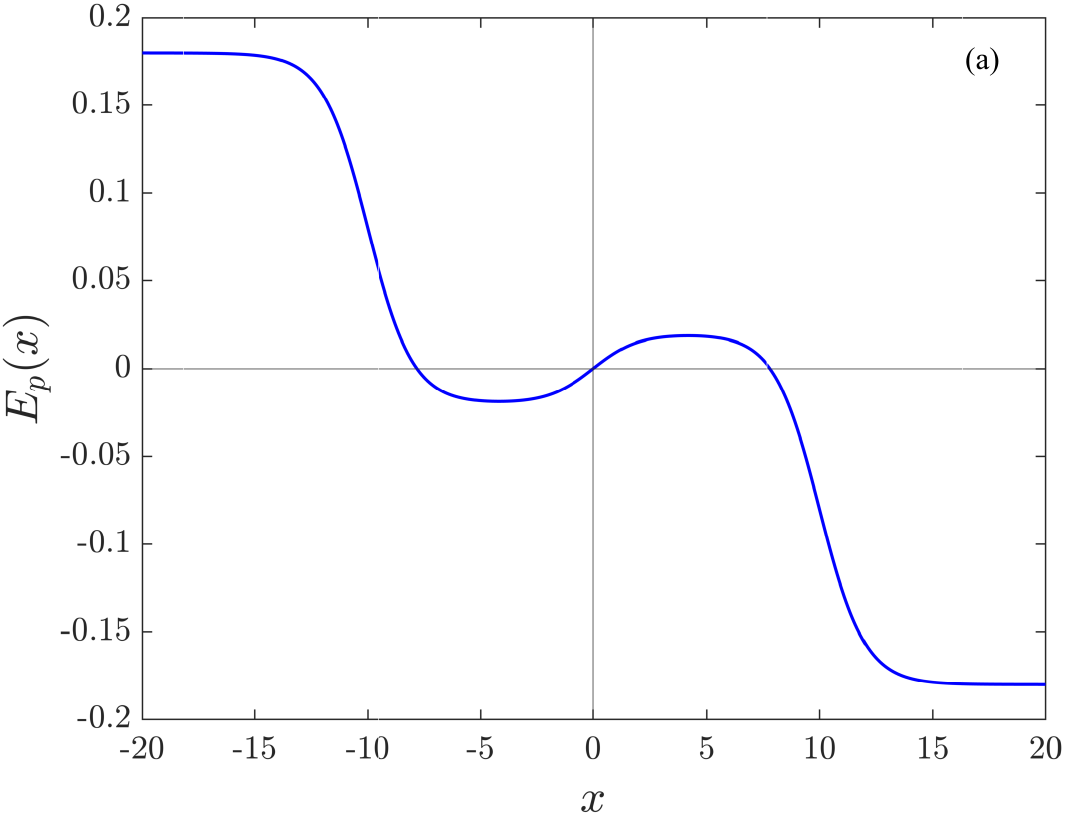}\includegraphics[width=0.5\linewidth]{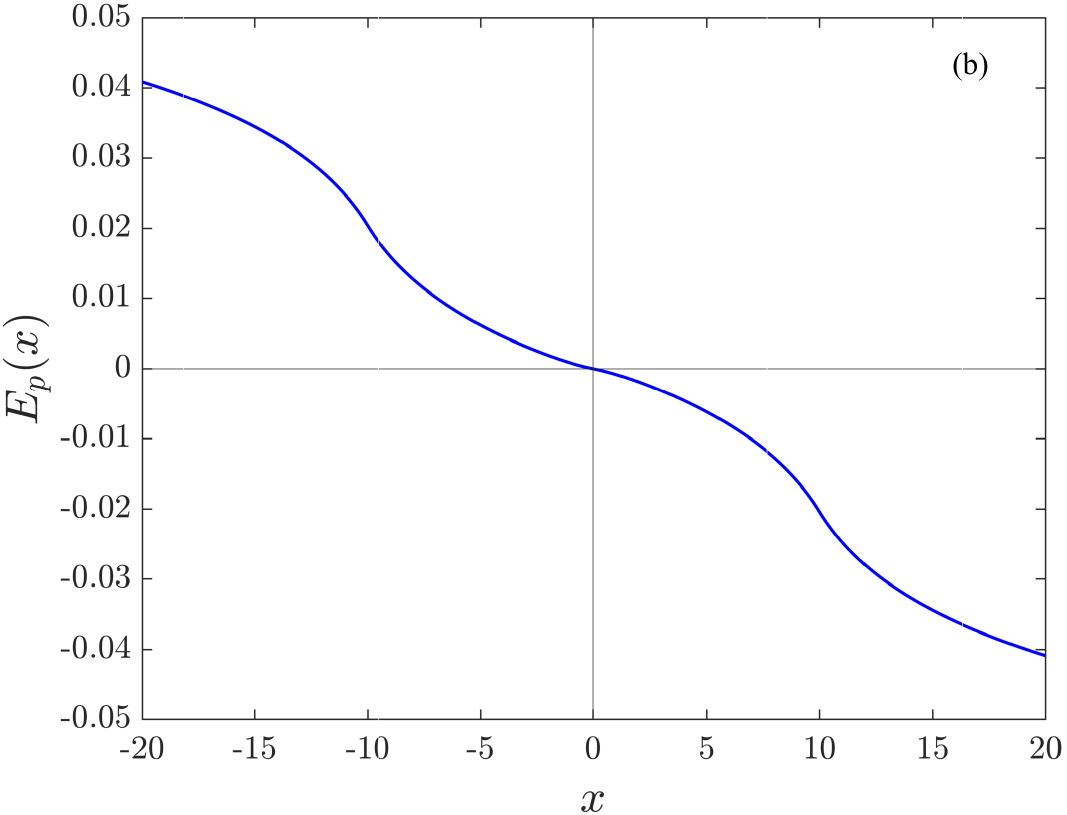}
    \caption{Behavior of quantity $E_{P}(x)$ for systems with
tunneling and without tunneling. \textbf{(a)} $E_{P}(x)$ constructed for
system \eqref{eqNS15} with $F(x)=-0.1\ \mbox{sech}^{2}(x+10)+0.02\ \mbox{sech}^{2}%
(x)-0.1\ \mbox{sech}^{2}(x-10)$ and $n=1$. At the same time
this is the typical behavior of $E_{P}(x)$ for a systems with an $F(x)$ as
that shown in Fig. \ref{6NS} when the kink remains trapped inside the
potential well.  \textbf{(b)} $E_{P}(x)$ constructed for system %
\eqref{eqNS15} with the same $F(x)$ as in \textbf{(a)} and with $n=9$. Also,
this is the typical behavior of $E_{P}(x)$ for a system with an $F(x)$ as
that shown in Fig. \ref{6NS} when the kink is able to tunnel through the
barrier.}
    \label{22NS}
\end{figure}

Now we will discuss tunneling in terms of function $E_{P}(x)$. For an external field $F(x)$ like that shown in Fig. \ref{6NS}, function $E_{P}(x)$ is a superposition of three components: an antikink-like structure with a center close to the point $x=x_{1}$, another antikink structure with a center close to $x=x_{2}$, and a kink-like structure at point $x=0$ (see Fig. \ref{17NS}). Even when the formulas that create the bell-shaped structures in $F(x)$ near the points $x_{1}$, $0$, and $x_{2}$ are not exponentially localized functions, the components of $E_{P}(x)$ are antikink-like or kink-like structures (see figures \ref{17NS} and \ref{22NS}). In general, there are only two possible very different regimes: \textit{i)} Function $E_{P}(x)$ has a local minimum and a local maximum in the interval $x_{1}<x<x_{2}$, or \emph{ii)} the function $E_{P}(x)$ is a monotonically decreasing function. These cases can be seen in Fig. \ref{22NS}. It is clear that in case of Fig.~\ref{22NS}(a), the kink will remain trapped inside the potential well. Meanwhile, in the case of Fig.~\ref{22NS}(b), the kink is able to move forward all the time despite the presence of a barrier. For a given system like Eq.~\eqref{eqNS15}, the analytical tunneling condition that decides if we are in the case (a) or in case (b) of Fig.~\ref{22NS} is given by the inequalities \eqref{eqNS16} and \eqref{eqNS16b}. 

\begin{figure}
    \centering
    \scalebox{0.5}{\includegraphics{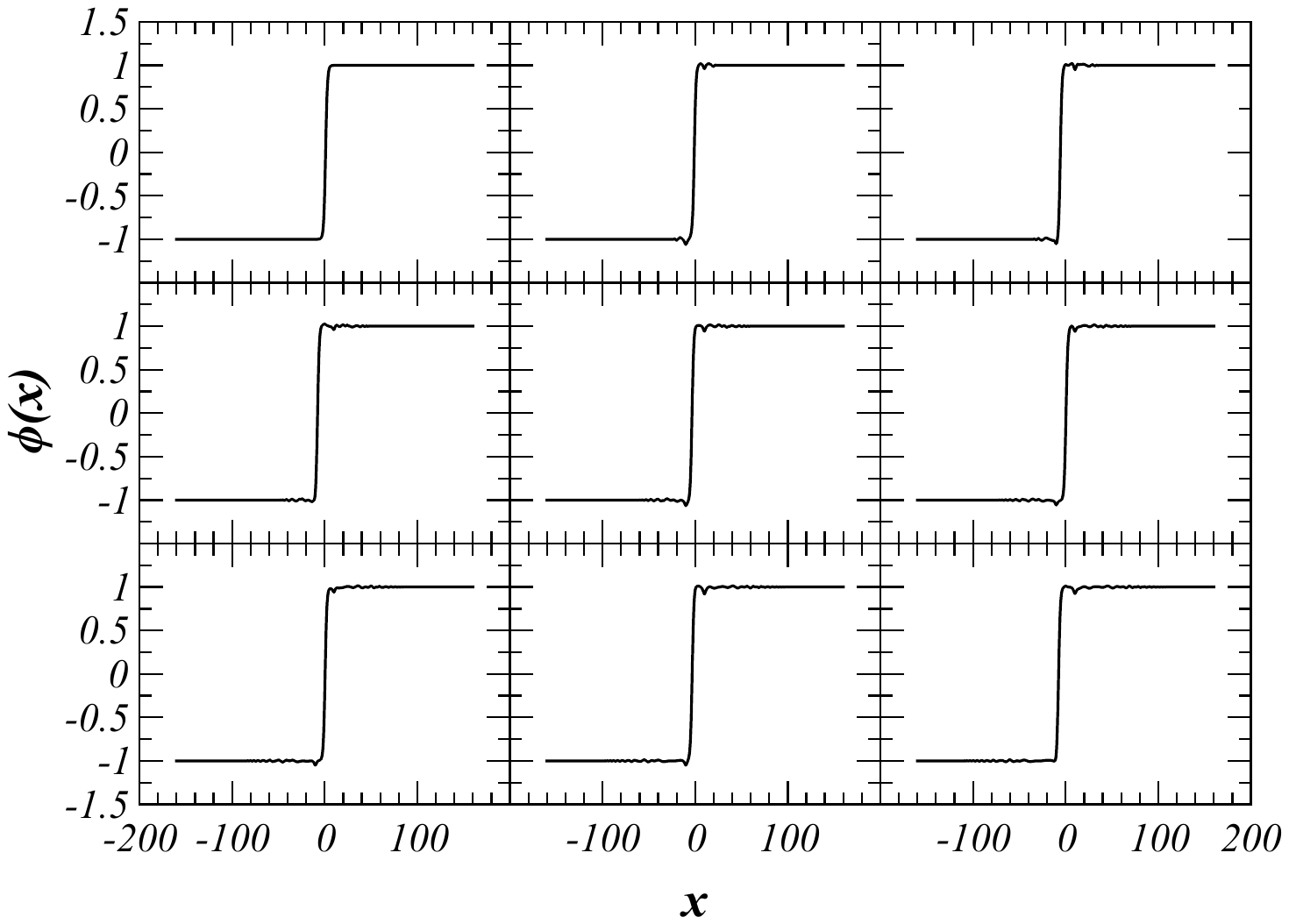}}
    \caption{The  $\phi ^{4}$-kink remains trapped all the time
inside the potential well, where $x_{cm}(t)$ is always negative ($x_{cm}<0$%
). Compare this behavior with Fig. \ref{23NS}.}
    \label{23pNS}
\end{figure}

\begin{figure}
    \centering
    \includegraphics[width=0.5\linewidth]{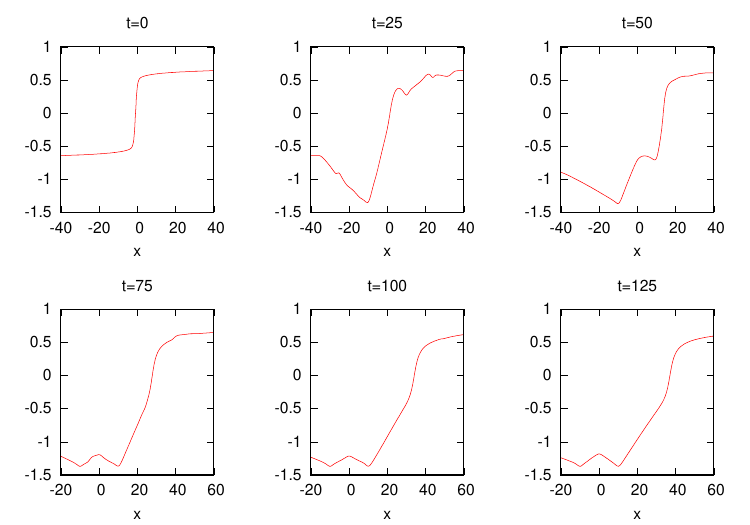}\includegraphics[width=0.5\linewidth]{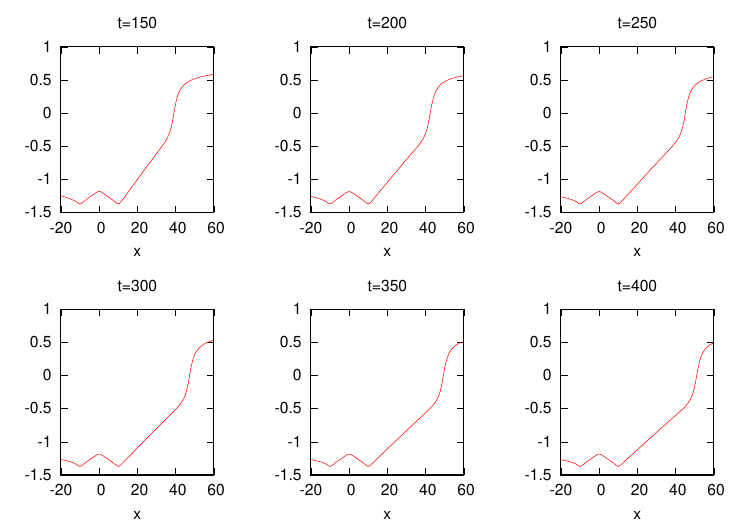}
    \caption{The long-range kink ($n=9$) is tunneling through a barrier that
exists between the points $x=0$ and $x=10$. Compare this outcome with that
shown in Fig. 23, where the $\phi ^{4}$-kink is unable to penetrate the same
barrier.
}
    \label{23NS}
\end{figure}

The functions $E_{P}(x)$ that can be seen in Fig. \ref{22NS} \ have been
constructed for the following two systems: $\phi _{tt}+0.1\phi _{t}-\phi
_{xx}+0.5\phi \left( \phi ^{2}-1\right) ^{2n-1}=F(x)$, $F(x)=-0.1\ \mbox{sech}%
^{2}(x+10)+0.02\ \mbox{sech}^{2}(x)-0.1\ \mbox{sech}^{2}(x-10)$, for $n=1$ (Fig. %
\ref{22NS}(a)) and for $n=9$ (Fig. \ref{22NS}(b)). Conditions \eqref{eqNS16} and \eqref{eqNS16b}
yield that for $n=1$ there is not tunneling whereas for $n=9$ there is
tunneling. See the outcomes of the simulation in figures \ref{23pNS} and \ref{23NS}. 

Here is a striking combination of results: the tunneling conditions %
\eqref{eqNS16} and \eqref{eqNS16b}, the quantity $E_{P}(x)$ (see Fig. \ref{22NS}), and the
numerical experiments (figures \ref{23pNS} and \ref{23NS}). All of them are in agreement with
the conclusion that the $\phi ^{4}$-kink is trapped inside the potential
well created by $F(x)=-0.1\ \mbox{sech}^{2}(x+10)+0.02\ \mbox{sech}^{2}(x)-0.1%
\ \mbox{sech}^{2}(x-10)$. However, the long-range ($n=9$) kink is tunneling
through the 10-wide barrier, thus escaping the 10-wide potential well.

The condition given in Eq.~\eqref{Eq:36} is valid for all Klein-Gordon equations. 

\begin{figure}
    \centering
    \includegraphics[width=0.59\linewidth]{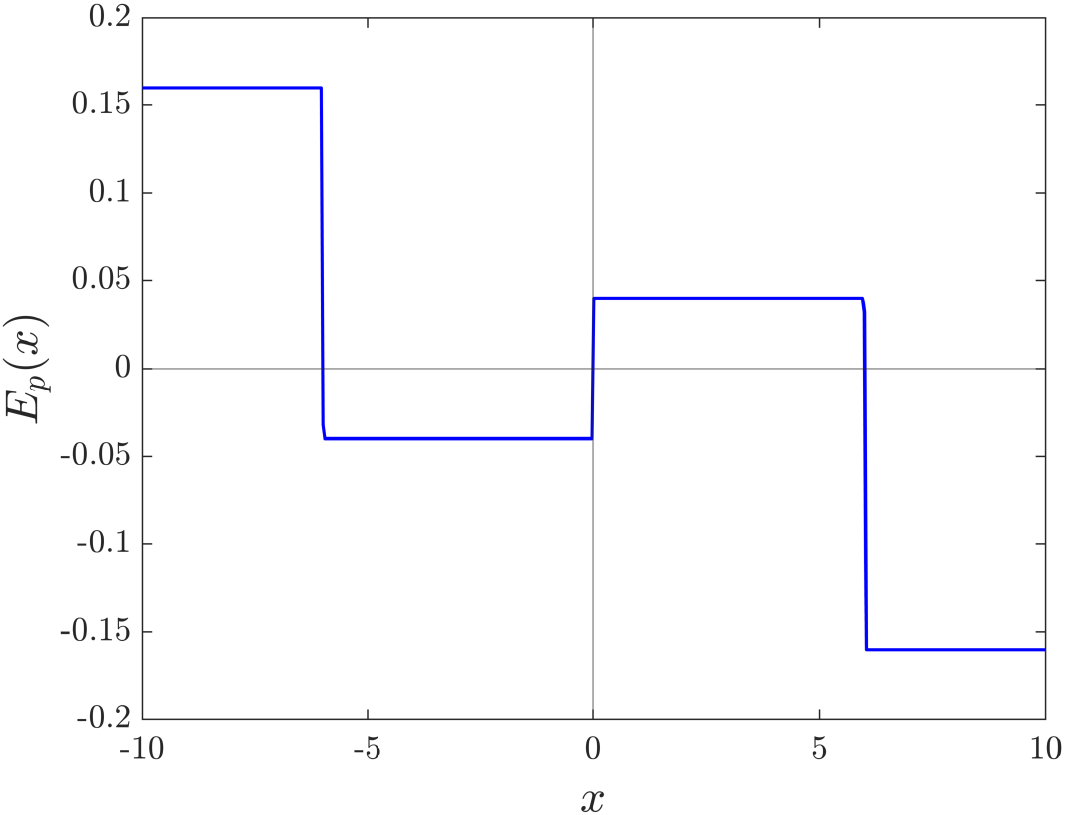}
    \caption{What happens if the kink is a truly point particle? The quantity $%
E_{P}(x)$ will always have a square-like potential well and a square-like
barrier. This means tunneling is impossible.
}
    \label{25NS}
\end{figure}

In this work, we have focused on perturbations created by exponentially localized impurities. In these systems, tunneling processes are very difficult. In the intervals $x_{1}+\varepsilon<x<\varepsilon$, and $0<x<x_{2}-\varepsilon$, the function $F(x)$ is exponentially small
(exponentially zero). Normal particle-like solutions do not participate in these tunneling processes (see Fig.~\ref{25NS}). Furthermore, we must say that if the function that creates the negative
bell-shaped structures in $F(x)$ close to the points $x=x_{1}$ and $x=x_{2}$ are more extended perturbations, condition \eqref{Eq:36} leads to even more spectacular propagation phenomena. These new phenomena will be discussed in forthcoming works.

\begin{figure}
    \centering
    \includegraphics[width=0.5\linewidth]{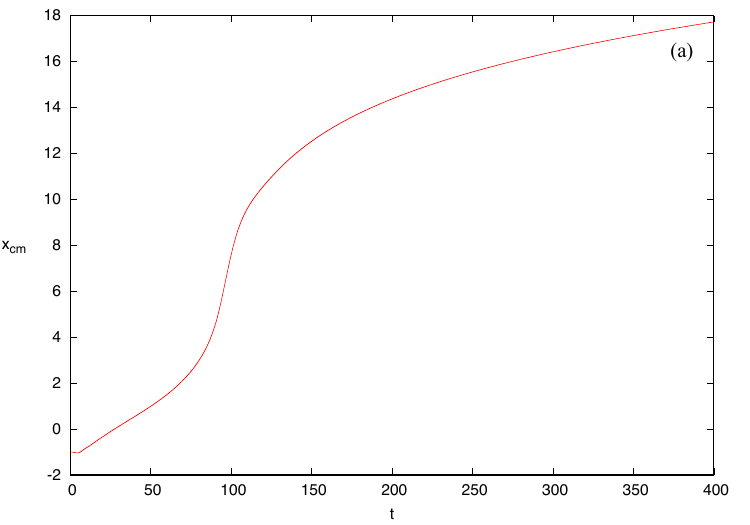}\includegraphics[width=0.5\linewidth]{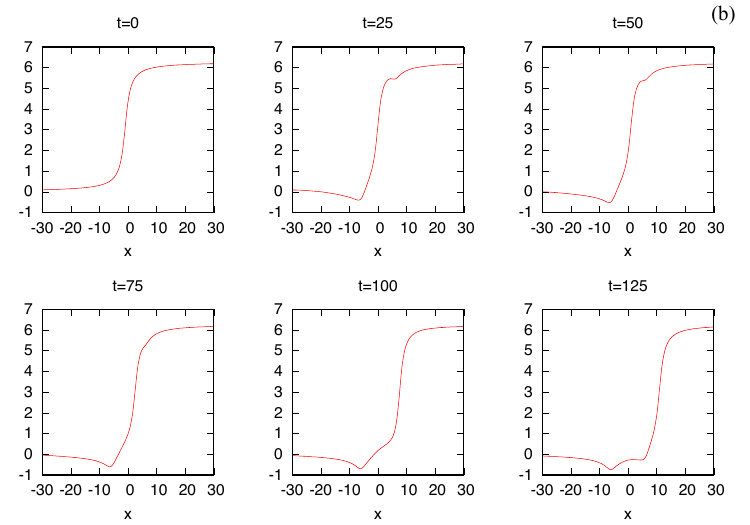}
    \caption{Simulation of Eq.~\eqref{Eq:4.17} which is the same system as the Eq.~\eqref{eqNS14}
    whose dynamics is shown in figures \ref{14NS}(b) and \ref{15NS}, but now we are doing it with damping.}
    \label{30NS}
\end{figure}

Now, we will briefly address the effects of damping, mass, and kink
excitation. Here, we are repeating the numerical simulation shown in figures \ref{14NS} and \ref{15NS} with damping. The conditions are the following
\begin{equation}
\frac{\partial^2\phi}{\partial t^2} + (0.3) \frac{\partial\phi}{\partial t} - \frac{\partial^2\phi}{\partial x^2} + 2\sin^3\left(\frac{\phi}{2}\right) \cos\left(\frac{\phi}{2}\right)=F(x),
\label{Eq:4.17}
\end{equation}
\begin{equation}
\phi(x,0) = 2\arctan\left[\left(\frac{\sqrt{2}}{2}\right)(x+1)\right]+\pi,
\end{equation}
\begin{equation}
\frac{\partial\phi}{\partial t}(x,0) = 0,
\end{equation}
\begin{equation}
F(x)=-0.25 \ \mbox{sech}^{2}(x+6) + 0.02 \ \mbox{sech}^{2}(x) - 0.25 \ \mbox{sech}^{2}(x-6).
\end{equation} 
The results are shown in Fig. \ref{30NS}. Note that it takes the kink more time to tunnel and the motion after the barrier is slowing down. If the sine-Gordon kink is trapped in the potential well when there is no damping, with the presence of damping the sine-Gordon kink will be stuck there with more reason.

The dynamics observed in Fig. \ref{30NS} can be observed also in the presence of strong damping ($b=1$).
\begin{equation}
\frac{\partial^2\phi}{\partial t^2} + \frac{\partial\phi}{\partial t} - \frac{\partial^2\phi}{\partial x^2} + 0.5 \phi\left(\phi^2-1\right)^{17} = F(x).
\end{equation}

\begin{figure}
    \centering
    \includegraphics[width=0.5\linewidth]{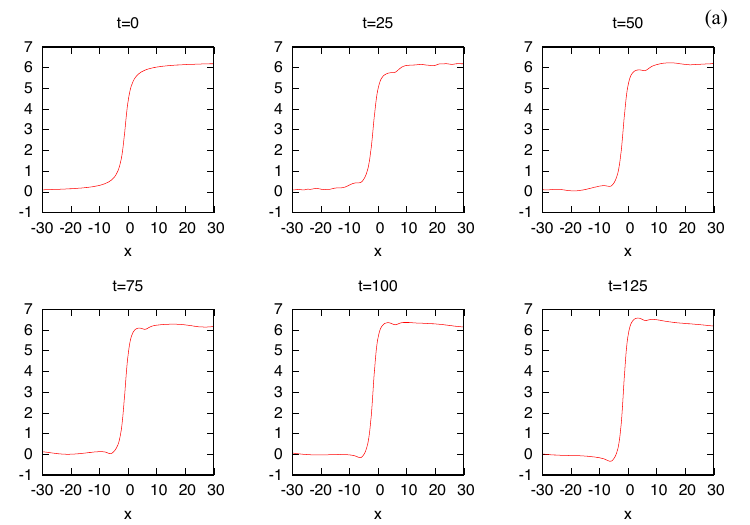}\includegraphics[width=0.5\linewidth]{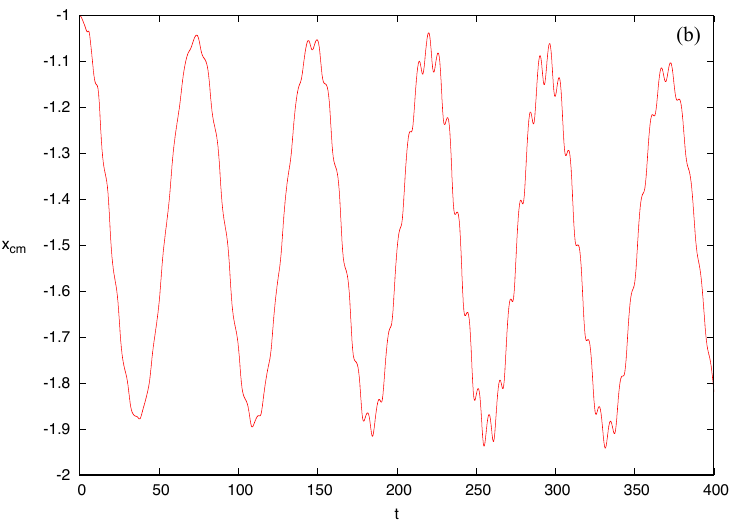}
    \caption{The sine-Gordon equation has been simulated with a kink-like
initial condition that has much smaller mass ($m \approx 4.442$). However, the tunneling does
not occur. The small mass does not help the sine-Gordon equation.
 }
    \label{32NS}
\end{figure}

What happens if we put a kink-like structure as an initial condition with less mass than the sine-Gordon kink? The simulation that we are proposing is
\begin{equation}
\frac{\partial^2\phi}{\partial t^2} - \frac{\partial^2\phi}{\partial x^2} + \sin(\phi)=F(x),
\end{equation}
\begin{equation}
\phi(x,0) = 2\arctan\left[\left(\frac{\sqrt{2}}{2}\right)(x+1)\right]+\pi,
\end{equation}
\begin{equation}
\frac{\partial\phi}{\partial t}(x,0) = 0,
\end{equation}
\begin{equation}
F(x)=-0.25 \ \mbox{sech}^{2}(x+6) + 0.02 \ \mbox{sech}^{2}(x) - 0.25 \ \mbox{sech}^{2}(x-6).
\end{equation}
Fig. \ref{32NS} shows that the kink is not able to tunnel through the 6-wide barrier. The small mass does not help the sine-Gordon equation.

\begin{figure}
    \centering
    \includegraphics[width=0.5\linewidth]{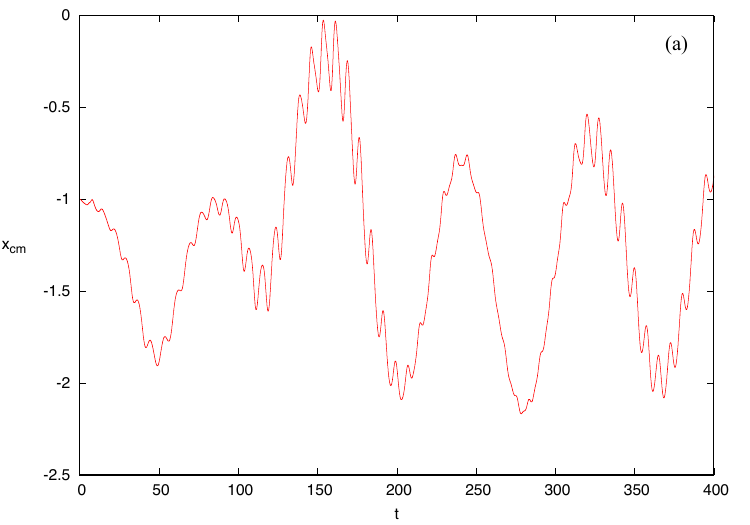}\includegraphics[width=0.5\linewidth]{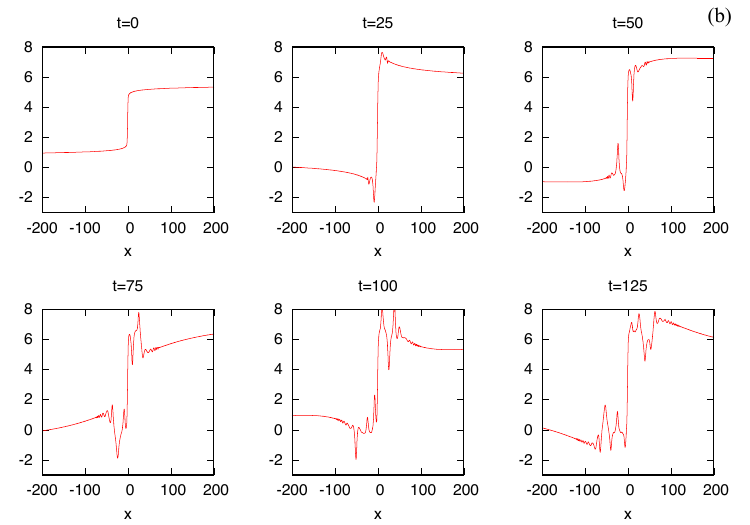}
    \caption{The kink can become very excited due to the fact that the initial
condition is very far from the exact kink solution \ Nevertheless, this does
not help the tunneling porcess.  \textbf{(a)} The position of the kink's
center of mass is always inside the potential well with $x_{cm}(t)<0$ for
any $t$. \textbf{(b)} Large shape deformations can be observed.
 }
    \label{33NS}
\end{figure}

\begin{figure}
    \centering
    \includegraphics[width=0.5\linewidth]{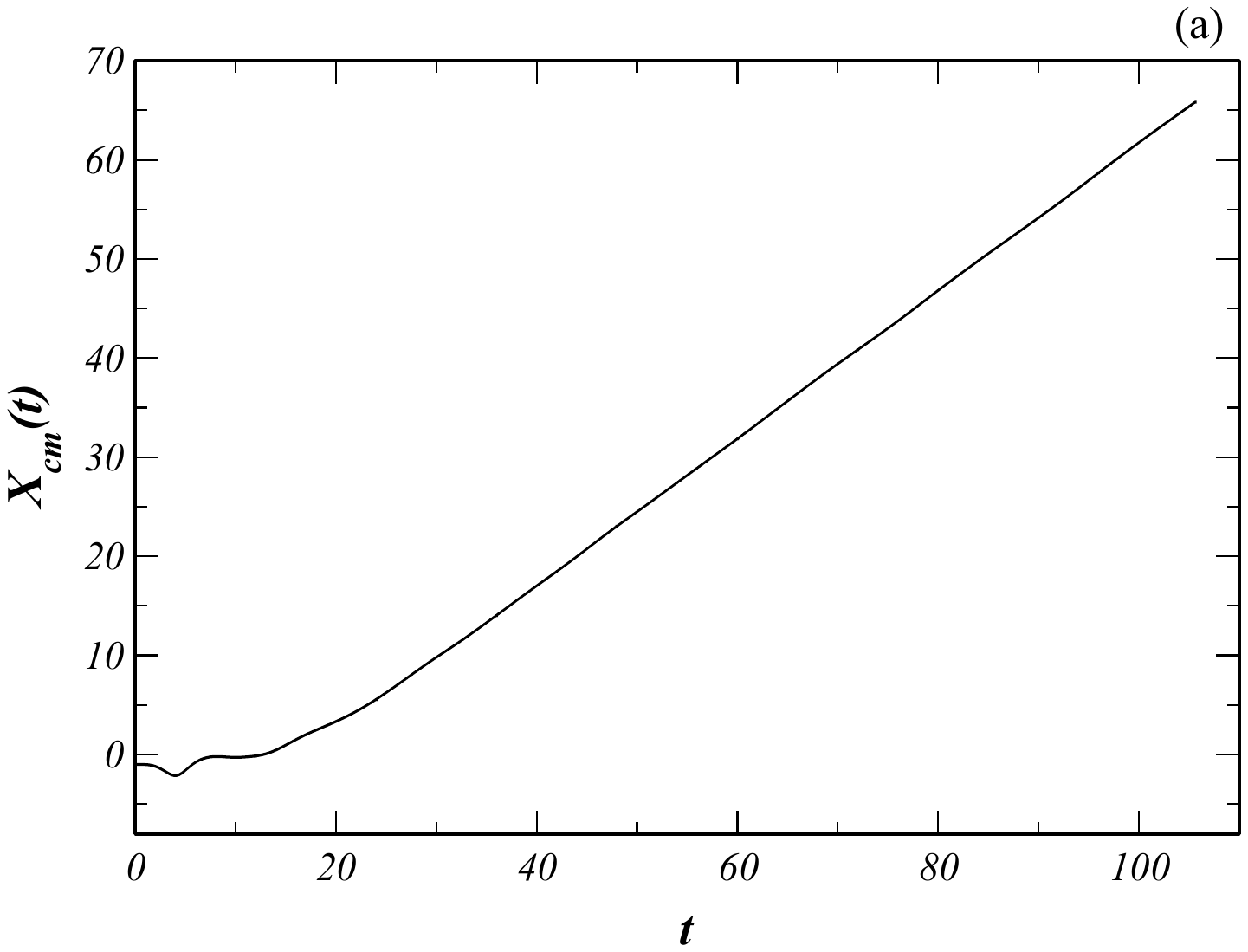}\includegraphics[width=0.5\linewidth]{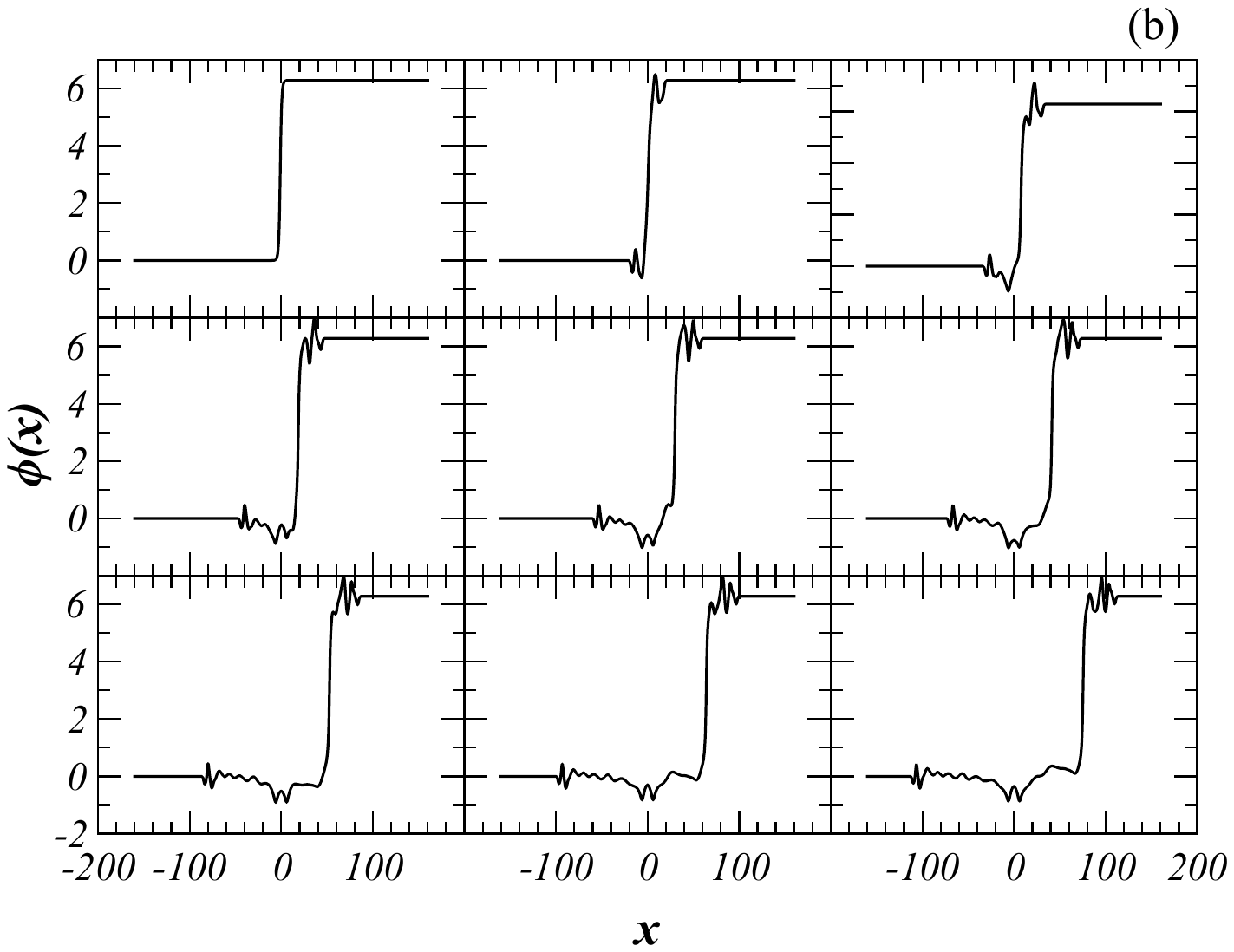}
    \caption{Tunneling in system~\eqref{Eq:4.30} using the sine-Gordon kink as
the initial condition. The mass of this initial kink is $m=8$. \textbf{(a)}
Dynamics of the kink's center of mass. \textbf{(b)} Spatiotemporal dynamics
of the kink in system \eqref{Eq:4.30}. The kink is tunneling through the
barrier despite the much larger mass of the initial kink.
}
    \label{34NS}
\end{figure}

Now we suggest the following simulation
\begin{equation}
\frac{\partial^2\phi}{\partial t^2} - \frac{\partial^2\phi}{\partial x^2} + \sin(\phi)=F(x),
\end{equation}
\begin{equation}
\phi(x,0) = 2\arctan\left[\tanh(x+1) |x+1|^{1/8}\right]+\pi,
\end{equation}
\begin{equation}
\frac{\partial\phi}{\partial t}(x,0) = 0,
\end{equation}
\begin{equation}
F(x)=-0.25 \ \mbox{sech}^{2}(x+6) + 0.02 \ \mbox{sech}^{2}(x) - 0.25 \ \mbox{sech}^{2}(x-6).
\end{equation}
The shape of the initial condition is so far from the sine-Gordon kink exact solution that the system generates enormous shape deformations. We are doing this without damping. Thus, the deformations are not deleted by the dissipative dynamics. This can be seen in Fig. \ref{33NS}. The kink is very excited. But all this does not help the tunneling process.

Could a greater mass thwart the tunneling? We will use the sine-Gordon kink as the initial condition for our 
equation~\eqref{eqNS14}
with $n=2$.

The simulation is the following
\begin{equation}
\label{Eq:4.30}
\frac{\partial^2\phi}{\partial t^2} - \frac{\partial^2\phi}{\partial x^2} + 2\sin^3\left(\frac{\phi}{2}\right) \cos\left(\frac{\phi}{2}\right)=F(x),
\end{equation}
\begin{equation}
\phi(x,0) = 4\arctan\left[\exp(x+1)\right],
\end{equation}
\begin{equation}
\frac{\partial\phi}{\partial t}(x,0) = 0,
\end{equation}
\begin{equation}
F(x)=-0.25 \ \mbox{sech}^{2}(x+6) + 0.02 \ \mbox{sech}^{2}(x) - 0.25 \ \mbox{sech}^{2}(x-6).
\end{equation}
The mass of the initial structure is $m=8$. This is a very large mass. The spatiotemporal dynamis is shown in Fig. \ref{34NS}. The kink is tunneling through the 6-wide barrier. The mass is not what determines the outcome. The nonlinear partial differential equation~\eqref{Eq:4.30}  ``knows'' its kink is long-range and its dynamics will behave accordingly.

\section{Numerical simulations}
\label{Sec:Tunneling}

In this section, we will show that the tunneling of long-range kinks can be highly enhanced by increasing their long-range character. The nonlocal properties of soliton dynamics \cite{Gonzalez1998, Gonzalez1987, Gonzalez1999, Gonzalez1992} can be observed in all their extraordinary splendor during the propagation of long-range kinks in heterogeneous systems. As quantum tunneling is considered a non-equilibrium process, we will see that long-range kink tunneling is also a non-equilibrium process.

\begin{figure}
    \centering
    \scalebox{0.4}{\includegraphics{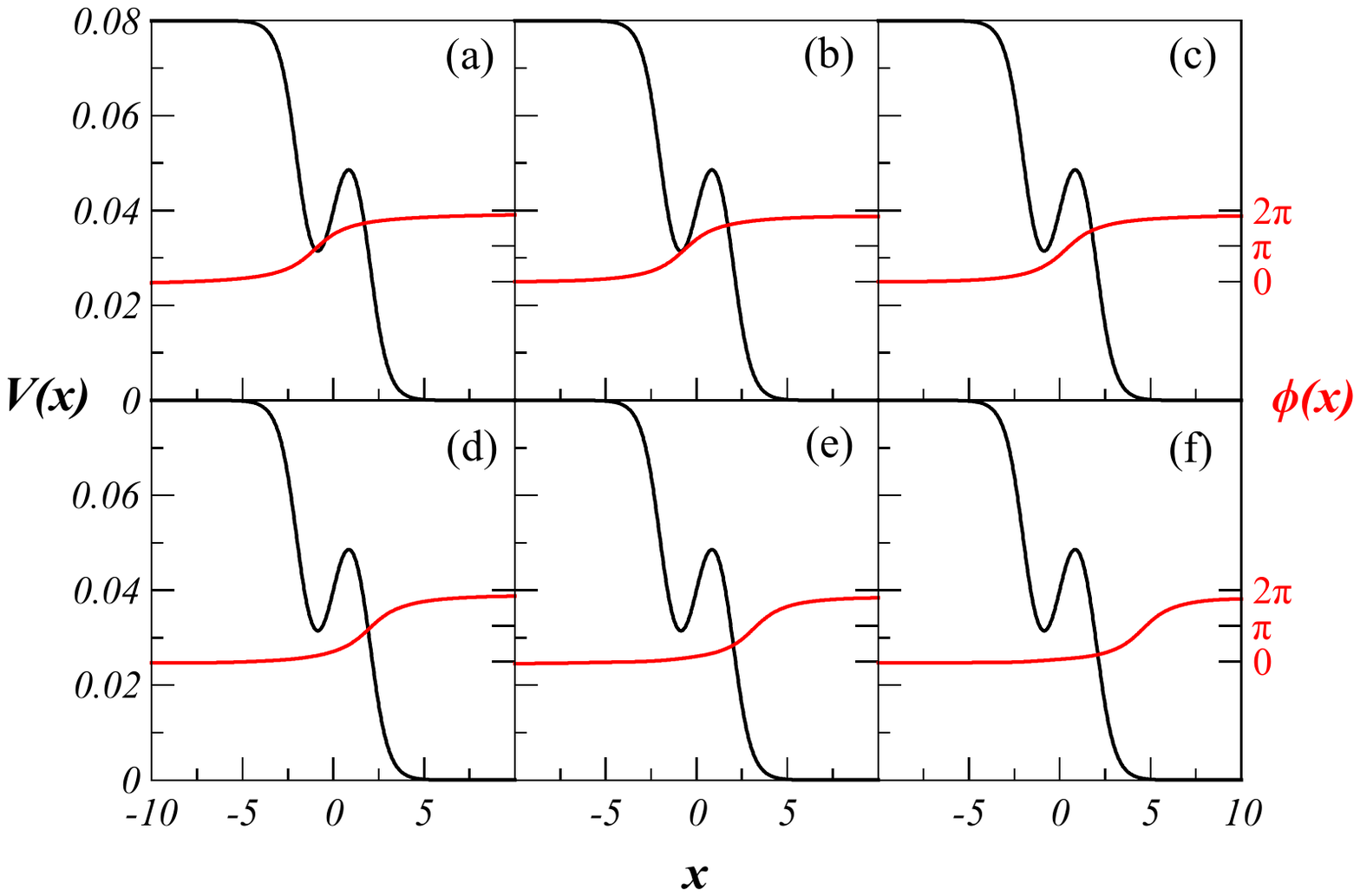}}
    \caption{A long-range kink tunneling through a potential barrier. Results obtained from the numerical simulation of Eq.~\eqref{Eq:30} for $b=0$ with $n=2$ and the initial conditions given in Eq.~\eqref{Eq:31} under the heterogeneous force of Eq.~\eqref{Eq:32} with $x_0=2$, $\alpha=0.01$, and $\beta=0.007$.}
    \label{Fig:07}
\end{figure}

Figure~\ref{Fig:07} shows a long-range kink tunneling through a potential barrier. This simulation was performed using Eq.~\eqref{Eq:30} with $n=2$. The initial conditions were
\begin{subequations}
\label{Eq:31}
\begin{equation}
    \label{Eq:31a}
\phi(x,0)=2\arctan\left[\left(\frac{\sqrt{2}}{2}\right)(x+1)\right]+\pi,
\end{equation}
\begin{equation}
    \label{Eq:31b}
\partial_t\phi(x,0)=0.
\end{equation}
\end{subequations}
The heterogeneous field was
\begin{equation}
    \label{Eq:32}
F(x)=-\alpha\ \mbox{sech}^2(x-x_1)+\beta\ \mbox{sech}^{2}(x)-\alpha\ \mbox{sech}^2(x-x_2),
\end{equation}
with $x_1=-2$, $x_2=2$, $\alpha=0.01$, and $\beta=0.007$.

\begin{figure}
    \centering
    \scalebox{0.35}{\includegraphics{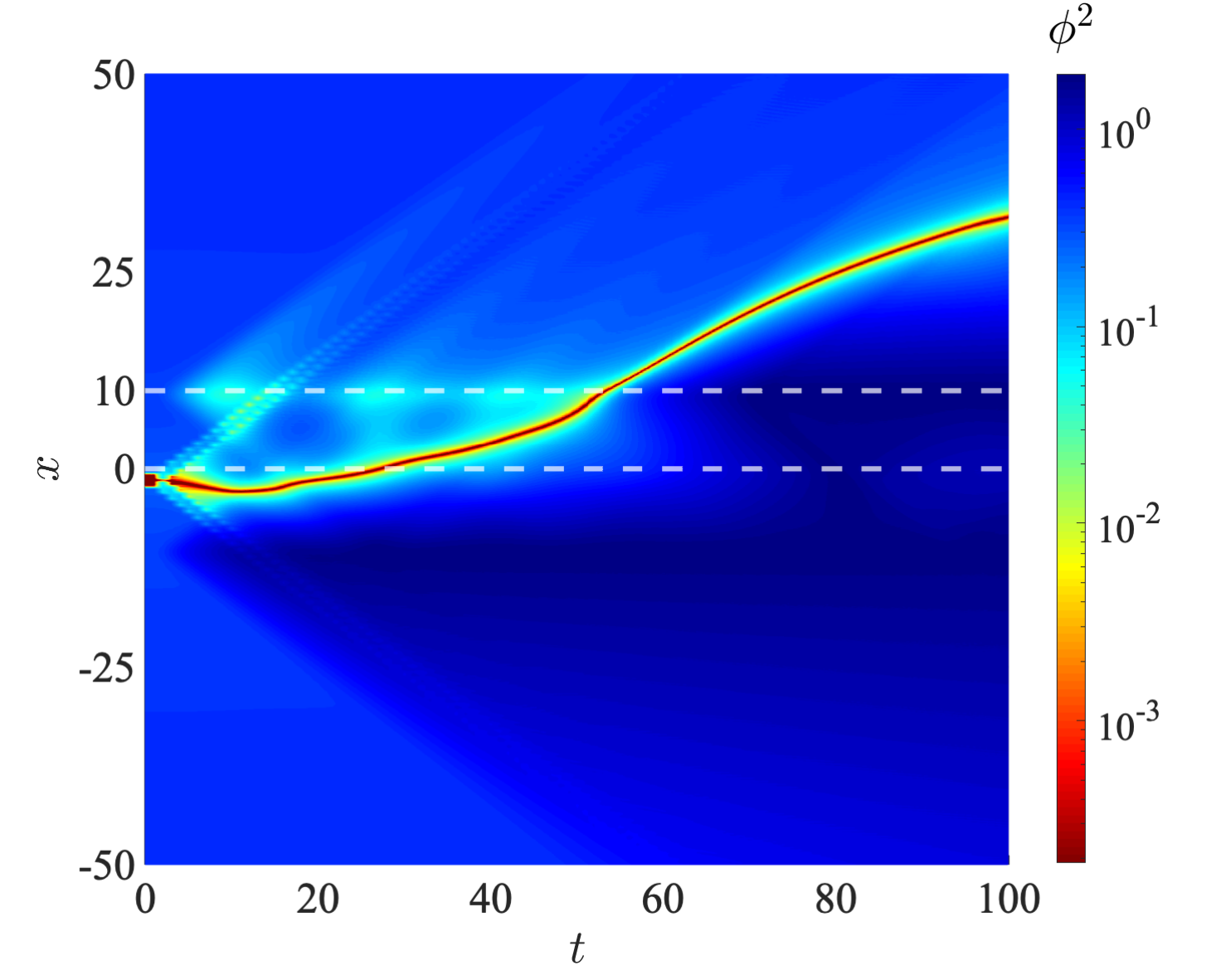}}
    \caption{Numerical simulation of the tunneling of a long-range kink in Eq.~\eqref{Eq:29} for $b=0.2$ with $n=9$, under the force of Eq.~\eqref{Eq:32} with $\alpha=0.1$, $\beta=0.02$, and $\vert x_1\vert=\vert x_2\vert=10$.}
     \label{Fig:09}
\end{figure}

Figure~\ref{Fig:09} shows the tunneling of a long-range kink through a wider barrier of width $w=10$. Here we are simulating Eq.~\eqref{Eq:29} with $n=9$, under the force of Eq.~\eqref{Eq:32} with $\alpha=0.1$, $\beta=0.02$, and $-x_1=x_2=10$. 

\begin{figure}
    \centering
    \scalebox{0.4}{\includegraphics{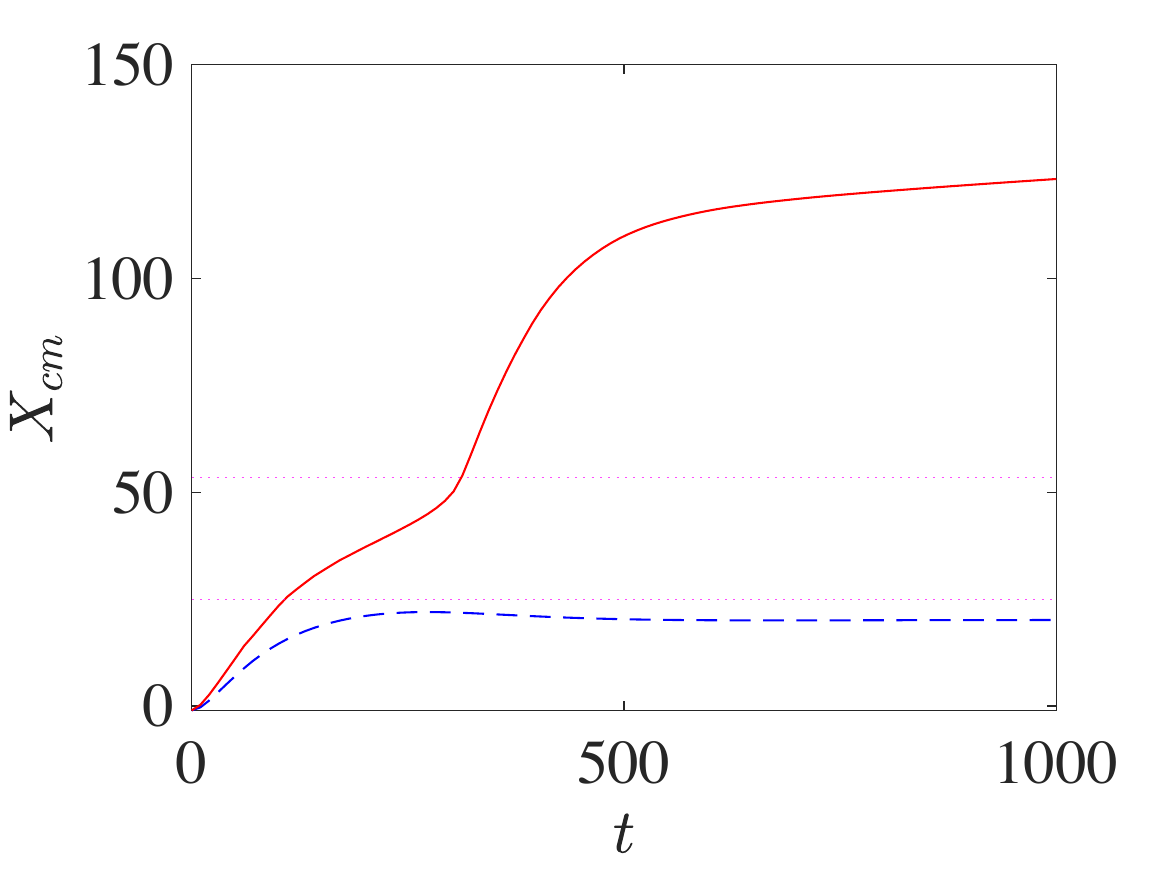}}
    \caption{Position of the center of mass of two long-range kinks with $n=2$ (blue dashed lines) and $n=9$ (red solid line). Numerical simulations obtained from Eq.~\eqref{Eq:29} for $b=0.01\,.$ }
    \label{Fig:10}
\end{figure}

\begin{figure}
    \centering
    \scalebox{0.65}{\includegraphics{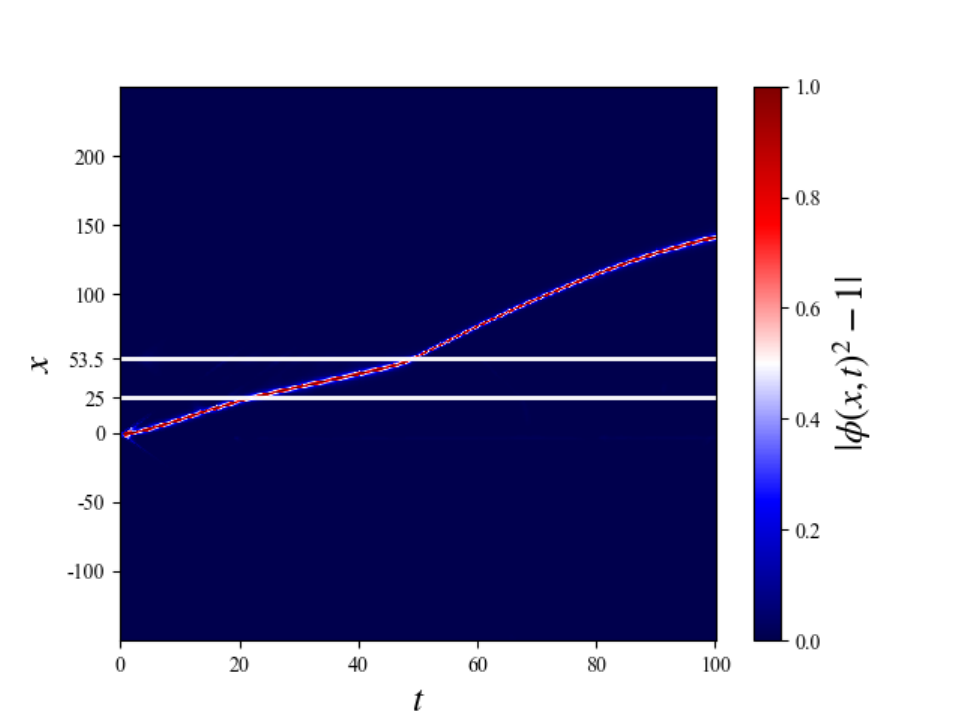}}
    \caption{Spatiotemporal dynamics of a long-range kink tunneling through a barrier. Numerical simulations obtained from Eq.~\eqref{Eq:29} for $b=0.01$.}
    \label{Fig:11}
\end{figure}

The difference between two long-range kinks with different extents is shown in Fig.~\ref{Fig:10}. A long-range kink with $n=2$ from Eq.~\eqref{Eq:29} collides with an extensive barrier. The long-range kink rebounds after colliding with the barrier. It is not able to penetrate such a strong obstacle. Nevertheless, the $n=9$ long-range kink (also from Eq.~\eqref{Eq:29}) is tunneling through this very long wall, and we can see it moving on the other side of the barrier. The spatiotemporal dynamics of an $n=9$ long-range kink can be observed in Fig.~\ref{Fig:11}. Within the context of Josephson junctions, notice that conventional fluxons would never penetrate a barrier like this. On the other hand, long-range fluxons created in a Josephson junction with nonlocal electrodynamics could cross this kind of barriers.

\begin{figure}
    \centering
    \scalebox{0.45}{\includegraphics{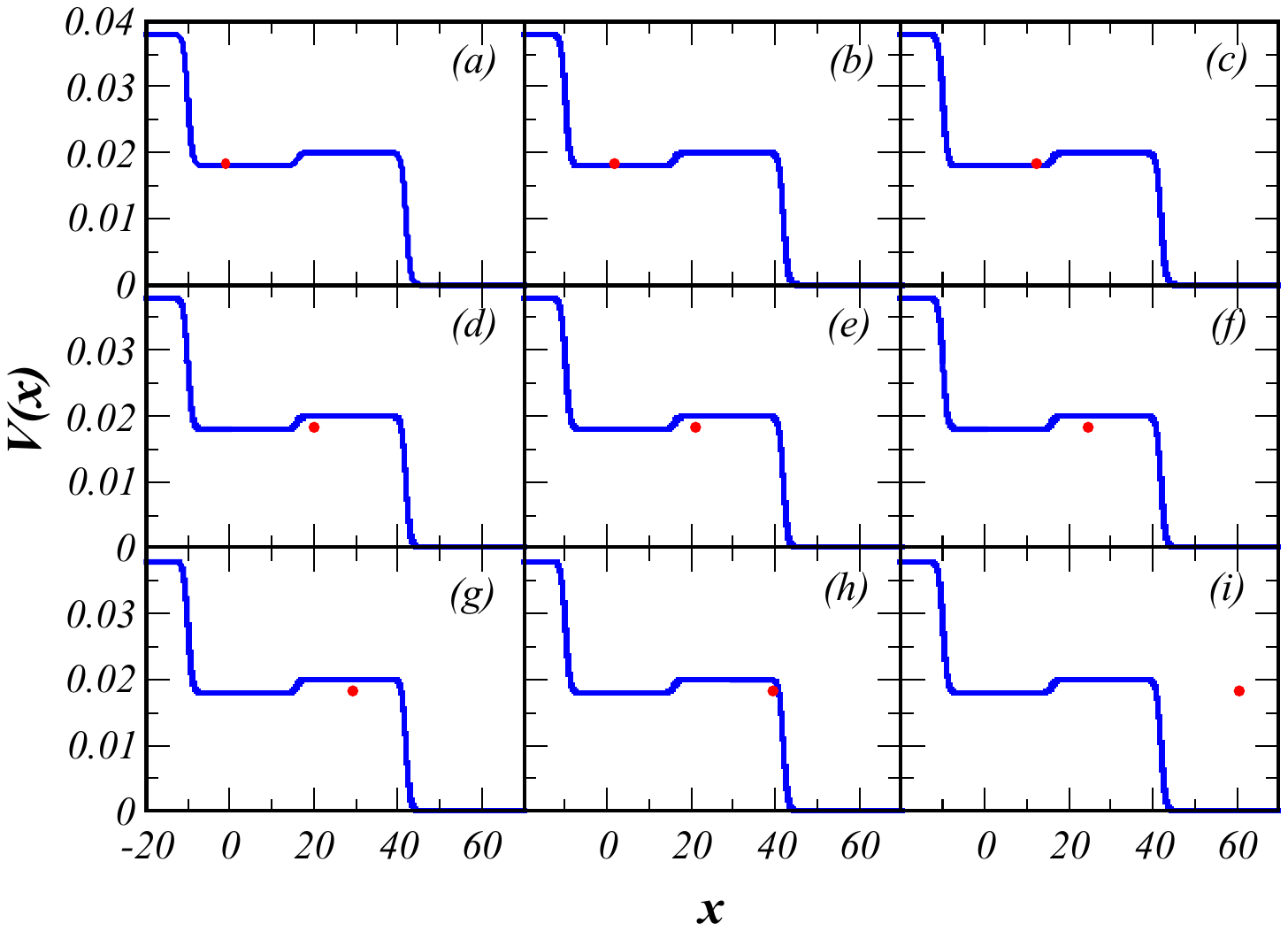}}
    \caption{Numerical simulations of the dynamics of a long-range kink, represented here with a red dot, moving in a potential with a wide barrier. The kink can tunnel through the barrier. }
    \label{Fig:12}
\end{figure}

Figure~\ref{Fig:12} shows the motion of a long-range kink in a potential $V(x)$ with a very long barrier. Conventional kinks, like those from the $\phi^4$ and sine-Gordon models, and all the other long-range kinks with shorter extents, cannot penetrate this barrier.

\subsection{Simulation parameters, equations and conditions}

\begin{enumerate}
\item 
$$\frac{\partial^2\phi}{\partial t^2} - \frac{\partial^2\phi}{\partial x^2} + 2 \sin^3\left(\frac{\phi}{2}\right) \cos\left(\frac{\phi}{2}\right)=F(x),$$
$$\phi(x,0)=2\arctan\left[\left(\frac{\sqrt{2}}{2}\right)(x+1)\right]+\pi,$$
$$\frac{\partial\phi}{\partial t}(x,0) = 0,$$
$$F(x)= 0.02\ \mbox{sech}^2(x),$$
damping coefficient $b=0$. The results are shown in Fig. \ref{5NS}(b).

\item 
$$\frac{\partial^2\phi}{\partial t^2} - \frac{\partial^2\phi}{\partial x^2} + \sin\phi=F(x),$$
$$\phi(x,0)=4\arctan\left[\exp(x+1)\right],$$
$$\frac{\partial\phi}{\partial t}(x,0) = 0,$$
$$F(x)= -0.25\ \mbox{sech}^2(x-15), \quad b=0.$$
The results are shown in Fig. \ref{10NS}(a).

\item 
$$\frac{\partial^2\phi}{\partial t^2} - \frac{\partial^2\phi}{\partial x^2} + 2 \sin^3\left(\frac{\phi}{2}\right) \cos\left(\frac{\phi}{2}\right)=F(x),$$
$$\phi(x,0)=2\arctan\left[\left(\frac{\sqrt{2}}{2}\right)(x+1)\right]+\pi,$$
$$\frac{\partial\phi}{\partial t}(x,0) = 0,$$
$$F(x)= -0.25\ \mbox{sech}^2(x-15), \quad b=0.$$
The results are shown in Fig. \ref{10NS}(b).

\item 
$$\frac{\partial^2\phi}{\partial t^2} - \frac{\partial^2\phi}{\partial x^2} + \sin\phi=F(x),$$
$$\phi(x,0)=4\arctan\left[\exp(x+1)\right],$$
$$\frac{\partial\phi}{\partial t}(x,0) = 0,$$
$$F(x)= -0.25\ \mbox{sech}^2(x+15), \quad b=0.$$
The results are shown in Fig. \ref{12NS}(a).

\item 
$$\frac{\partial^2\phi}{\partial t^2} - \frac{\partial^2\phi}{\partial x^2} + 2 \sin^3\left(\frac{\phi}{2}\right) \cos\left(\frac{\phi}{2}\right)=F(x),$$
$$\phi(x,0)=2\arctan\left[\left(\frac{\sqrt{2}}{2}\right)(x+1)\right]+\pi,$$
$$\frac{\partial\phi}{\partial t}(x,0) = 0,$$
$$F(x)= -0.25\ \mbox{sech}^2(x+15), \quad b=0.$$
The results are shown in Fig. \ref{12NS}(b).

\item 
$$\frac{\partial^2\phi}{\partial t^2} - \frac{\partial^2\phi}{\partial x^2} + \sin\phi=F(x),$$
$$\phi(x,0)=4\arctan\left[\exp(x+1)\right],$$
$$\frac{\partial\phi}{\partial t}(x,0) = 0,$$
$$F(x)= -0.25\ \mbox{sech}^2(x+20), \quad b=0.$$
The results are shown in Fig. \ref{13NS}(a).

\item 
$$\frac{\partial^2\phi}{\partial t^2} - \frac{\partial^2\phi}{\partial x^2} + 2 \sin^3\left(\frac{\phi}{2}\right) \cos\left(\frac{\phi}{2}\right)=F(x),$$
$$\phi(x,0)=2\arctan\left[\left(\frac{\sqrt{2}}{2}\right)(x+1)\right]+\pi,$$
$$\frac{\partial\phi}{\partial t}(x,0) = 0,$$
$$F(x)= -0.25\ \mbox{sech}^2(x+20), \quad b=0.$$
The results are shown in Fig. \ref{13NS}(b).

\item 
$$\frac{\partial^2\phi}{\partial t^2} - \frac{\partial^2\phi}{\partial x^2} + \sin\phi=F(x),$$
$$\phi(x,0)=4\arctan\left[\exp(x+1)\right],$$
$$\frac{\partial\phi}{\partial t}(x,0) = 0,$$
$$F(x)=-0.25\ \mbox{sech}^2(x+6) + 0.02\ \mbox{sech}^2(x) 
- 0.25\ \mbox{sech}^2(x-6), \quad b=0.$$
The results are shown in Fig. \ref{14NS}(a) and Fig. \ref{15NS} (black line).

\item 
$$\frac{\partial^2\phi}{\partial t^2} - \frac{\partial^2\phi}{\partial x^2} + 2 \sin^3\left(\frac{\phi}{2}\right) \cos\left(\frac{\phi}{2}\right)=F(x),$$
$$\phi(x,0)=2\arctan\left[\left(\frac{\sqrt{2}}{2}\right)(x+1)\right]+\pi,$$
$$\frac{\partial\phi}{\partial t}(x,0) = 0,$$
$$F(x)=-0.25\ \mbox{sech}^2(x+6) + 0.02\ \mbox{sech}^2(x) 
- 0.25\ \mbox{sech}^2(x-6), \quad b=0.$$
The results are shown in Fig. \ref{14NS}(b) and Fig. \ref{15NS} (red line).

\item 
$$\frac{\partial^2\phi}{\partial t^2} + (0.1) \frac{\partial\phi}{\partial t} - \frac{\partial^2\phi}{\partial x^2} + 0.5\, \phi\left(\phi^2 -1\right)=F(x),$$
$$\phi(x,0)=\tanh\left[0.5 (x+1)\right],$$
$$\frac{\partial\phi}{\partial t}(x,0) = 0,$$
$$F(x)= -0.1\ \mbox{sech}^2(x+10) + 0.02\ \mbox{sech}^2(x) 
- 0.1\ \mbox{sech}^2(x-10), \quad b=0.1\,.$$
The results are shown in Fig. \ref{23pNS}.

\item 
$$\frac{\partial^2\phi}{\partial t^2} + (0.1) \frac{\partial\phi}{\partial t} - \frac{\partial^2\phi}{\partial x^2} + 0.5\, \phi\left(\phi^2 -1\right)^{17}=F(x),$$
$$\phi(x,0)=\frac{2}{\pi}\arctan\left[\tanh(x+1)|x+1|^{1/8}\right],$$
$$\frac{\partial\phi}{\partial t}(x,0) = 0,$$
$$F(x)= -0.1\ \mbox{sech}^2(x+10) + 0.02\ \mbox{sech}^2(x) 
- 0.1\ \mbox{sech}^2(x-10), \quad b=0.1\,.$$
The results are shown in Fig. \ref{23NS}.

\item 
$$\frac{\partial^2\phi}{\partial t^2} + (0.3) \frac{\partial\phi}{\partial t} - \frac{\partial^2\phi}{\partial x^2} + 2 \sin^3\left(\frac{\phi}{2}\right) \cos\left(\frac{\phi}{2}\right)=F(x),$$
$$\phi(x,0)=2\arctan\left[\left(\frac{\sqrt{2}}{2}\right)(x+1)\right]+\pi,$$
$$\frac{\partial\phi}{\partial t}(x,0) = 0,$$
$$F(x)=-0.25\ \mbox{sech}^2(x+6) + 0.02\ \mbox{sech}^2(x) 
- 0.25\ \mbox{sech}^2(x-6), \quad b=0.3\,.$$
The results are shown in Fig. \ref{30NS}.

\item 
$$\frac{\partial^2\phi}{\partial t^2} - \frac{\partial^2\phi}{\partial x^2} + \sin\phi=F(x),$$
$$\phi(x,0)=2\arctan\left[\left(\frac{\sqrt{2}}{2}\right)(x+1)\right]+\pi,$$
$$\frac{\partial\phi}{\partial t}(x,0) = 0,$$
$$F(x)=-0.25\ \mbox{sech}^2(x+6) + 0.02\ \mbox{sech}^2(x) 
- 0.25\ \mbox{sech}^2(x-6), \quad b=0.$$
The mass of the normal sine-Gordon kink is $m_{SG}=8$. The mass of the initial kink used in this simulation is $m_{in}\approx 4.442$
The results are shown in Fig. \ref{32NS}.
The smaller mass does not help the tunneling. See discussion in the main text.

\item 
$$\frac{\partial^2\phi}{\partial t^2} - \frac{\partial^2\phi}{\partial x^2} + \sin\phi=F(x),$$
$$\phi(x,0)= 2\arctan\left[\tanh(x+1)|x+1|^{1/8}\right]+ \pi,$$
$$\frac{\partial\phi}{\partial t}(x,0) = 0,$$
$$F(x)=-0.25\ \mbox{sech}^2(x+6) + 0.02\ \mbox{sech}^2(x) 
- 0.25\ \mbox{sech}^2(x-6), \quad b=0.$$
The results are shown in Fig. \ref{33NS}.

\item 
$$\frac{\partial^2\phi}{\partial t^2} - \frac{\partial^2\phi}{\partial x^2} + 2 \sin^3\left(\frac{\phi}{2}\right) \cos\left(\frac{\phi}{2}\right)=F(x),$$
$$\phi(x,0)=4\arctan\left[\exp(x+1)\right],$$
$$\frac{\partial\phi}{\partial t}(x,0) = 0,$$
$$F(x)=-0.25\ \mbox{sech}^2(x+6) + 0.02\ \mbox{sech}^2(x) 
- 0.25\ \mbox{sech}^2(x-6), \quad b=0.$$
The mass of the initial kink used in this simulation is $m_{in} \approx 8$, which is much larger than the one used in the simulation shown in Fig. \ref{14NS}(b) and Fig. \ref{15NS}.
The results are shown in Fig. \ref{34NS}. See discussion in the main text.

\item 
$$\frac{\partial^2\phi}{\partial t^2} - \frac{\partial^2\phi}{\partial x^2} + 2 \sin^3\left(\frac{\phi}{2}\right) \cos\left(\frac{\phi}{2}\right)=F(x),$$
$$\phi(x,0)=2\arctan\left[\left(\frac{\sqrt{2}}{2}\right)(x+1)\right]+\pi,$$
$$\frac{\partial\phi}{\partial t}(x,0) = 0,$$
$$F(x)=-0.01\ \mbox{sech}^2(x+2) + 0.007\ \mbox{sech}^2(x) 
- 0.01\ \mbox{sech}^2(x-2), \quad b=0.$$
The results are shown in Fig. \ref{Fig:07}.

\item 
$$\frac{\partial^2\phi}{\partial t^2} + (0.2) \frac{\partial\phi}{\partial t} - \frac{\partial^2\phi}{\partial x^2} + 0.5\, \phi\left(\phi^2 -1\right)^{17}=F(x),$$
$$\phi(x,0)=\phi(x,0)=\tanh\left[0.5 (x+1)\right],$$
$$\frac{\partial\phi}{\partial t}(x,0) = 0,$$
$$F(x)= -0.1\ \mbox{sech}^2(x+10) + 0.02\ \mbox{sech}^2(x) 
- 0.1\ \mbox{sech}^2(x-10), \quad b=0.2\,.$$
The results are shown in Fig. \ref{Fig:09}.

\item 
$$\frac{\partial^2\phi}{\partial t^2} + (0.01) \frac{\partial\phi}{\partial t} - \frac{\partial^2\phi}{\partial x^2} + 0.5\, \phi\left(\phi^2 -1\right)^{3}=F(x),$$
$$\phi(x,0)=\frac{2}{\pi}\arctan(x+1),$$
$$\frac{\partial\phi}{\partial t}(x,0) = 0,$$
$$F(x)=-0.03\ \mbox{sech}^2(x+3.5) + 0.005\ \mbox{sech}^2(x-25) 
- 0.03\ \mbox{sech}^2(x-53.5), \quad b=0.01,\; n=2.$$
The width of the barrier is $L_B=28.5\,.$ The results are shown in Fig. \ref{Fig:10} (blue dashed line).

\item 
$$\frac{\partial^2\phi}{\partial t^2} + (0.01) \frac{\partial\phi}{\partial t} - \frac{\partial^2\phi}{\partial x^2} + 0.5\, \phi\left(\phi^2 -1\right)^{17}=F(x),$$
$$\phi(x,0)=\frac{2}{\pi}\arctan\left[\tanh(x+1)|x+1|^{1/8}\right],$$
$$\frac{\partial\phi}{\partial t}(x,0) = 0,$$
$$F(x)=-0.03\ \mbox{sech}^2(x+3.5) + 0.005\ \mbox{sech}^2(x-25) 
- 0.03\ \mbox{sech}^2(x-53.5), \quad b=0.01,\; n=9.$$
The width of the barrier is $L_B=28.5\,.$ The results are shown in Fig. \ref{Fig:10} (red solid line).

\item 
$$\frac{\partial^2\phi}{\partial t^2} + (0.01) \frac{\partial\phi}{\partial t} - \frac{\partial^2\phi}{\partial x^2} + 0.5\, \phi\left(\phi^2 -1\right)^{9}=F(x),$$
$$\phi(x,0)=\frac{2}{\pi}\arctan\left[\tanh(x+1)|x+1|^{1/4}\right],$$
$$\frac{\partial\phi}{\partial t}(x,0) = 0,$$
$$F(x)=-0.03\ \mbox{sech}^2(x+3.5) + 0.005\ \mbox{sech}^2(x-25) 
- 0.03\ \mbox{sech}^2(x-53.5), \quad b=0.01,\; n=5.$$
The width of the barrier is $L_B=28.5\,.$ The results are shown in Fig. \ref{Fig:11}.

\item 
$$\frac{\partial^2\phi}{\partial t^2} + (0.01) \frac{\partial\phi}{\partial t} - \frac{\partial^2\phi}{\partial x^2} + 0.5\, \phi\left(\phi^2 -1\right)^{9}=F(x),$$
$$\phi(x,0)=\frac{2}{\pi}\arctan\left[\tanh(x+1)|x+1|^{1/4}\right],$$
$$\frac{\partial\phi}{\partial t}(x,0) = 0,$$
$$F(x)=-0.03\ \mbox{sech}^2(x+10) + 0.005\ \mbox{sech}^2(x-16) 
- 0.03\ \mbox{sech}^2(x-42), \quad b=0.01,\; n=5.$$
The potential well is defined between the points $x=-10$ and $x=16$. The barrier is defined between the points $x=16$ and $x=42$. The results are shown in Fig. \ref{Fig:12}. The masses of all the kinks used in all simulations are shown in Tables~\ref{Tab:01} and \ref{Tab:02}, inside Section \ref{Sec:The models}.

\end{enumerate}

\section{Kink propagation in disordered media}
\label{Sec:Disorder}

Consider the heterogeneous kink-bearing nonlinear partial differential equation \eqref{Eq:03}, with
\begin{equation}
    \label{Eq:38}
    F(x)=\sum_{i=1}^{N}\alpha_i\ \mbox{sech}^2\left[\beta_i(x-x_i)\right],
\end{equation}
where $\alpha_i$, $\beta_i$ and $x_i$ can be random numbers. We assume that these parameters do not take values that lead to instabilities of the shape modes of kinks so that the kink will conserve its integrity as an entity. Thus, Eq.~\eqref{Eq:03} can describe the dynamics of a kink in a disordered system. 

If the condition of Eq.~\eqref{Eq:36} is satisfied for every basic block structure containing a negative local minimum, a positive local maximum, and another negative minimum (as in Fig.~\ref{Fig:Illustrating}), the long-range kink can pass freely through the disordered zone as if there are no obstacles there.

\begin{figure}
    \centering
    \scalebox{1}{\includegraphics{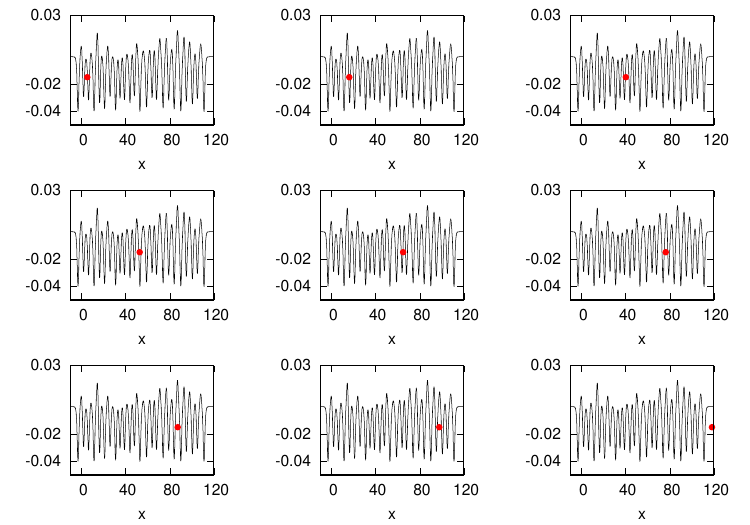}}
    \caption{Numerical simulation of a long-range kink moving through an array of impurities.}
    \label{Fig:last}
\end{figure}

Figure~\ref{Fig:last} shows a long-range kink ($n=5$) moving through an array of impurities. This kink moves almost with constant velocity. Remarkably, the kink does not feel any obstacle in its way.

\section{Applications: Experiments and technology}
\label{Sec:Applications}

Our results have applications in spintronics, Josephson junction circuits, new superconductivity technology, quantum information technology, and quantum computing. In Ref.~\cite{Brooke2001}, the experimental macroscopic quantum tunneling of a domain wall is reported. They discuss their results in the context of the quantum tunneling escape rate
\begin{equation}
    \label{Eq:41}
    \Gamma=f_m\,e^{-2W\sqrt{\frac{2mE_0}{\hbar^2}}},
\end{equation}
where $W$ is the width of the barrier, $E_0$ is the height of the barrier, and $m$ is the domain wall mass. The potential is similar to that shown in Fig.~\ref{Fig:07}. When we carefully study the original data from Ref.~\cite{Brooke2001}, we observe that the domain wall behaves like our long-range kink. They report on a disordered magnetic system for which it is possible to adjust the parameters of the system with a knob in the laboratory. The authors can control the domain wall mass, the extent of the kink, and many other physical quantities. When they change some fields, they can broaden the domain walls (at some critical values of the parameters, the domain wall can fill the entire system, and its mass approaches zero). If suitable experimental conditions are introduced into $LiH_0F_4$, the domain wall dynamics can pass from classical (large $m$) to quantum mechanical (small $m$). They can observe the trapped domain walls at the fixed pinning centers (the disorder can be ideal for pinning domain walls). The kink tunneling mechanism can serve to depin the domain walls despite the existence of the disorder. As parameters that increase the extent of the kink and decrease the domain wall mass are changed appropriately, the measured tunneling rate has been observed to increase. The general mobility of the domain wall through the disorder will increase as the extent of the kink is increased exactly in the same way as our long-range kinks.

The superconductors, the superinsulators, and the transitions that lead to these states of matter have been active areas of research for the last several decades \cite{Rotschild2006, Grimshaw2022, Arkhipov2022, Hermon1996, SanchezManzano2022, AharonovBohm1959, Delsing1994,  Haviland1996, Sambandamurthy2004, Sambandamurthy2005, Dubi2007, Baturina2007, Fistul2008, Haviland1989}. Nevertheless, some fundamental questions remain unsolved. In particular, the nature of the insulating phase, which shows unconventional transport properties, is unclear. One exciting scenario suggests that this phase incorporates superconducting fluctuating islands embedded in an insulating matrix. These phenomena lead to very exciting physics governed by the interplay between disorder and superconductivity.

Josephson junction arrays, i.e., networks of superconducting islands weakly coupled by tunnel junctions, are among the physical systems used to study the aforementioned phenomena. One of the parameters controlling the superconductor-insulator transition in Josephson arrays is the ratio $E_J/E_C$, where $E_J$ is the Josephson coupling energy, and $E_C$ corresponds to the energy needed to add an extra charge to a neutral island. Another relevant parameter is the dimensionless resistance $R/R_Q$, where $R_Q=h/(2e)^2$. In addition, $L$ will be the length of the system. Several works in the literature have studied Josephson arrays, where superconducting grains (or islands) are connected through tunnel junctions. The dynamics of excess cooper pairs in the array are described using charge kinks created by the polarization of the grains \cite{Hermon1996}.

Let us consider a ring-shaped 1-D array of serially coupled Josephson junctions biased by an external flux through its center. Typically, the width of the junction is $d\simeq2\hbox{nm}$. Using results from Ref.~\cite{Hermon1996} in combination with our findings in this work, we can make the following statement. If the kink extent is larger than the circumference of the array, quantum phenomena such as persistent currents are predicted. As we estimate that the characteristic kink extent in non-local Josephson arrays can be of the order of several millimeters, this is a macroscopic quantum phenomenon. This is just another manifestation of long-range quantum kinks moving smoothly through different media. A long-range Cooper-pair charge kink with a very large extent implies an extremely long-range Josephson coupling. This could explain the recently measured experimental data where extremely long-range Josephson coupling was observed \cite{SanchezManzano2022}. In some situations, the charge kink becomes lighter than the plasmons \cite{Hermon1996}. Recall that the mass of the kink depends on its extent. In that case, the kink takes the role of the fundamental quantum of the system. However, it loses its correspondence to the classical particle configuration. In a sense, the charge kinks become ``too quantum''. Thus, any classical theory cannot describe the long-range charge kink (with a large kink extent and very small mass). All the processes linked to the kink will be very macroscopic quantum phenomena. The persistent current oscillations phenomenon manifests the Aharonov-Bohm effect, where a charged particle encircles a flux tube \cite{AharonovBohm1959}. Weak spatial heterogeneity in the array, e.g., non-identical grains of junctions or disordered grains, leads to persistent current oscillations as a function of the external flux. The amplitude of the persistent current of one charge kink decreases as the heterogeneity increases. This ``damping'' of the current can be minimized if the charge kink is long-range. Under certain conditions, the long-range Cooper pair charge kink would not feel the heterogeneity.

Several experiments \cite{Haviland1989, Delsing1994, Hermon1996, Haviland1996, Sambandamurthy2004, Sambandamurthy2005, Dubi2007, Baturina2007, Fistul2008} have shown that in thin superconducting films, disorder creates a droplet-like electronic texture (superconducting islands immersed into a standard matrix). Tuning disorder can drive the system from superconducting to insulating behavior. Consider an array of superconducting islands, each coupled to its nearest neighbors by Josephson weak links.  Several parameters are relevant for the quantum phase transitions: the Josephson coupling energy of the two adjacent superconducting islands ($E_J$), the charging energy ($E_C$), the length of the sample $L$, and the dimensionless conductance
\begin{equation}
    \label{Eq:42}
    g=\frac{2\pi\hbar}{e^2R}.
\end{equation}

The inequalities $E_J\gg E_C$ and $g\gg1$ lead to the insulator-superconductor transition. The activation of Cooper pair charge kinks mediates the charge transfer. According to some experimental and theoretical evidence \cite{Haviland1989, Delsing1994, Hermon1996, Haviland1996, Sambandamurthy2004, Sambandamurthy2005, Dubi2007, Baturina2007, Fistul2008} and references quoted therein and our results in this work, the long-range Cooper-pair charge kink can spread over several superconducting islands, and the long-range Cooper pair kinks can be the ultimate charge carriers. Thus, they can lead to a transition to superconductivity. The extent of the kink is proportional to the screening length in the system. In some cases, they are equal. At the same time, much of the picture we have about these materials holds, provided the screening length in the system exceeds the sample size. So, to satisfy these conditions, the extent of the kink must exceed the sample size. Hence, this must be a very long-range Cooper-pair charge kink. The Cooper pair is transferred across the effective junction in the form of a charge kink spread over the array \cite{Mooij1990}. Only a long-range kink can be spread over the whole array. So, the Cooper-pair charge kink must be a long-range kink.

The conditions for the superconducting and insulator states can be formulated in terms of the super propagation or localization of the Cooper-pair charge kinks. If the disorder exponentially localizes the charge kinks, we observe the superinsulator state \cite{Baturina2007}. On the other hand, if the charge kinks undergo a regime of super propagation, where the kinks do not feel the disorder, then we can observe the superconducting state. Consider a Josephson junction array as that presented in Ref.~\cite{Hermon1996}. Our prediction is the following. For $\pi(E_J/E_C)^{1/2}=2.5$, we start with a kink with a very small extent ($L_{SE}\gtrapprox0$). Then, we keep increasing the extent of the kink until $L_{SE}\approx L$. This process will lead to an insulator-superconductor transition. We estimate that this will hold even in the presence of disorder and dissipation. The inverse process will lead to a superconductor-insulator transition.

\section{Discussion}
\label{Sec:Discussion}

Our results show that the solitary waves are long-range kinks when one of the following conditions is fulfilled:
\begin{itemize}
    \item The potential $\mathcal{U}(\phi)$ possesses the property $\left.\mbox{d}^2\mathcal{U}/\mbox{d}\phi^2\right|_{\phi=\phi_{1,3}}\simeq0$.

    \item Nonlocal electrodynamics rules in the physical system.

    \item The interaction between the elements of the lattice is described by a Kac-Baker potential.

    \item The extent of the soliton is of the same order of magnitude as the length of the system or sample.
\end{itemize}
This means these kink-like objects can move relatively easily, even in a medium with impurities and heterogeneity. For example, if these kinks are charged objects, this enhanced propagation can lead to an enormous increase in conduction. 

DNA kinks have been shown to exhibit long-range interactions \cite{Mello1998, Gonzalez1994, Gonzalez2002}. Thus, they can have increased mobility. This effect can have biological relevance.

There are magnetic systems where long-range kinks have been observed, and recently, long-range skyrmions have been studied \cite{Kosevich1990, Shnir2021}. In Ref.~\cite{Wallraff2003}, the authors report the experimental observation of macroscopic quantum tunneling of a fluxon in a long Josephson junction. These findings have been controversial \cite{Gronbech-Jensen2004, Blackburn2016}. Our results show that a normal fluxon can tunnel through a barrier with a width of the order of $20\mu\ \mbox{m}$-$100\mu\ \mbox{m}$. On the other hand, long-range fluxons supported on the non-local electrodynamics of some ultranarrow Josephson junctions can cross much wider barriers ($L_B\simeq5\mbox{mm}$). 

Several experimental works have reported the giant enhancement of the quantum tunneling rate in stacks of Josephson junctions \cite{Jin2006, Koyama2008}. Some researchers have invoked the presence of non-local electrodynamics. However, these works have failed to use the fact that these fluxons are long-range kinks. Our findings indicate that there is still much more room for improving the macroscopic tunneling rate in these stacks of Josephson junctions. Ordinary point particles and kinks cannot cross very wide barriers. The fact that standard quantum tunneling can only be observed in systems with very narrow barriers has been used to discard quantum effects as an explanation of some transport measurements in charge density wave materials \cite{RojoBravo2016}. The argument is due to experimental evidence, where quantum current oscillations have been observed in samples as long as several millimeters or even a couple of centimeters. Our results demonstrate that long-range quantum kinks can tunnel through very wide barriers or very long structures filled with obstacles. 

Charged Cooper-pair kinks have been subject to intense experimental study. If we can engineer long-range Cooper-pair kinks, this could help the development of new materials for superconducting technology. In addition, the recently reported extremely long-range Josephson coupling \cite{SanchezManzano2022} could indicate that long-range kinks can play an important role in modern quantum technology.

The exact condition for an effective translational symmetry would be
\begin{equation}
    \label{Eq:35Eff}
    \int_{-\infty}^{\infty}\mbox{d}x\,F(x)\phi_s(x-x_0)=\mbox{constant}.
\end{equation}
However, condition~\eqref{Eq:35Eff} does not need to be satisfied exactly. That is, exact tuning is not necessary. Condition~\eqref{Eq:35Eff}  can also be replaced by
\begin{equation}
    \label{Eq:36Eff}
    \frac{\mbox{d}E_p(x_0)}{\mbox{d}x_0}\simeq0.
\end{equation}
When the long-range kink tunneling condition is satisfied for every structure with a negative minimum, a positive maximum, and another negative minimum in $F(x)$, the kink behaves as if there are no obstacles in its way. The values of the parameters of $F(x)$ for which this phenomenon can occur form a continuum set of points in the parameter space. This phenomenon is robust.

\begin{figure}
    \centering
    \scalebox{0.8}{\includegraphics{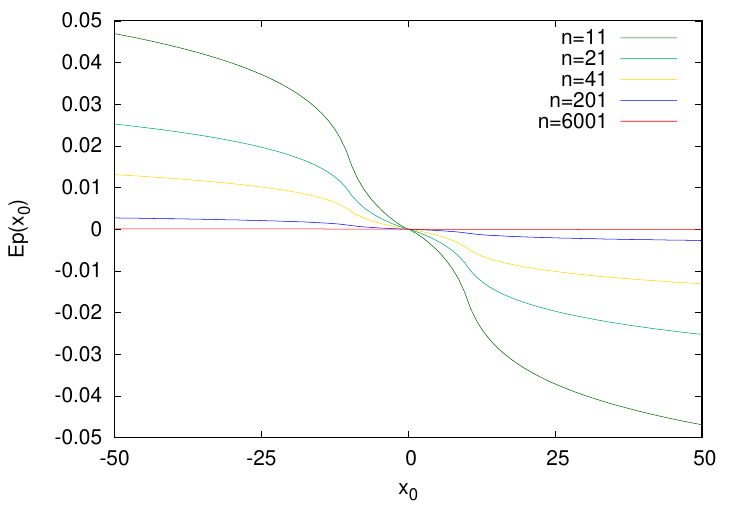}}
    \caption{Effective translational symmetry. Potential energy for a long-range kink with $n\gg1$ in a medium with five localized impurities. Here, $x_0$ is the position of the kink.}
    \label{Fig:TrasnlationalSymmetry}
\end{figure}

Effective translational symmetry means that flat potential energy is not required. The function $F(x)$ in Fig.~\ref{Fig:last} is an example that describes a completely disordered medium. The tunneling condition is satisfied for every region containing two minimums and one maximum. However, the exact condition~\eqref{Eq:35Eff} does not hold. Nevertheless, the long-range kink moves approximately with constant velocity from point $x=25$ to point $x=100$, an interval where the medium is completely disordered. Figure~\ref{Fig:TrasnlationalSymmetry} shows the effective potential energy for a long-range kink with $n\gg 1$ moving in a medium with five localized impurities located at points $x=-6$, $x=-2$, $x=0$, $x=2$, and $x=6$. This is a striking example of an almost exactly-satisfied condition for ``effective translational symmetry''.

Combining the BPS property with the long-range property can lead to perfect long-range kink propagation. For instance, consider a real system where the BPS condition is approximately satisfied. Consider also that the system supports the existence of long-range kinks. Then, the kink may move freely through the medium despite the impurities. Moreover, the low-velocity adiabatic motion condition can be dropped.


\section{Conclusions}
\label{Sec:Conclusions}

We have investigated very general perturbed Klein-Gordon equations that can support kink-like solutions. We have found general conditions for the existence of long-range kinks. External fields and heterogeneity in the parameters of the system can create potential wells and barriers for the motion of the kinks. Under certain conditions, the kink can tunnel through the barrier even when the initial energy of the kink is less than the height of the energy barrier. When the moving object is a long-range kink, it can cross very wide barriers. Some long-range kinks can move in a completely disordered medium as if there are no obstacles there. The phenomena reported in this work can be observed and used to design new technologies in several physical systems, like conducting polymers, charge density waves, long Josephson junctions, Josephson junction arrays, DNA, proteins, superconductors, superinsulators, topological superconductors, and insulators, among others.

\begin{acknowledgments}
This work started when J.F.M. was a lecturer and researcher at Laboratoire Gulliver, École Supérieure de Physique et de Chimie Industrielles (ESPCI-PSL), Paris, France. J.F.M. thanks Olivier Dauchot, Alexandre Allauzen, and Laboratoire Gulliver for their hospitality while part of this research was done. Project supported by the Competition for Research Regular Projects, year 2023, code LPR23-06, Universidad Tecnológica Metropolitana. L.E.G. acknowledges financial support from the Venezuelan Mincyt and FONACIT through Project CFP 2024000323.   
\end{acknowledgments}



\providecommand{\noopsort}[1]{}\providecommand{\singleletter}[1]{#1}%

\end{document}